%% file: main.tex
\documentclass[onecolumn]{aastex702}
\usepackage{amsmath,booktabs,longtable,array,placeins,needspace,float}
\hypersetup{hypertexnames=false}
\newcommand{\obj}{3I/ATLAS}
\newcommand{\afrho}{A(0^\circ)f\rho}

\renewcommand{\today}{October 20, 2026}
\shorttitle{Keck and Gemini monitoring of 3I/ATLAS}
\shortauthors{Bolin et al.}
\journalinfo{}
\graphicspath{{figures/}}
\begin{document}
\raggedbottom
\title{Comprehensive Visible and Near-infrared Keck and Gemini Monitoring and Characterization of Interstellar Comet 3I/ATLAS during its Encounter with the Solar System}
\correspondingauthor{Bryce T. Bolin}
\input{authors.tex}

\begin{abstract}
We present 14 optical imaging and eight optical spectroscopic epochs of interstellar comet \obj\ from Keck/LRIS and Gemini/GMOS, two near-infrared (NIR) imaging epochs from Keck/MOSFIRE, and four NIR spectroscopic epochs from Gemini/Flamingos-2 and GNIRS, obtained between 2025 July and 2026 May. Mean GMOS colors of $g-r=0.732\pm0.029$, $r-i=0.237\pm0.025$, and $i-z=0.109\pm0.022$ mag show no significant change across perihelion. Phase-corrected red-band activity follows $Af\rho\propto r_h^{-2.10\pm0.14}$ inbound and $r_h^{-3.34\pm0.43}$ outbound. Transverse-width and sunward-plume speeds imply dust losses of $2.7\times10^9$ and $1.0\times10^{10}$ kg over 2025 July 3--2026 May 15 for 1-$\mu$m--1-mm grains, equivalent to global layers of 0.26 and 0.99 m under the adopted grain and nucleus assumptions. The concentrated coma, compact sunward feature, and antisolar tail seen on 2025 August 27 broadened by 2025 November 26; extended dust persisted through 2026 February 17. C$_2$/CN nearly doubled from $0.257\pm0.014$ on 2025 September 14--15 to $0.481\pm0.080$ on 2025 November 26, while C$_3$/CN fell from $0.327\pm0.024$ on 2025 August 28 to $0.214\pm0.009$ on 2025 September 14--15. MOSFIRE NIR imaging on 2025 July 11 gives $Y-J=0.466\pm0.143$, $J-H=0.239\pm0.052$, $H-K_s=0.454\pm0.150$, and $J-K_s=0.693\pm0.202$ mag. Spectra give $J-H=0.24$ mag on 2025 September 5--6 and $J-H=0.27$--0.28 and $J-K_s=0.21$--0.22 mag on 2025 December 1--2 over the retained passband intervals. Model-calibrated imaging on 2026 February 6 gives near-solar $J-K_s=0.303\pm0.051$ mag. No water-ice absorption is robust to the tested continuum and calibration choices in the NIR spectra. These observations provide a lasting record of the comet's evolution through the Solar System and a benchmark for studying and monitoring future interstellar objects.
\end{abstract}

\section{Introduction}
Interstellar objects (ISOs) carry material formed in other planetary systems into the Solar System \citep{FitzsimmonsInterstellarObjects2024}. Their ejection links this population to planetary scattering and the solid inventory of their parent systems \citep{Raymond2018Ejection,MoroMartin2018Origin,Rice2026Ejection,Lee2026Mass}. Radial drift, migrating ice lines, and volatile redistribution affect the material available for ejection \citep{Yunerman2026Disks}; white-dwarf pollution provides independent evidence for volatile-rich extrasolar planetesimals \citep{Sahu2025Icy}. Irradiation modifies ISO surfaces, while solar heating releases gas and dust that can obscure the nucleus \citep{Fitzsimmons2018Oumuamua,Guzik2020Borisov}. Observations across an encounter help separate this evolution from compositional differences, although size, activity, and survey sensitivity bias the small detected sample.

1I/`Oumuamua showed a red continuum and a large-amplitude, irregular lightcurve consistent with a strongly nonspherical body in non-principal-axis rotation \citep{Meech2017,Bolin2018,Drahus2018Tumbling}. Its featureless optical/NIR spectrum resembled primitive Solar System surfaces \citep{Fitzsimmons2018Oumuamua}, and its nongravitational acceleration occurred without a resolved dust coma \citep{Micheli2018Oumuamua}. 2I/Borisov had a coma and tail with comet-like optical colors \citep{JewittLuu2019Borisov,Guzik2020Borisov,Bolin2020Borisov}. Although the coma complicated nucleus and rotation measurements \citep{BolinLisse2020}, \textit{HST} profiles constrained its radius to $<0.5$ km for geometric albedo 0.04 \citep{Jewitt2020BorisovNucleus}. Ultraviolet and millimeter spectra revealed high CO/H$_2$O abundance \citep{Bodewits2020BorisovCO,Cordiner2020BorisovCO}. Optical--millimeter observations favored compact grains exceeding approximately 1 mm and found evolution in CO/H$_2$O across perihelion \citep{Yang2021BorisovPebbles}. Ni I was also detected in its cold coma \citep{GuzikDrahus2021Ni}. Comparing the surface-dominated spectrum of 1I with 2I's coma requires accounting for grain size, phase angle, and activity.

\citet{Denneau2025} announced the ATLAS discovery of \obj, followed by early characterization by \citet{Bolin2025Discovery} and \citet{Seligman2025Discovery}. The comet showed a red continuum and resolved coma, with weak or undetected canonical gas bands near discovery \citep{Opitom2025MUSE,Puzia2025SOAR,AlvarezCandal2025XSHOOTER,FuenteMarcos2025GTC}. Rubin observations cover 2025 June 21--July 20 \citep{Chandler2026Rubin}; ATLAS resolves changes in the inbound growth law \citep{Tonry3IATLASPhotometry2025}, and TESS, HATPI, and smaller telescopes extend the inbound baseline \citep{MartinezPalomera2025TESS,Hartman2026HATPI,Wesolowski2026Photometry}. Optical monitoring tracks the changing continuum, morphology, and gas emission \citep{Belyakov2025Optical,Rahatgaonkar2025,Beniyama2025,SantanaRos2026,Ahuja2026}, including the CN and Ni coma observed with UH88/SNIFS and Keck/KCWI \citep{Hoogendam2026SNIFS,Hoogendam2025KCWI,Hoogendam2026Post}. JCMT constrains parent molecules \citep{Hinkle2026JCMT}, while infrared observations probe ice, volatiles, and dust mineralogy \citep{Yang2025,Belyakov2026Volatiles,Belyakov2026Dust}. The red visible continuum flattens in the NIR \citep{Medler2026,Yang2025}, becoming neutral or slightly blue in IRTF spectra \citep{Kareta2025IRTF}. Polarimetry reveals an unusually deep, narrow negative branch \citep{Gray2025Polarization}.

We use Keck and Gemini imaging and spectroscopy before and after perihelion to measure \obj's morphology, colors, dust activity, and gas production and constrain water ice. Comparisons use a common physical aperture, phase correction, and orbital-plane geometry, following the 2I/Borisov and P/2019 LD$_2$ studies \citep{Bolin2020Borisov,Bolin2021LD2}. Prediscovery observations of 2I demonstrate the value of extending the baseline beyond efficient water sublimation \citep{Ye2020Borisov}. We compare our measurements with preliminary reports \citep{ATel17363,ATel17503,ATel17561,ATel17719,ATel17841,ATel18098}; the December 22 GMOS spectra also appear in Appendix A of \citet{Belyakov2026Dust}.

\section{Observations}\label{sec:observations}
The campaign comprises 14 optical imaging epochs (13 GMOS, one LRIS), eight optical spectroscopic epochs (seven GMOS, one LRIS), two MOSFIRE imaging epochs, and four NIR spectroscopic epochs (two Flamingos-2, two GNIRS; Table~\ref{tab:observations}). These counts include excluded observations and treat each session, including the August 27--28 imaging and other sequences crossing UTC midnight, as one epoch. Alternative extractions, orders, and combinations share exposures.

\subsection{Optical observations}\label{sec:optical_observations}
GMOS-S observations were obtained under GS-2025B-DD-103 (PI: Elena Sabbi) on 2025 August 27--28 and GS-2025B-DD-102 (PI: Bryce T. Bolin) on September 5, 6, 8, and 14--15. GMOS-N observations used GN-2025B-DD-102 (PI: Bryce T. Bolin) on 2025 November 26, December 19 and 22, and 2026 January 20--21 and February 5 and 7; February 17 and March 27 used GN-2025B-FT-113 (PI: Bryce T. Bolin).\footnote{Program and PI assignments follow exposure records in the Gemini Observatory Archive, \url{https://archive.gemini.edu/}.} LRIS spectroscopy used Keck program 2025B W353 (PI: Jeffrey Cooke); imaging on 2026 May 15 UTC used the target-of-opportunity program 2026A N041 (PI: Ian Wong). Liverpool IO:O data \citep{Graykowski2026} were obtained under PATAG AZ22A01 (PI: F. Romanov).

We obtained $u,g,r,i,Z$ images with GMOS-S, $g,r,i,Z$ with GMOS-N, and $G,R,I$ with LRIS. The GMOS setup resembles those used for 2024 YR$_4$ and 2024 PT$_5$ \citep{Bolin2025YR4,Bolin2025PT5}. GMOS long-slit spectra used 540- and 550-nm central settings. We evaluate the 2025 September 1 LRIS spectroscopic geometry at 00:00 UTC because reduced headers lack exposure timestamps; LRIS imaging uses mean exposure midpoints.

The electronic GMOS log contains timestamps and airmasses for 148 exposures contributing to 49 date/filter stacks. We use exposure-weighted midpoints, assigning parent-stack midpoints where subset membership is unknown. The 14 spectral settings retain midpoints, integration times, and slit position angles. GMOS-S used B480+\_G5327 and GMOS-N B480+\_G5309, both in first order with open filters and a 1-arcsec slit. Their spatial scales are 0.1600 and 0.1614 arcsec pixel$^{-1}$, respectively. The 34 spectral exposures total 10,180 s.

The 2025 August 27 Gemini South run included the public Shadow the Scientists (StS) session ``A Close Encounter with an Interstellar Visitor from Gemini South.''\footnote{\href{https://noirlab.edu/public/news/noirlab2525/}{NSF NOIRLab report}, published 2025 September 4.} The 2025 November 26 GMOS-N observations accompanied ``Exploring Comet 3I/ATLAS: A Rare Encounter with an Interstellar Visitor,'' led by B.T.B., Laura-May Abron, and Matthew Belyakov.\footnote{\href{https://www.maunakeaobservatories.org/events/partner-event-exploring-comet-3iatlas-a-rare-encounter-with-an-interstellar-visitor}{Maunakea Observatories announcement}.}

\paragraph{Imaging reduction and calibration}
The GMOS count-rate stacks are comet-aligned on a north-up, east-left grid with 0.16-arcsec pixels (Figures~\ref{fig:optical_images} and \ref{fig:optical_north}). Science, variance, quality, exposure-count, and exposure-map extensions are retained. Each calibrated derivative uses its own exposure time. Diagonal variances omit interpolation, background, and calibration covariance. Field-source minor-axis FWHM spans 0.63--1.68 arcsec for GMOS-S and 0.68--0.97 arcsec for GMOS-N in the measured subset.

GMOS $gri$ calibration uses PS1 stars evaluated at $g-r=0.6$. Comet-specific passband corrections are unmeasured; $u$ and $Z$ also have sparse-standard or catalog-transformation limitations. Native calibrations are retained electronically; accepted measurements are transformed to approximate SDSS AB equivalents below. Calibration or aperture-quality failures exclude March and affected February images.

Saturated-flat flags and repeated resampling produced patches in the September 5, 6, 8, and 14 GMOS-S $r$ displays. Figures~\ref{fig:optical_images} and \ref{fig:dust} use aligned instrumental stacks, with valid-pixel averages in $3\times3$ bins (0.48 arcsec) on September 5, 8, and 14. Bins with fewer than three valid pixels remain black. September 6 uses a mask-normalized Gaussian average with $\sigma=0.40$ arcsec (FWHM 0.94 arcsec) and at least 5\% valid kernel weight. These bilinearly rendered displays have nonuniform coverage; profile measurements and Figure~\ref{fig:enhancement} use unsmoothed data.

GMOS intensity images, dust-model backgrounds, and February MOSFIRE images use a median background from 20--29 arcsec and an inverse-hyperbolic-sine stretch. The upper level is $I_{\rm upper}=\max(1.05I_{\rm peak},8\sigma_{\rm sky})$, with $I_{\rm peak}$ measured within 3 arcsec and $\sigma_{\rm sky}$ the background scatter; softening is $\max(2\sigma_{\rm sky},0.005I_{\rm upper})$. Figure~\ref{fig:enhancement} uses a fixed ratio scale with unity in white.

The LRIS imaging setup resembles those used for 1998 KY$_{26}$ and Neptunian Trojans \citep{Bolin2025KY26,Bolin2023Trojans}. LPipe \citep{PerleyLPipe2019} reductions were registered and summed. May data use approximate PS1 $g,i$ AB and Cousins $R$ Vega calibrations. Photometry uses a 1.7765-arcsec radius (10,000 km) and 4--6-arcsec sky annulus. Blank apertures receive identical sky subtraction and exclude masked or uncovered pixels. Sky-variance, saturation-mask, and exposure maps support quality checks, although covariance is incomplete. Adopting the larger of the blank-aperture scatter and saved magnitude error gives native $g,R,i$ uncertainties of 0.121, 0.133, and 0.197 mag. Shared color-term propagation gives SDSS $g,r,i$ errors of 0.140, 0.116, and 0.197 mag, excluding the unmeasured comet-specific correction.

\paragraph{Spectroscopic reduction and calibration}
GMOS solar-analog ratios use air wavelengths and normalization over 545--555 nm. We adopt the 40-pixel full-width extraction, testing alternative apertures and sky regions. Gas calculations use the full rectangular extraction dimensions.

LRIS blue/red pairs provide all-exposure, coma, and coma--tail spectra. We convert native $F_\nu$ to $F_\lambda$ before division by the CALSPEC synthetic solar spectrum \citep{Bohlin2014}. The 320--560- and 560--1000-nm arms meet at 560 nm. Robust linear continua fitted to 1-nm median bins over 545--559 and 561--575 nm, excluding 556--558 nm and immediate edges, give an all-extraction red-arm multiplier $k_R=1.1915\pm0.0287$. Its uncertainty comes from residual bootstraps in contiguous 2-nm blocks; 10-, 15-, and 20-nm fitting neighborhoods give 1.1771--1.2160. We multiply the red arm by $k_R$ and normalize both solar-divided arms over 545--555 nm. Original and adjusted fluxes retain the shared scale uncertainty. Air and vacuum solar templates are tested separately because the native wavelength convention is undocumented. Absolute calibration and wavelength-dependent slit losses remain sources of uncertainty in the arm match.

LRIS used 400/3400 and 400/8500 dispersers. We adopt a 0.7-arcsec slit for production rates, although reduced headers do not verify its width. J.C.'s reduction notes identify three tail-aligned exposures in coma--tail and two parallactic-angle exposures in coma, combined into the all-exposure spectrum. A fourth tail-aligned exposure was rejected for possible stellar contamination. Numerical slit position angles are unavailable. G191B2B provides flux calibration; the observed solar twin HD150469 was not divided from the delivered spectra. The 70-pixel extraction-boundary separation corresponds to 9.45 arcsec at the assumed unbinned blue scale of 0.135 arcsec pixel$^{-1}$. Unknown binning motivates a doubled-height sensitivity test.

\subsection{Infrared observations}\label{sec:nir_observations}\label{sec:mosfire}
MOSFIRE observations on 2025 July 11 UTC used Keck engineering program 2025A E498 (PI: Josh Walawender); 2026 February 6 UTC used 2026A N045 (PI: Ian Wong).\footnote{Keck program assignments follow the public schedule, whose night dates are in Hawaii Standard Time: \url{https://www2.keck.hawaii.edu/observing/keckSchedule/}.} Flamingos-2 observations on 2025 September 3 and September 5--6 used GS-2025B-FT-206, and GNIRS on December 1--2 used GN-2025B-FT-110 (both PI: Bryce T. Bolin).

MOSFIRE broadband setups resemble the $Y,J$ observations of 2024 YR$_4$ \citep{Bolin2025YR4} and $Y,J,H,K$ observations of 2024 PT$_5$ \citep{Beniyama2025PT5}. July data cover $Y,J,H,K_s$ and February $J,K_s$. Complete-aperture individual exposures provide July $Y-J$ through sidereal calibration transfer. Flamingos-2 obtained JH and HK spectra; GNIRS obtained two cross-dispersed blocks on each night. July MOSFIRE geometry uses an exposure-weighted midpoint; February uses the first retained MJD-OBS because no stack midpoint is available.

NIR exposure start times and airmasses are identified in Table~\ref{tab:observations}. GNIRS used the 32/mmSB\_G5533 grating, XD filter, 0.68-arcsec slit, and 0.1517-arcsec pixels. Flamingos-2 used JH\_G5801 and HK\_G5802 grisms with an 8-pixel (1.43-arcsec) slit. LRIS, Flamingos-2, and GNIRS lack independent seeing measurements; image-quality percentile categories are retained electronically.

\paragraph{MOSFIRE imaging reduction and calibration}
MOSFIRE stacks are shown in ADU s$^{-1}$ (Figure~\ref{fig:nir_images}). July $J,H,K_s$ each contain 27 exposures totaling 707.0, 785.6, and 785.6 s; the $Y$ display combines 21 of 27 exposures (611.0 s). Masks exclude nonfinite science or variance, nonzero quality flags, and coverage below three exposures. The 12-pixel $Y$ target aperture has 78.4\% valid coverage; missing regions are shown in dark gray and excluded from photometry.

Seven common unsaturated stars initially gave inconsistent July $Y-J$ colors ($-0.17$ to $+0.62$ mag). We transfer calibration from the $J$-band A0-like reference star, unsaturated in 8.72874-s sidereal $Y$ exposures but saturated in 29.0958-s comet exposures. Three bridge stars provide five accepted transfer pairs across two comet exposures. Same-frame bridge/anchor ratios cancel transparency without predicted bridge-star colors.

Sidereal images are offset-subtracted and flat-fielded using the existing $Y$ flat, bad-pixel mask, astrometry, and 26,000-DN threshold. Local background planes use 25--35-pixel sky annuli; stellar aperture radii are 8 pixels, with 12-pixel growth checks. Complete comet apertures have a 12-pixel radius (2.1576 arcsec) and 60--80-pixel sky annulus. We multiply the science-frame comet/bridge ratio by the sidereal bridge/anchor ratio, divide by the comet/anchor $J$ ratio, and apply $(Y-J)_a=0.2663$ mag. We combine the three per-bridge medians without missing-pixel or nonlinear corrections. The incomplete $Y$ mosaic is used only for display.

For $J,H,K_s$, ephemeris-centered apertures have 12-pixel radii (2.1576 arcsec at 0.1798 arcsec pixel$^{-1}$); 8, 16, and 20 pixels (1.4384, 2.8768, and 3.5960 arcsec) test aperture sensitivity. Fractional weights use $5\times5$ subpixels. A robust sky plane uses a source-masked 60--80-pixel annulus, with alternative annuli testing background sensitivity. Pixels require zero quality flags and at least three exposures; nested apertures are correlated.

February zero points average predictions for two stars with solar-like Pan-STARRS colors, using the CALSPEC solar spectrum \citep{Bohlin2014} scaled to each star's $r$ AB magnitude. Their unknown infrared spectral mismatch is excluded from quoted uncertainties. July calibration fits a reddened Castelli--Kurucz atmosphere \citep{CastelliKurucz2003} to one A0-like star's Pan-STARRS $grizy$ photometry with a 0.03-mag error floor. The adopted model has $T_{\rm eff}=9000$ K, $\log g=4.5$, solar metallicity, and $A_V=2.60$ mag under the \citet{Cardelli1989} law ($R_V=3.1$); model variations test sensitivity. Stellar apertures use 8-pixel radii without total-flux correction. We reject a blended G-like candidate. The adopted star's raw $J$ peak, 25,326 DN, approaches the 26,000-DN rejection threshold.

Synthetic photometry uses manufacturer MOSFIRE filter curves and CALSPEC solar and Vega spectra. Missing atmospheric and full-system responses limit calibration to approximate native MOSFIRE colors. Solar values are $(Y-J,J-H,H-K_s,J-K_s)_\odot=(0.2382,0.3002,0.0561,0.3563)$ mag. Following \citet{Morate2021}, $R_b/R_J=10^{0.4[(J-b)_{\rm comet}-(J-b)_\odot]}$; reported colors include the solar term, and $J$-normalized reflectances have correlated errors.

Direct 2MASS calibration failed quality checks: February zero points scatter by 0.57 mag in $J$ and 0.73 mag in $K_s$, with unstable aperture growth, incomplete core coverage, and structured residuals. July candidates lack a consistent astrometric translation. We adopt the model-dependent calibrations above. Field-source minor-axis FWHM spans 0.50--1.44 arcsec in July and 1.15--1.42 arcsec in February $J$; no satisfactory similar-brightness February $K_s$ star survives. Band-by-band profiles are given in Section~\ref{sec:mosfire_profiles}.

\paragraph{Flamingos-2 and GNIRS spectral reduction}
Figure~\ref{fig:nir_spectra} shows Flamingos-2 JH alongside the retained H spectrum from HK and groups GNIRS blocks by night. F2 integrations are 1020 s in HK on September 3 and 340 s in JH on September 5/6. GNIRS blocks contain 1708 and 1220 s on December 1 and 1464 s each on December 2 (2928 s per night). F2 cross-setting colors are excluded because settings were observed on different nights. GNIRS orders are normalized to order 4, using each nonoverlapping segment once.

We use a 40-pixel F2 extraction radius (7.2 arcsec at the nominal pixel scale of 0.18 arcsec pixel$^{-1}$); radii of 20, 60, and 80 pixels provide correlated extraction checks. The primary extraction spans 14.4 arcsec along the slit. Reduction notes identify the second columns as solar-divided spectra, normalized here over 1.57--1.63 $\mu$m. Signed third columns are excluded from uncertainty estimates. We discard JH samples below 1.0 $\mu$m and retain only HK's 1.45--1.80-$\mu$m H segment, removing its spurious short-wavelength extension and red upturn. The cuts eliminate F2 $K_s$ coverage and its 2-$\mu$m ice diagnostic. Deep atmospheric intervals remain masked.

Empirical errors use robust scatter about local linear trends in 0.01-$\mu$m bins, divided by the square root of an effective sample count accounting for positive autocorrelation and at least three native pixels per independent element. Smooth calibration errors remain unquantified. GNIRS nightly bins are inverse-variance averages with a half-block-separation error floor; individual blocks test repeatability and set a separate color-error floor.

\FloatBarrier
\input{observation_table.tex}

\section{Results}\label{sec:results}
\subsection{Optical results}\label{sec:optical_results}
\subsubsection{Optical imaging overview}\label{sec:optical_imaging_overview}
Figures~\ref{fig:optical_images} and \ref{fig:optical_north} show the north-up, east-left GMOS-S and GMOS-N/LRIS sequences, with dates, filters, total exposures, scales, and projected directions marked. Orange denotes the Sun and blue the heliocentric velocity throughout the imaging figures. The sequence shows the coma broadening and the tail becoming more prominent. Measurements use unsmoothed arrays; Section~\ref{sec:optical_observations} describes the masks and display smoothing.

\input{generated/figure_optical_images.tex}
\input{generated/figure_optical_north.tex}
\FloatBarrier

\subsubsection{Common-aperture photometry and optical colors}\label{sec:photometric_results}
We measure photometry in a 10,000-km-radius aperture (20,000-km diameter), subtending 1.78--7.67 arcsec (Table~\ref{tab:conditionalphot}), with fractional pixel weights at its boundary. Two calibrated apertures lack less than 0.15\% of their area; we estimate the missing flux from valid pixels at equal cometocentric radius. The uncertainty includes the larger of half this correction and its difference from a diametrically opposed-pixel estimate. Larger gaps remain uncorrected. Tables list valid areas and quality flags.

Sky subtraction removes some extended coma, so the photometry measures aperture excess above the local background. We test sky annuli, centering, neighboring stars, masks, and coverage. Statistical errors use the diagonal variance; calibration and background uncertainties are recorded separately. Colors use same-night measurements in a common photometric system and physical aperture: $u-g$, $g-r$, $r-i$, $g-i$, $i-z$, $J-K_s$, $Y-J$, $J-H$, and $H-K_s$ where available. We distinguish $K_s$ from $K$.

We transform visible photometry to approximate SDSS AB equivalents \citep{Tonry2012}. For $x=(g-r)_{\rm P1}$, $g_{\rm SDSS}-g_{\rm P1}=0.013+0.145x+0.019x^2$, $r_{\rm SDSS}-r_{\rm P1}=-0.001+0.004x+0.007x^2$, $i_{\rm SDSS}-i_{\rm P1}=-0.005+0.011x+0.010x^2$, and $z_{\rm SDSS}-z_{\rm P1}=0.013-0.039x-0.012x^2$. Monte Carlo errors include magnitude uncertainties, stellar residuals of 0.008, 0.004, 0.004, and 0.010 mag, respectively, and shared-band covariance. Without simultaneous $g-r$, we adopt $x=0.60\pm0.15$ for this correction. SDSS $u$ is unchanged. The electronic tables retain native fluxes and passbands. Applying stellar transformations to an emission-bearing comet introduces an unquantified systematic error. MPC-scaled $r$ is an SDSS-like proxy with a 0.20-mag error floor and is excluded from colors. Independently calibrated spectra and gas rates retain their flux scales.

Table~\ref{tab:conditionalphot} contains 29 campaign measurements on 12 dates, including two MPC-normalized $r$ estimates, and 27 aperture-normalized Liverpool measurements from 15 published epochs \citep{Graykowski2026}. The imaging log accounts for all 17 dates and 57 independent filter stacks. Missing-area fractions in November $Z$ and December $r$ are 0.138\% and 0.056\%, with recovered flux fractions of 0.069\% and 0.140\%. Other stacks lack zero points or contain large masks or bright-star artifacts. The August 28 and September 8 $u$ measurements give $A(0)f\rho=610$ and 1271 cm, subject to gas contamination and sparse-standard calibration. Section~\ref{sec:nir} presents MOSFIRE photometry.

Matched $gri$ photometry is available on August 27, February 7, and February 17. Figure~\ref{fig:activity} includes all accepted $g,r,R$ measurements; Figure~\ref{fig:crosssection} omits the two MPC-normalized $r$ points from display. Both include errors, some smaller than the symbols.

\FloatBarrier
\input{photometry_table.tex}

Discovery photometry gives SDSS $g-r=0.84\pm0.05$, $r-i=0.16\pm0.03$, and $i-z=-0.02\pm0.07$ \citep{Bolin2025Discovery}; the August Gemini report gives $g-r=0.68\pm0.01$ and $r-i=0.32\pm0.01$ \citep{ATel17363}. Our three matched GMOS epochs, transformed as above (Figure~\ref{fig:color_color}), give weighted means $g-r=0.732\pm0.029$ and $r-i=0.237\pm0.025$ mag. Constant-color fits yield $\chi^2=2.85$ and 0.60 for two degrees of freedom ($p=0.24$ and 0.74), indicating no significant change. \citet{SantanaRos2025} reported increasing activity and reddening during July 2--29; their longer homogeneous campaign found stable reddish colors \citep{SantanaRos2026}. Coverage, passband, and aperture differences complicate direct comparison.

HATPI found no significant short-timescale variability after August 6 \citep{Hartman2026HATPI}. Its broad-$G$ brightening index differs from our red-band, phase-corrected $Af\rho$ slope in passband, phase treatment, and aperture weighting.

Matched $i-z$ measurements on 2025 September 5 and 8 and 2026 February 7 and 17 give an inverse-variance weighted mean $0.109\pm0.022$ mag and $\chi^2=0.88$ for three degrees of freedom. Errors assume independent epochs and exclude shared comet-specific passband errors. September uses the assumed $g-r$ transformation color because matched $r$ photometry is unavailable; February uses measured $g-r$.

The approximate SDSS colors on August 27, February 7, and February 17 are $g-r=0.755\pm0.050$, $0.665\pm0.050$, and $0.777\pm0.050$, and $r-i=0.262\pm0.043$, $0.216\pm0.043$, and $0.234\pm0.043$. February also gives $r-z=0.290\pm0.043$ and $0.347\pm0.043$. Calibration and aperture exclusions limit the matched sample (Figure~\ref{fig:colors_time}). The July 15 PS1 colors $g-r=0.603\pm0.031$ and $r-i=0.210\pm0.031$ \citep{Beniyama2025} transform to $0.706\pm0.037$ and $0.209\pm0.031$ mag. Post-perihelion averages of $g-r=0.52\pm0.09$ and $r-i=0.13\pm0.08$ \citep{Ahuja2026} are bluer, although aperture and passband differences limit a compositional interpretation.

Rubin commissioning photometry gives $g-r=0.657\pm0.013$ and $r-i=0.235\pm0.018$ mag \citep{Chandler2026Rubin}, with similar $r-i$ and bluer $g-r$ than our means. UH88/SNIFS spectra at $r_h=4.4$--2.5 au give mostly stable synthetic PS1 colors $g-r=0.69$--0.75 and $r-i=0.26$--0.30 mag \citep{Hoogendam2026SNIFS}. These results support a persistently red continuum across differing passbands and extraction apertures.

Figure~\ref{fig:color_color} includes 6,738 distinct SDSS Moving Object Catalog asteroids \citep{Ivezic2002} with $2<a<3.5$ au, $q>1.67$ au, and errors $\leq0.05$ mag in $g,r,i,z$. Using $a^*=0.89(g-r)+0.45(r-i)-0.57$, we define C-like colors by $a^*<0$, V-like colors by $a^*\geq0$ and $i-z<-0.15$, and S-like colors by the remaining $a^*\geq0$ objects. These photometric groups approximate taxonomy. We also plot 31 active-comet observations \citep{Solontoi2012} and usable median colors from \citet{Ofek2012}, retaining their published apertures and magnitude estimators.

We adopt the ISO colors compiled by \citet{Bolin2025Discovery}: $(g-r,r-i,r-z)=(0.84,0.31,0.41)$ with marginal errors $(0.05,0.09,0.10)$ mag for 1I \citep{Meech2017}, and $(0.63,0.20,-0.03)$ with $(0.05,0.05,0.059)$ mag for 2I \citep{Bolin2020Borisov}; covariances are unavailable. The discovery $g-r$ and $r-i$ average July 3 and 6, while $r-z$ uses July 6 alone. Figure~\ref{fig:colors_time} includes the campaign-average $g-r=0.58\pm0.08$ of \citet{Graykowski2026}, with the reported dispersion as its uncertainty; that study provides no mean $r-i$.

\input{generated/figure_color_color.tex}
\input{generated/figure_colors_time.tex}

\subsubsection{Dust activity and scattering cross section}
For a calibrated bandpass flux ratio, the conventional aperture activity parameter \citep{Ahearn1984} is
\begin{equation}
Af\rho(\alpha)=\frac{4r_h^2\Delta_{\mathrm{cm}}^2}{\rho_{\mathrm{cm}}}
10^{-0.4(m_b-m_{\odot,b})},
\label{eq:afrho}
\end{equation}
Here $r_h$ is heliocentric distance in au, $\Delta_{\mathrm{cm}}$ is observer distance, and $\rho_{\mathrm{cm}}$ is aperture radius, both in cm. Comet and solar magnitudes share a passband and magnitude system. We adopt solar AB magnitudes at 1 au from Table 3, column 6, of \citet{Willmer2018}: SDSS $m_{\odot,u}=-25.18$, $m_{\odot,g}=-26.47$, $m_{\odot,r}=-26.93$, $m_{\odot,i}=-27.05$, and $m_{\odot,z}=-27.07$ mag. The Halley--Marcus dust phase function \citep{Schleicher1998,Schleicher2010HM}, normalized to $\Phi_{\mathrm{HM}}(0)=1$, gives
\begin{equation}
\afrho=Af\rho(\alpha)/\Phi_{\mathrm{HM}}(\alpha).
\end{equation}
Molecular emission contributes to $g$-band activity; dust estimates use $r$, including the May Cousins $R$ photometry transformed to approximate SDSS $r$. Native measurements remain in the electronic table.

Figure~\ref{fig:activity} follows the LD$_2$ presentation \citep{Bolin2021LD2}, with constant-$\afrho$ curves evaluated at the observing geometry, geometry-normalized brightness on the second ordinate, and heliocentric distance on the upper abscissa. We plot $g-0.732$ beside $r$ using the mean GMOS color; the shift retains gas contamination. P200/APO \citep{Bolin2025Discovery} and Liverpool IO:O Sloan photometry \citep{Graykowski2026} extend the sample. The 2-m Liverpool Telescope contributes 13 $r$ and 14 $g$ epochs from 2025 July 17 to 2026 February 21. We omit its December 6 and 7 $r$ measurements and retain $g$. Published magnitudes and geometry are adopted, with $t-T_p$ recomputed from UTC dates; broad-$G$ measurements are excluded from the $r$ sample.

Following \citet{Graykowski2026}, we fit phase-corrected red-band activity. At fixed physical aperture, their law $m(0)=H_0+5\log_{10}\Delta+2.5n\log_{10}r_h$ and Equation~\ref{eq:afrho} imply $\afrho\propto r_h^{2-n}$, or $b=2-n$. Their broad-$G$ indices give $b=-1.14\pm0.16$ before the September brightening and $-1.58\pm0.09$ outbound. We fit
\begin{equation}
\afrho=A_3\left(\frac{r_h}{3\,\mathrm{au}}\right)^b,
\label{eq:afrho_scaling}
\end{equation}
where $A_3$ is activity at 3 au. Separate inbound and outbound fits use red-band measurements, including Liverpool, in $\log_{10}\afrho$ with $\sigma_{\log A}=0.4\sigma_m$. Covariances are multiplied by $\max(1,\chi^2/\nu)$ for excess scatter. Our inbound series ends on September 14, before the brightening reported by \citet{Graykowski2026}; their broad-$G$ results provide slope comparisons.

Published P200/APO SDSS and Liverpool Sloan magnitudes receive no further passband transformation. We normalize angular apertures to 10,000 km assuming steady-coma $F\propto\rho$: $m_{10000}=m_\rho-2.5\log_{10}(10000/\rho_{\rm km})$. Liverpool's apertures already approximate 10,000 km \citep{Graykowski2026}; tabulated angular radii and rounded distances give 9955--10,065 km and corrections of $-0.005$ to $+0.007$ mag. Their zero points are retained, with no additional camera or seeing correction. All facilities use the same Halley--Marcus phase law. For LRIS, we solve $R_C-r_{\rm P1}=-0.137-0.108x-0.029x^2$ \citep{Tonry2012} with measured $g_{\rm P1}-R_C$, then transform $r_{\rm P1}$ to SDSS. Errors include the 0.015-mag $R_C$ and tabulated $r$ transformation residuals, with the comet-specific uncertainty noted above. Sensitivity tests exclude LRIS or allow a separate Liverpool normalization.

Figure~\ref{fig:activity} compares growth laws normalized to our $A_3$ at 3 au. Fixed 10,000-km $R$ photometry gives $n=3.8\pm0.3$, hence $b=-1.8\pm0.3$, for a linear phase coefficient $0.04\pm0.02$ mag deg$^{-1}$ \citep{JewittLuu2025Development}. \citet{Tonry3IATLASPhotometry2025} find cross-section growth changing from $r_h^{-3.9}$ to $r_h^{-1.1}$ near 3.3 au in fixed angular apertures with a $G=0.15$ phase function. Their purple dot-dashed law and the black dotted Jewitt--Luu curve span the published coverage. The common normalization compares growth shapes across differing apertures, passbands, and phase laws.

The outbound \textit{HST} index $n=4.5\pm0.3$ gives $b=-2.5\pm0.3$ \citep{Hui2026}, shallower than our $-3.34\pm0.43$ by $1.6\sigma$. Its teal dash-dot curve is normalized to outbound $A_3$ over December 12--January 22 ($r_h=2.108$--3.348 au). \textit{HST} F350LP photometry uses 200--1000-km apertures and a linear-exponential phase correction.

\input{generated/figure_activity.tex}

We calculate an effective scattering cross section from a phase-corrected absolute magnitude,
\begin{align}
H_b &= m_b-5\log_{10}(r_h\Delta)+2.5\log_{10}\Phi_{\mathrm{HM}}(\alpha),\\
C_b &= \frac{\pi(\mathrm{au}_{\mathrm{km}})^2}{p_b}
10^{0.4(m_{\odot,b}-H_b)}.
\label{eq:cross}
\end{align}
Here $b$ denotes the band, $m_b$ is apparent magnitude, and $H_b$ is its value at $r_h=\Delta=1$ au and zero phase angle. Distances are in au, $\alpha$ is the Sun--comet--observer angle, and $\Phi_{\mathrm{HM}}(0)=1$. Cross section $C_b$ is in km$^2$, $p_b$ is dust geometric albedo, $\mathrm{au}_{\mathrm{km}}=149597870.7$ km, and $m_{\odot,b}$ is solar apparent magnitude at 1 au \citep{Willmer2018}. The 2I/Borisov expression \citep{Bolin2020Borisov} becomes $C_V\simeq1.5\times10^6p_V^{-1}10^{-0.4H_V}$ km$^2$ in $V$; each band requires its corresponding solar magnitude. We adopt $p=0.10$, with $C\propto p^{-1}$ (Figure~\ref{fig:crosssection}).

\input{generated/figure_crosssection.tex}

\paragraph{Cross-section growth, maximum, and fading.}
Following \citet{Bolin2020Borisov}, we fit $C_r=a+b(t-t_{\rm ref})$ to inbound growth and late outbound fading, where $t=t_{\rm UTC}-T_p$ in days and $t_{\rm ref}$ is the weighted mean epoch. The extrapolated zero is $t_0=t_{\rm ref}-a/b$. We weight by photometric errors and inflate covariance by $\max(1,\chi^2/\nu)$ for residual scatter, with $\nu$ degrees of freedom. Errors assume the adopted linear model, phase law, albedo, and calibration.

The 11 inbound measurements give $dC_r/dt=8.15\pm0.27$ km$^2$ day$^{-1}$, $\chi^2/\nu=12.5$, and a zero at $t=-144.8\pm1.5$ days (UT 2025 June 6; Figure~\ref{fig:crosssection}). Excluding Liverpool gives $6.29\pm0.28$ km$^2$ day$^{-1}$ and a 2025 May 29 zero. TESS detections during May 7--June 2 at $r_h=6.35$--5.46 au show evidence for weak activity \citep{MartinezPalomera2025TESS}; clear ZTF nightly-stack detections begin May 15 \citep{Ye2025Prediscovery}. The start of TESS coverage and first clear ZTF detection precede our zero by about 30 and 22 days, respectively. \citet{Ye2025Prediscovery} inferred sustained outflow by $r_h\sim9$ au under their grain-speed and residence-time assumptions. Our zero therefore describes the later growth interval; activity began earlier. Applying the 2I/Borisov onset-extrapolation method \citep{JewittLuu2019Borisov} depends on the interval sampled.

A linear fit to all 12 outbound points gives $\chi^2/\nu\simeq385$. The 2026 February 7, February 17, and May 15 GMOS/LRIS points give $dC_r/dt=-4.10\pm0.35$ km$^2$ day$^{-1}$ and a zero at $t=+214.9\pm6.1$ days (UT 2026 June 1), with $\chi^2/\nu=7.5$ for one degree of freedom. Figure~\ref{fig:crosssection} extends this trend about 17 days beyond the last image. Including the two late Liverpool epochs gives $-4.75\pm1.06$ km$^2$ day$^{-1}$ and a zero near 2026 May 30 ($t=+212.5\pm16.8$ days), with greater scatter. Continued linear fading would remove the fitted cross section by late May to early June. The sparse baseline and residual scatter limit this extrapolation: aperture cross section depends on dust production and escape, so the coma could persist longer if fading slows or production resumes.

Equal-epoch means in seven-day bins anchored on perihelion, interpolated with a shape-preserving cubic, peak at $C_r\simeq3.15\times10^3$ km$^2$ at $t=+52.4$ days. The largest individual measurement is Liverpool's December 19 value, $4.64\times10^3$ km$^2$ at $+50.7$ days. Tests with 3- and 14-day bins and half-bin shifts give peaks at $+50.7$--$52.4$ days and $(3.15$--$4.64)\times10^3$ km$^2$. The 72.7-day gap from $-44.5$ to $+28.2$ days permits an earlier unobserved peak. Interpolated $C_r(T_p)$ is $2.47\times10^3$ km$^2$ for the cubic and $2.26\times10^3$ km$^2$ for a straight line between the bounding epochs.

\subsubsection{Heliocentric dependence of phase-corrected dust activity}\label{sec:afrho_scaling}
The 11 inbound measurements at $r_h=4.43$--2.11 au give $A_3=680\pm21$ cm and $b=-2.10\pm0.14$; 12 outbound measurements at 1.70--7.12 au give $A_3=1940\pm189$ cm and $b=-3.34\pm0.43$. Outbound normalization is $2.85\pm0.29$ times higher. Before covariance inflation, $\chi^2/\nu=47.0$ and 154.3, respectively. Confidence bands describe the fitted mean; variability, aperture assumptions, and inter-facility calibration contribute additional scatter.

The May LRIS point extends the baseline to $t-T_p=+197.8$ days; omitting it gives outbound $b=-3.33\pm0.47$. Excluding Liverpool gives $b=-1.83\pm0.05$ inbound and $-2.56\pm0.12$ outbound; Liverpool alone gives $-1.54\pm0.15$ and $-3.44\pm0.19$. Free Liverpool offsets give $-1.64\pm0.11$ and $-3.10\pm0.18$, with Liverpool higher by 0.122 and 0.263 dex, respectively, and substantial residual scatter. These sensitivity-test offsets are not applied to the photometry or adopted fits. The electronic table provides measurements, covariances, and facility/MPC/LRIS exclusion tests.

Our outbound slope is steeper than the broad-$G$ value $b=-1.58\pm0.09$ \citep{Graykowski2026}, with gas and facility differences complicating interpretation. The May point alone provides no evidence for another break. Since $\afrho$ measures albedo-weighted cross section per aperture radius, grain residence times, sizes, and phase behavior affect its evolution. The higher outbound normalization and decline trace dust persistence and escape; mass loss also requires a grain and velocity model.

The \textit{HST} phase model \citep{Hui2026} has linear slope $0.026\pm0.006$ mag deg$^{-1}$, opposition enhancement $0.19\pm0.01$ mag, and e-folding width $3\pm1^\circ$. Phase behavior and orbital-plane projection make January especially geometry-sensitive. The \textit{HST} radius $1.3\pm0.2$ km sets our shell-depth scale and is consistent with the earlier 2.8-km upper bound \citep{Jewitt2025HST}.

At 10,000 km, \citet{Liu2026DustIce} find cross-section indices $-1.86\pm0.27$ inbound and $-2.42\pm0.06$ outbound, using our activity-index convention with a linear phase correction. Their shallower outbound slope reinforces the sensitivity to sampling, phase correction, and relative calibration.

\subsubsection{Coma morphology and dust trajectories}\label{sec:morphology}
We divide $g$ and $r$ images by their azimuthal-median radial profiles \citep{CIE2013,Samarasinha2014}, using 0.32-arcsec annuli and detector masks. For four affected September $r$ stacks, both bands use matched 0.48-arcsec native-area bins and annuli, excluding insufficient coverage. Figure~\ref{fig:enhancement} omits March and uses no spatial filter. Red, blue, and white denote excess, deficit, and unity relative to the median, with north up and east left. These ratios locate asymmetries but do not separate gas and dust. Profiles use the original masked 0.16-arcsec arrays. Centering tests shift one pixel in each coordinate and compare angular medians at 2--6 and 6--10 arcsec, excluding residual stars and incomplete coverage.

\input{generated/figure_enhancement.tex}

We generate a Finson--Probstein reference grid \citep{FinsonProbstein1968} by propagating particles under solar gravity reduced by radiation pressure,
\begin{equation}
\ddot{\boldsymbol r}_d=-(1-\beta)GM_\odot\frac{\boldsymbol r_d}{|\boldsymbol r_d|^3},
\end{equation}
Here $\beta$ is the radiation-pressure/solar-gravity ratio. Particles begin at the comet's position and velocity, forming a zero-ejection-speed grid; synchrones share release time and syndynes share $\beta$. Figure~\ref{fig:dust} projects trajectories onto the sky and labels the signed orbital-plane angle $\delta_E$ at each midpoint. The grid spans $\beta=0.001$--1 and ages 5--240 days. For compact spheres, $a_{\mu\mathrm{m}}\simeq0.574Q_{\mathrm{pr}}/[\rho_d(\mathrm{g\,cm^{-3}})\beta]$ \citep{Burns1979}, or 0.6--600 $\mu$m for $Q_{\mathrm{pr}}=1$ and $\rho_d=1$ g cm$^{-3}$. Sizes also depend on ejection velocity and release history. Inbound syndynes overlap within 0.5--1.3 degrees at 10 arcsec, preventing unique size or age estimates.

\input{generated/figure_dust.tex}

Molecular emission can broaden $g$ relative to dust-dominated $r$ \citep{ATel17561}. We normalize azimuthal-median profiles at 0.8--1.6 arcsec and compare their 3--6-arcsec means. The $g/r$ shape ratio is 1.01 in August and 1.05, 1.23, 1.06, and 1.10 on September 5, 6, 8, and 14. From November 26 through February 17 it lies within 0.4--3.2\% of unity. PSF and background differences prevent converting these ratios to gas fractions.

On August 27, a tail at position angle $105^\circ$ reaches the 6--10-arcsec annulus, while the opposite enhancement near $285^\circ$ is confined to the inner coma. This compact sunward feature weakens through September as the tail strengthens (Figures~\ref{fig:optical_images} and \ref{fig:optical_north}).

By November 26 the coma is broader, developing an asymmetric envelope in December--January and a persistent projected tail in February. The $r$ profile at 3--6 arcsec, normalized at 0.8--1.6 arcsec, rises from 0.10 on August 27 to 0.26--0.28 in November--February. These angular profiles sample changing physical radii and lack PSF matching. Similar post-perihelion $g$ and $r$ profiles show that the envelope includes red continuum. Stars and limited surface-brightness sensitivity hinder tracing the outer coma in March and May.

\citet{JewittLuu2025Development} observed a July sunward fan developing a prominent antisolar tail by mid-August, attributing the delay to approximately 100-$\mu$m grains moving at 5 m s$^{-1}$ under their onset and dust assumptions. Our sequence is consistent with continued release and radiation-pressure redistribution, although changing geometry projects different populations and directions. The antisolar position angle moves from approximately $103^\circ$ inbound to $292$--$294^\circ$ in November--December and $94^\circ$ in February. Foreshortening is strongest on January 20 ($\alpha=1.4^\circ$), and the perihelion gap prevents dating a sudden morphological transition.

Shifting the August 27 center by 0.16 arcsec preserves both inner maxima in $g$ and $r$; the weaker feature moves one 10-degree bin. It fades in September and is compatible with the sunward structure near $280^\circ$ reported by \citet{SantanaRos2026}; the stronger side follows the tail. A distinct sunward source remains tentative, as do source identifications for the structures seen by \citet{Ahuja2026}. Residual stars also generate angular maxima, especially in February.

Persistent extended dust is compatible with sustained activity. Recovering the release history requires fitting surface brightness across epochs, as for P/2016 J1 \citep{Moreno2017} and 2I/Borisov \citep{BolinLisse2020}; overlapping syndynes alone do not constrain production rates.

\FloatBarrier

\subsubsection{Dust ejection speeds and mass loss}\label{sec:dustmass}
\paragraph{Dust speeds from transverse widths and the sunward plume}\label{sec:tail_speed}
At small orbital-plane angles, tail widths constrain dust motion normal to the orbit \citep{Jewitt2019Gault,Bolin2021LD2}. Nine GMOS $r$ epochs provide initial profiles at $|\delta_E|<1^\circ$. In the unsmoothed, masked 0.16-arcsec arrays, we fit a Gaussian plus linear background to 0.32-arcsec transverse bins in 2-arcsec-wide strips at antisolar offsets of 3--21 arcsec. Initial cuts require $>70\%$ valid coverage per bin, at least 40 valid bins, amplitude significance $>5$, fractional width error $<0.3$, and a centroid within 3.8 arcsec of the axis.

We then require fully unmasked footprints, reduced $\chi^2\leq5$, and no compact contaminants. This excludes affected September and February 7 profiles, strongly foreshortened January profiles, and the innermost February 5 cut with a suspected star. The 13 retained profiles comprise seven at 9--21 arcsec on August 27, two at 5--7 arcsec on February 5, and four at 3--9 arcsec on February 17. These determine the plotted speeds and mass interpolation; the electronic log lists rejected profiles.

For transverse FWHM $\theta_W$, antisolar offset $\theta_l$, and sunward half-maximum offset $\theta_D$ in arcsec, the projected lengths at geocentric distance $\Delta$ in meters are
\begin{equation}
 \begin{aligned}
 W_{\rm sky}&=\Delta\frac{\theta_W}{206265\,\mathrm{arcsec\,rad^{-1}}},\\
 l_{\rm sky}&=\Delta\frac{\theta_l}{206265\,\mathrm{arcsec\,rad^{-1}}},\\
 D_{\rm sky}&=\Delta\frac{\theta_D}{206265\,\mathrm{arcsec\,rad^{-1}}}.
 \end{aligned}
 \label{eq:width_to_length}
\end{equation}
Here 206265 is arcsec per radian and $\Delta=1.495978707\times10^{11}(\Delta/{\rm au})$ m; $W_{\rm sky}$ is a full width and $D_{\rm sky}$ a half-width.

At grain age $\tau$, the half-width speed is $v_\perp\simeq W_{\rm sky}/(2\tau P_\perp)$, where $P_\perp=|\widehat{\boldsymbol n}\cdot\widehat{\boldsymbol c}|$ projects the orbital normal onto the sky-plane cut direction. We derive the normal from archived heliocentric positions and velocities. Constant local radiation pressure gives $\tau=[2l_{\rm sky}/(\beta g_\odot\sin\alpha)]^{1/2}$ and
\begin{equation}
 v_\perp=U_\perp\sqrt{\beta}\simeq
 \frac{W_{\rm sky}}{2P_\perp}
 \left(\frac{\beta g_\odot\sin\alpha}{2l_{\rm sky}}\right)^{1/2},
 \label{eq:tail_velocity}
\end{equation}
Here $\alpha$ is phase angle and $g_\odot=0.00593(r_h/{\rm au})^{-2}$ m s$^{-2}$ is solar gravitational acceleration. The factor of two converts FWHM to half-width; $\sin\alpha$ deprojects distance for a radial tail. Figure~\ref{fig:tail_width_examples} reports $U_\perp$; speed requires $\beta$. For spherical grains,
\begin{equation}
 \beta=\frac{0.574\,Q_{\rm pr}}{[\rho_d/(1000~{\rm kg\,m^{-3}})]\,[a/(1~\mu{\rm m})]},
 \label{eq:beta_grain_speed}
\end{equation}
where $a$ is grain radius, $\rho_d$ density, and $Q_{\rm pr}$ radiation-pressure efficiency \citep{Burns1979}. Adopting $\rho_d=1000$ kg m$^{-3}$ and $Q_{\rm pr}=1$ gives $\beta=0.00396$ at an illustrative radius of 145 $\mu$m. Retained epochs have $P_\perp>0.999$, although release ages remain ambiguous for curved or old trajectories. Median $U_\perp$ is 19.5, 14.5, and 18.0 m s$^{-1}$ on August 27, February 5, and February 17. Profile-to-profile ranges are 15.7--21.7, 14.2--14.7, and 17.5--19.2 m s$^{-1}$; corresponding 145-$\mu$m median speeds are 1.23, 0.91, and 1.13 m s$^{-1}$.

For the 2025 August 27 GMOS-S $r$ cut at 15 arcsec (Figure~\ref{fig:tail_width_examples}), FWHM $8.66\pm0.46$ arcsec gives $W_{\rm sky}=1.626\times10^7$ m. With $l_{\rm sky}=2.817\times10^7$ m, $r_h=2.637$ au, $\alpha=22.27^\circ$, and $P_\perp=0.99982$, this gives $v_\perp=19.48\sqrt{\beta}$ m s$^{-1}$, or 1.23 m s$^{-1}$ at $\beta=0.00396$.

\input{generated/figure_tail_width_examples.tex}

\textit{HST} gives $v_\perp\simeq7.5\sqrt{\beta}$ m s$^{-1}$ on July 21 with the same half-width convention \citep{Jewitt2025HST}, lower than our medians by factors of 2.6, 1.9, and 2.4. Applying our $\beta=0.574/a_{\mu\mathrm{m}}$ rather than their $1/a_{\mu\mathrm{m}}$ gives an \textit{HST} speed of 0.47 m s$^{-1}$ at 145 $\mu$m. Offsets, seeing, coma contributions, and grain ages differ, limiting an acceleration interpretation. Our zero-speed grid spans 0.1--0.4 arcsec in-plane at these offsets; subtracting this extent in quadrature would require a brightness-weighted dust model.

We apply the sunward-coma method \citep{JewittComaProfiles1987}, using HWHM as in \citet{Bolin2025Discovery}. We extract 0.64-arcsec strips around a subpixel photocenter, testing 0.32- and 0.96-arcsec widths and $\pm0.08$-arcsec shifts in each coordinate. Unmasked footprints and peaks exceeding ten times the outer-sky scatter retain ten of twelve $r$ epochs: September 6 has a masked core and March a weak, irregular profile. August 28 $u$ and December 19 $g$ remain diagnostic only. Clean sunward cores on September 5, 8, 14 and February 7 are usable despite contaminated outer-tail profiles. Unlike the discovery analysis, these stacks lack a validated seeing correction.

Radiation pressure gives $v_{\rm sun}^2=2\beta g_\odot D$ for turnaround distance $D$ \citep[][Equation~4]{JewittComaProfiles1987}. Using projected HWHM gives a projected-extent speed; a Sun-aligned trajectory gives $D=D_{\rm sky}/\sin\alpha$. Thus,
\begin{equation}
 \begin{aligned}
 v_{\rm sun,sky}&=S_{\rm sky}\sqrt{\beta}
     =\sqrt{2\beta g_\odot D_{\rm sky}},\\
 v_{\rm sun,\parallel}&=S_\parallel\sqrt{\beta}
     =\sqrt{\frac{2\beta g_\odot D_{\rm sky}}{\sin\alpha}}.
 \end{aligned}
 \label{eq:sunward_speed}
\end{equation}
The coefficients $S_{\rm sky}$ and $S_\parallel$ span 33.5--70.5 and 70.3--105.3 m s$^{-1}$ (Figure~\ref{fig:sunward_width_examples}); January's $\alpha=1.4^\circ$ geometry is excluded from the axial estimate. At 145 $\mu$m, the speeds are 2.1--4.4 and 4.4--6.6 m s$^{-1}$. The axial model equates HWHM with turnaround distance, although no intrinsic edge is detected there. Trailed or multiply imaged stars prevent per-epoch PSF deconvolution. Open symbols therefore show observed-width estimates with strip-width and centering sensitivities. PSF broadening raises the inferred speed within the adopted model.

The August 27 GMOS-S $r$ profile (Figure~\ref{fig:sunward_width_examples}) has HWHM 0.646 arcsec, or 1214 km. Errors propagate diagonal variance through median extraction and peak normalization in 10,000 Monte Carlo realizations, excluding resampling covariance and sky systematics. For $D_{\rm sky}=1.214\times10^6$ m, $S_{\rm sky}=45.5$ and $S_\parallel=73.9$ m s$^{-1}$, giving 2.86 and 4.65 m s$^{-1}$ at $\beta=0.00396$.

\input{generated/figure_sunward_width_examples.tex}

The discovery relation $v_e=\sqrt{2D_t\beta g_1}/r_h$ agrees with \citet{JewittComaProfiles1987} for acceleration $g_1$ at 1 au, $r_h$ in au, and $D_t$ in meters. Its 1850-km extent, $g_1=0.006$ m s$^{-2}$, and $r_h=4.43$ au give $33.6\sqrt{\beta}$ m s$^{-1}$, correcting the published $1.1\sqrt{\beta}$ m s$^{-1}$. This uses the published extent without further deprojection. \textit{HST}'s sunward coefficient, 50 m s$^{-1}$ \citep{Jewitt2025HST}, is lower than our axial values by factors of 1.4--2.1. Epoch, resolution, profile, and projection differences limit this comparison. The directional contrast is consistent with anisotropic dayside emission, motivating separate mass estimates.

Comparison at fixed $v/\sqrt{\beta}$ avoids mixing size--$\beta$ prescriptions. Our illustrative 145-$\mu$m conversion ($\beta=0.00396$) gives sunward speeds of 3.1 m s$^{-1}$ for \textit{HST} and 6.3 m s$^{-1}$ for Tianwen-1 \citep{Jewitt2025HST,Ren2026Tianwen}.

Tianwen-1 images from September 30--October 3 give $v=100\sqrt{\beta}$ m s$^{-1}$ \citep{Ren2026Tianwen}, within our axial coefficient range 70.3--105.3 m s$^{-1}$. Their $\beta=10^{-3}$--$10^{-2}$ gives our axial speeds of 2.2--10.5 m s$^{-1}$, overlapping their 3--10 m s$^{-1}$, with differing epochs, geometry, and extent definitions. The approximately 5-m s$^{-1}$ estimate for 100-$\mu$m grains \citep{JewittLuu2025Development} combines the \textit{HST} normalization with delayed tail development and $\beta=1/a_{\mu\mathrm{m}}$; it shares the \textit{HST} velocity constraint.

\citet{Liu2026DustIce} obtain an inbound sunward coefficient $190\pm3$ m s$^{-1}$ from a deprojected 10\%-of-peak boundary, beyond our HWHM. Their January 18 perpendicular coefficient $79\pm6$ m s$^{-1}$ uses projected tail distance. Our radial-tail correction gives $(79\pm6)\sqrt{\sin2.5^\circ}=16.5\pm1.3$ m s$^{-1}$, within our range. PSF, geometry, and their $\beta=1/a_{\mu\mathrm m}$ prescription remain different.

\citet{SantanaRos2025} report 0.012--1.2 m s$^{-1}$, overlapping our perpendicular 0.9--1.2 m s$^{-1}$ at 145 $\mu$m, but provide no size-to-speed mapping for a common-radius comparison. The morphology model of \citet{Moreno2026Dust} uses $v=500\sqrt{\beta}\,r_h^{-1/2}\cos z$ m s$^{-1}$, with release distance $r_h$ in au and solar zenith angle $z$. Uniform averaging over the illuminated hemisphere gives $v/\sqrt{\beta}=250r_h^{-1/2}$ m s$^{-1}$, or 125--177 m s$^{-1}$ at 4--2 au. This terminal-speed prescription exceeds our directional scales and uses grains from 10 $\mu$m to an upper cutoff of 1 or 10 cm, density 800 kg m$^{-3}$, and a different release geometry.

\input{generated/figure_tail_speed.tex}

\paragraph{Dust mass loss and equivalent surface erosion}
We convert phase-corrected $r$-band activity to dust mass loss using Equation~5 of \citet{Bolin2025Discovery}, generalized to size-dependent speed,
\begin{equation}
 \dot M_d=\frac{4\pi\rho_d}{3p_r}\,\langle av\rangle\,A(0)f\rho,
 \qquad
 \langle av\rangle=\frac{\int a^3v(a,t)n_{\rm ap}(a)\,da}{\int a^2n_{\rm ap}(a)\,da}.
 \label{eq:dustmass}
\end{equation}
Here $\rho_d=1000$ kg m$^{-3}$ is grain density, $p_r=0.10$ is dust geometric albedo, $a$ is radius, $v(a,t)$ is radial speed, and $n_{\rm ap}(a)$ is the aperture grain-number distribution. All quantities use SI units, including $A(0)f\rho$ in meters. The conversion assumes optically thin, approximately steady outflow; gas, nucleus light, fragmentation, and changing residence times introduce systematic errors.

We adopt $n_{\rm ap}\propto a^{-3}$ over $a_{\min}=1~\mu$m to $a_{\max}=1$ mm, encompassing the 10--100-$\mu$m grains favored by MIRI \citep{Belyakov2026Dust}. The upper cutoff is unconstrained by our images. The area-weighted radius is $(a_{\max}-a_{\min})/\ln(a_{\max}/a_{\min})=145~\mu$m. With $\beta=B/a$, $B=5.74\times10^{-7}$ m, and $v(a,t)=C(t)\sqrt{\beta}$,
\begin{equation}
 \langle av\rangle=\frac{2C(t)\sqrt{B}\,[\sqrt{a_{\max}}-\sqrt{a_{\min}}]}{\ln(a_{\max}/a_{\min})}.
 \label{eq:dust_velocity_moment}
\end{equation}
For isotropic Gaussian velocities, $C=2U_\perp/\sqrt{\pi\ln2}=1.355U_\perp$ converts one-dimensional HWHM to mean three-dimensional speed. The Sun-aligned prescription uses $C=S_\parallel$, giving a separate mass estimate.

The 23 accepted red-band measurements span 2025 July 3--2026 May 15 ($t-T_p=-118.18$ to $+197.80$ days), including P200, APO, Liverpool, and transformed LRIS data. We interpolate $A(0)f\rho$ and each velocity coefficient independently in time and integrate their product,
\begin{equation}
 M_d=\int_{t_1}^{t_2}\dot M_d(t)\,dt,
\end{equation}
with time in seconds. Photometry is not extrapolated; velocity coefficients are held at the nearest endpoint outside their measured baseline, excluding \textit{HST} as an anchor. Figure~\ref{fig:dustmass} shows rates and cumulative masses.

The transverse model gives $M_d=2.71\times10^9$ kg, rates of 6.2--360 kg s$^{-1}$, and effective speeds $\langle av\rangle/a_{\rm eff}=0.91$--1.23 m s$^{-1}$. Substituting profile extrema gives $(2.42$--$2.91)\times10^9$ kg. Holding the February or August coefficient across the 161-day velocity gap gives $2.40$ or $3.07\times10^9$ kg. The nine axial coefficients give $1.05\times10^{10}$ kg. Directional and PSF assumptions dominate profile-repeatability errors.

Across the 72.7-day perihelion gap, interpolating $[A(0)f\rho]r_h^2$ and dividing by daily $r_h^2$ preserves the endpoints while allowing a solar-flux-shaped enhancement. It gives $3.16\times10^9$ and $1.23\times10^{10}$ kg for the transverse and sunward models; an unobserved outburst remains unconstrained. Transverse aperture crossing times of 94--127 days weaken the steady-flow approximation. If $a^{-3}$ describes source production instead of the aperture distribution, residence-time weighting raises the estimates by 1.84.

The cross-section derivative gives net aperture dust accumulation: $M_{\rm ap}=(4/3)\rho_d a_{\rm eff}C_r$, with $C_r$ in m$^2$ and $a_{\rm eff}=145~\mu$m. The inbound slope gives $dM_{\rm ap}/dt=18.2\pm0.6$ kg s$^{-1}$ and $1.58\times10^8$ kg accumulated from the fitted zero to September 14. The seven-day maximum contains $6.08\times10^8$ kg. Continued grain escape makes total emission exceed the retained inventory.

Extending the seven-day profile to its zeros on 2025 June 6 and 2026 June 1 gives $2.79\times10^9$ and $1.08\times10^{10}$ kg, equivalent to 0.26- and 1.02-m shells. The same smoothed profile over July--May gives $2.76\times10^9$ and $1.07\times10^{10}$ kg. Smoothing and perihelion interpolation slightly raise these values over the principal unsmoothed budgets; extending the endpoints adds little. Earlier activity and later cessation remain unconstrained.

\input{generated/figure_dust_mass.tex}

We express the dust budget as an equivalent globally removed shell,
\begin{equation}
 \Delta R=R_0-\left(R_0^3-\frac{3M_d}{4\pi\rho_n}\right)^{1/3},
 \label{eq:dustshell}
\end{equation}
Here $R_0=1.3\pm0.2$ km is the \textit{HST} post-perihelion radius for nucleus albedo $p_V=0.04$ \citep{Hui2026}, and $\rho_n=500$ kg m$^{-3}$ is assumed nucleus density. Transverse and axial models give global dust-equivalent depths of 0.26 and 0.99 m, excluding volatile loss. A smaller active area implies deeper local erosion.

The larger MIRI estimates \citep{Belyakov2026Dust} use different velocity, grain-size, and activity-history assumptions. Their size-independent 50 m s$^{-1}$ exceeds our effective speeds of 0.91--1.23 m s$^{-1}$ (transverse) and 3.26--4.89 m s$^{-1}$ (axial), raising conversion factors by 41--55 and 10--15 at fixed grain distribution. On December 15, their 1-mm-cutoff rates of 1820--2770 kg s$^{-1}$ exceed our interpolated 341 and 1230 kg s$^{-1}$ by factors of 5.3--8.1 and 1.5--2.3.

For comparison, adopting 50 m s$^{-1}$ with our grain distribution, density, albedo, and piecewise-linear activity gives $1.26\times10^{11}$ kg over July 3--May 15 and a 12.02-m shell. Our principal budgets use image-based speeds. The common-speed model gives $1.68\times10^4$ kg s$^{-1}$ on December 15, still 6.1--9.2 times the MIRI 1-mm-cutoff rates. MIRI measures thermal $\epsilon f\rho$ within approximately 3700 km, whereas our optical aperture is 10,000 km. Emissivity, albedo, temperature, and radial grain distribution also affect the comparison \citep{Belyakov2026Dust}.

Their December-normalized $r_h^{-2}$ integration spans June 1--December 27, including perihelion; ours interpolates July--May. Their 2.5--56.5-m shells span radii of 1.1--1.5 km, temperature assumptions, and 1-mm--1-cm upper grain cutoffs; the centimeter extension raises constant-speed mass conversion by 7.5 \citep{Belyakov2026Dust}. Equation~\ref{eq:dustshell} gives 0.255 and 0.989 m for our nominal masses, confirming that mass conversion drives the discrepancy. Both estimates omit volatile loss and local recession. Projection, PSF broadening, anisotropy, and residence times limit our absolute mass estimates.

\FloatBarrier

\subsubsection{Optical spectra and molecular and atomic production}\label{sec:optical_spectroscopy}
Figures~\ref{fig:visible_spectra} and \ref{fig:gmos_flux} show six GMOS epochs and the September 1 LRIS spectra, excluding March. September and November GMOS fluxes use MPC normalizations. We display the noisy February 17 spectrum only at $\lambda\geq400$ nm, but retain the complete extraction for gas measurements. Its CN limit excludes the unknown blue-response systematic.

\input{generated/figure_visible_spectra.tex}
\input{generated/figure_gmos_flux.tex}

We derive CN and C$_2$ production with a forward Haser model. The steady, spherical daughter density is projected along the line of sight and convolved with a normalized Gaussian PSF $P$ of assumed FWHM 0.8 arcsec. For GMOS, the extraction operator $\mathcal{E}$ integrates the source rectangle and subtracts the linear sky fit; combined spectra use the 550-nm geometry and equally weighted nod positions. For LRIS, it integrates a centered $0.7\times9.45$-arcsec rectangle without a sky correction because the sky footprint is unavailable. The effective residence time and production rate are
\begin{equation}
 \tau_{j,\mathrm{eff}}=\mathcal{E}\!\left[\left(\frac{\Sigma_j}{Q_{\rm trial}}\right)*P\right],\qquad
 Q_j=\frac{4\pi\Delta_{\rm cm}^{2}F_j}{g_j\tau_{j,\mathrm{eff}}},
\end{equation}
where $\Sigma_j$ is column density, $F_j$ continuum-subtracted band flux, $g_j$ fluorescence efficiency, and $\Delta_{\rm cm}$ observer distance in cm. Physical-area integration gives $\tau_{j,\mathrm{eff}}$ in seconds, independent of the trial rate $Q_{\rm trial}$. We adopt $v=0.85(r_h/1\,\mathrm{au})^{-1/2}$ km s$^{-1}$. Independent rectangular integration agrees with \texttt{sbpy} \citep{Mommert2019} for matched parameters.

CN and C$_2$ parent/daughter lengths at 1 au are $(1.3\times10^4,2.1\times10^5)$ and $(2.2\times10^4,6.6\times10^4)$ km \citep{Ahearn1995}, scaled by $r_h^2$. CN fluorescence is interpolated in $r_h$ and heliocentric radial velocity using the Lowell implementation of \citet{Schleicher2010CN}; C$_2$ uses $4.5\times10^{-13}r_h^{-2}$ erg s$^{-1}$ molecule$^{-1}$. C$_3$ follows the December ATel prescription: $g_3=3.8\times10^{-13}r_h^{-2}$ erg s$^{-1}$ molecule$^{-1}$, consistent with \citet{Cochran2012}, parent/daughter lengths $(3100,145000)r_h^2$ km, and $v_3=0.5$ km s$^{-1}$. We integrate its Haser column over the source rectangle with \texttt{sbpy}, without PSF or sky corrections, using each epoch's Horizons geometry (Table~\ref{tab:haserparams}).

\FloatBarrier
\input{generated/table_haser_parameters.tex}

We integrate CN and C$_2$ solar-continuum residuals in 0.5-nm bins over 383.0--390.5 and 486--518.5 nm. A broadened solar spectrum with a local linear color term is fitted over 380--382.5 and 390.8--393 nm for CN, and 475--486 and 519--527 nm for C$_2$. Errors include rebinning covariance, continuum scatter, and the shared flux scale. C$_3$ uses
\begin{equation}
 F_3=\left[\left\langle F_\lambda\right\rangle_{397.5<\lambda<415}
 -\mathrm{median}\!\left(F_\lambda\big|_{379<\lambda<384\;\mathrm{or}\;393<\lambda<397}\right)\right](175\,\mbox{\AA}),
\label{eq:c3atel}
\end{equation}
with wavelengths in nm and $F_\lambda$ in erg cm$^{-2}$ s$^{-1}$ \AA$^{-1}$, before solar division. We combine the 540- and 550-nm arrays with inverse-variance weights in 0.25-nm bins, preserving September/November's native 0.5-nm information and covariance and omitting rejected bins. All epochs cover the full band; August covers 81\% of the 379--384-nm continuum window, starting near 380 nm, and all of 393--397 nm. Extraction heights are 6.400 arcsec for GMOS-S and 6.456 arcsec for GMOS-N, with 1-arcsec slits; LRIS uses $0.7\times9.45$ arcsec.

We calibrate the August spectrum with the spectrophotometric standard EG 274 \citep{HamuyStandards1992,HamuyStandards1994} and imaging. The December--February spectra use the reduced PHOENIX-based response \citep{HusserPHOENIX2013}, normalized to imaging. September 14/15 and November 26 use MPC $S_r$ photometry: station J47 reports median $r=14.70$ within 6.2 arcsec on September 14, and station 213 reports $r=12.60$ within 6.5 arcsec on November 26. GMOS image flux ratios transfer these circular-aperture magnitudes to the spectroscopic rectangles, giving slit-equivalent magnitudes 16.200 and 14.620. We normalize the solar-restored spectra through the PS1 $r$ response, with report scatter and a 0.20-mag calibration floor. The MPC observations precede the spectra by approximately 4 and 10 hr. Published gas rates are excluded from calibration, and the blue response remains uncertain.

We propagate C$_3$ noise through Equation~\ref{eq:c3atel} in 4096 realizations, including combination weights and September/November rebin covariance. The imaging or MPC normalization enters once as a correlated gray-scale term. Residual reduction covariance and the LRIS absolute-scale error are incompletely known. Formal errors exclude continuum choice, response shape, fluorescence, source geometry, and sky-subtraction systematics. We classify excesses above $3\sigma$ as formal detections; otherwise we report $\max(0,\widehat Q)+3\sigma_Q$, retaining signed residuals electronically. Dust color and solar absorption affect C$_3$ attribution, while local continuum fits can absorb molecular wings. We test this with the broadened CALSPEC Kurucz spectrum \texttt{sun\_mod\_001.fits} \citep{Bohlin2014} and a linear color term. Extended/red-only fits define the December/January ranges; the February limit spans all tested continua. These tests cover four imaging-scaled and two MPC-scaled GMOS epochs, both LRIS orientations, and their correlated all-exposure combination. February's 379--397-nm continuum retains the blue-response limitation.

For Ni I, we jointly fit 347--356 and 358--366 nm with a common template amplitude, solar continuum, and transitions from the corrected multi-level fluorescence model \citep{Bromley2026Erratum,Bromley2026NiGrid}, evaluated at the LRIS geometry and radial velocity. Following \citet{Hutsemekers2026}, we adopt a short source, long-lived atoms, and projected column $\propto1/\rho$, with speed 0.5 km s$^{-1}$ in the $0.7\times9.45$-arcsec rectangle. For rectangular half-sizes $a,b$, the residence time is $[a\,\mathrm{asinh}(b/a)+b\,\mathrm{asinh}(a/b)]/v$, without a PSF/sky correction. Gaussian FWHMs of 0.50, 0.65, and 0.80 nm test broadening. Blends and the unknown source scale limit interpretation. Table~\ref{tab:newq} reports this joint fit without assigning a single-band flux.

\FloatBarrier
\input{generated/table_newq.tex}
\input{generated/figure_gas_indices.tex}

CN production is $(1.85\pm0.07)\times10^{25}$, $(3.65\pm0.70)\times10^{25}$, $(3.81\pm0.88)\times10^{25}$, and $(1.36\pm0.16)\times10^{25}$ s$^{-1}$ in August, September, November, and December. September and November also show C$_2$. On November 26, CN and C$_2$ have signal-to-noise ratios 7.4 and 10.2 before the common scale error. The C$_3$ excess, $(1.655\pm0.319)\times10^{-13}$ erg cm$^{-2}$ s$^{-1}$ including that error, gives $Q$(C$_3$)$=(1.13\pm0.22)\times10^{25}$ s$^{-1}$. MPC scatter, aperture transfer, and the 0.20-mag floor give an 18.7\% normalization uncertainty. December's weaker 516.5-nm bandhead and integrated C$_2$ residual consistent with zero yield an upper limit.

Keck/KCWI detected CN, C$_2$, C$_3$, Ni, and Fe on November 16 \citep{Hoogendam2026Post}; \citet{Medler2026} detected CN and Ni before and after perihelion and weak outbound C$_3$. These identifications support November carbon-chain emission, although rate comparisons depend on epoch, extraction, calibration, and source model.

The LRIS coma spectrum gives $Q$(CN)=$(6.51\pm0.21)\times10^{24}$, $Q$(C$_3$)=$(4.64\pm0.13)\times10^{24}$, $Q$(C$_2$)$<1.28\times10^{24}$ s$^{-1}$, and $Q$(Ni I)=$(2.18\pm0.04)\times10^{23}$ atoms s$^{-1}$. Coma--tail values are $(7.96\pm0.33)\times10^{24}$, $(6.55\pm0.13)\times10^{24}$, $<1.88\times10^{24}$ s$^{-1}$, and $(2.48\pm0.04)\times10^{23}$ atoms s$^{-1}$, respectively. Both use the 0.7-arcsec slit and the species-specific models above. Errors exclude the unknown LRIS absolute scale, blends, and source geometry.

The December ATel CN rate, $6.49\times10^{26}$ s$^{-1}$ \citep{ATel17719}, contains an unresolved conversion inconsistency. Its written $g_{\rm CN}(1\,\mathrm{au})=3.7\times10^{-13}$ applied to the original Jy arrays gives $1.47\times10^{25}$ s$^{-1}$, close to our rate. The saved output corresponds instead to $g_{\rm CN}(1\,\mathrm{au})=8.4\times10^{-15}$ with other inputs fixed. Figure~\ref{fig:published_rates} retains the published value; the discrepancy has no secure physical interpretation. November ATel CN and C$_2$ rates exceed ours by factors of 18.9 and 23.3 \citep{ATel17561}.

\citet{Ahuja2026} report December 20 $Q$(CN)$=(1.855\pm0.232)\times10^{25}$ s$^{-1}$ at 1 km s$^{-1}$. Scaling only speed to our December 22 value, 0.551 km s$^{-1}$, gives $(1.02\pm0.13)\times10^{25}$ s$^{-1}$, a factor of 1.33 below our rate. Their unscaled November 27 CN and C$_2$ rates are $(5.54\pm0.051)\times10^{25}$ and $(5.62\pm0.329)\times10^{25}$ s$^{-1}$. Table~\ref{tab:publishedq} and Figure~\ref{fig:published_rates} retain published rates; epoch, aperture, and fluorescence differences remain. Our LRIS coma CN rate is close to the August 21 value $10^{24.81\pm0.01}$ s$^{-1}$ \citep{Rahatgaonkar2025}, eleven days earlier and with different model assumptions.

\FloatBarrier
\input{generated/table_publishedq.tex}
\par\medskip
ATel-prescription C$_3$ rates are $(6.04\pm0.36)\times10^{24}$ s$^{-1}$ on August 28 and $(7.79\pm1.50)\times10^{24}$ s$^{-1}$ on September 14/15. The correlated LRIS all-exposure value, $(5.72\pm0.13)\times10^{24}$ s$^{-1}$, lies between the two orientations and is retained electronically. Their difference includes orientation, time, and calibration effects.

On December 22, the uncorrected estimator gives $F_3=(7.53\pm0.34)\times10^{-14}$ erg cm$^{-2}$ s$^{-1}$, or $(1.54\pm0.07)\times10^{25}$ s$^{-1}$ for the $6.456\times1$-arcsec extraction. The ATel used a different spectrum and a 130-arcsec rectangle, reproducing $6.98\times10^{24}$ s$^{-1}$. The spectra differ in continuum shape and by a factor of 2.34 in band excess, motivating the solar-continuum tests below.

Our November C$_3$ rate is a factor of 1.59 below the ATel value $1.80\times10^{25}$ s$^{-1}$ \citep{ATel17561} and 1.66 times above the November 27 value $(6.79\pm0.77)\times10^{24}$ s$^{-1}$ \citep{Ahuja2026}; scaling their speed from 1 to 0.5 km s$^{-1}$ increases the latter ratio to 3.33. January 21's uncorrected estimator gives $(1.67\pm0.09)\times10^{25}$ s$^{-1}$, 28.8 times the ATel 17841 value $5.79\times10^{23}$ s$^{-1}$. December--February use model-response calibrations, with January anchored to imaging approximately 22 hr earlier. C$_3$ response-shape envelopes are approximately 19\%, 15\%, and 15\%, respectively; continuum placement and chromatic calibration limit these comparisons.

\paragraph{Post-perihelion C$_3$ continuum and aperture controls.}
We apply Equation~\ref{eq:c3atel} to gas-free solar continua to estimate contamination of the C$_3$ excess. A template convolved to 0.65 nm FWHM is fitted in amplitude and linear color to each GMOS setting. Anchors are 391--394 and 415--420 nm (local), 380--382, 440--445, and 475--480 nm (extended), or 440--445 and 475--480 nm (red only). Models use the data's native-bin combination weights. Anchors were chosen before comparing published rates: local fits risk absorbing molecular wings, whereas distant anchors extrapolate continuum color.

Gas-free controls give apparent C$_3$ fluxes of $(6.18$--$6.30)\times10^{-14}$ erg cm$^{-2}$ s$^{-1}$ in December and $(2.11$--$2.30)\times10^{-14}$ erg cm$^{-2}$ s$^{-1}$ in January, accounting for 82--84\% and 86--94\% of the estimator signals. Subtracting the extended/red-only solar fits gives 397.5--415-nm residual integrals of $(2.80$--$3.28)\times10^{-14}$ and $(5.89$--$6.51)\times10^{-15}$ erg cm$^{-2}$ s$^{-1}$. The $6.456\times1$-arcsec extraction, epoch geometry, and fixed-speed Haser model then give $Q$(C$_3$)$=(5.73$--$6.70)\times10^{24}$ molecules s$^{-1}$ on 2025 December 22 and $(4.00$--$4.42)\times10^{24}$ molecules s$^{-1}$ on 2026 January 21. December approaches the ATel value $6.98\times10^{24}$ molecules s$^{-1}$; January remains 6.9--7.6 times higher. The adopted ranges describe continuum-choice sensitivity, with additional statistical and calibration uncertainty. Local fits give much smaller residuals, especially in January, and reduced $\chi^2\simeq3$--8 indicates mismatch beyond statistical noise.

For fixed flux, the actual aperture increases $Q$ by factors of 5.08, 5.94, and 6.15 relative to a $130\times1$-arcsec rectangle in December, January, and February. Increasing extraction height from 3.228 to 12.912 arcsec changes the aperture-corrected estimator rate from $2.03$ to $1.18\times10^{25}$ s$^{-1}$ in December and $2.51$ to $1.01\times10^{25}$ s$^{-1}$ in January. This exceeds formal errors and may reflect dust contamination, seeing, sky subtraction, or chromatic calibration. At the adopted height, alternative sky regions change the estimates by 0.3\% and 2.0\%; central-wavelength settings differ by factors of 1.12 and 1.27. These effects explain only part of January's discrepancy.

February 17's residual changes sign with the continuum anchors. We adopt the envelope
$F_{3,\mathrm{lim}}=\max_j[\max(0,F_{3,j})+3\sigma_j]$
over the local, extended, and red-only fits, where $\sigma_j^2=\sigma_{\mathrm{stat},j}^2+(f_{\mathrm{gray}}F_{3,j})^2$ includes continuum-fit and integration noise and shared gray-scale fraction 0.030794. The red-only fit sets $F_{3,\mathrm{lim}}=3.33\times10^{-15}$ erg cm$^{-2}$ s$^{-1}$ and $Q$(C$_3$)$<5.71\times10^{24}$ molecules s$^{-1}$. This $3\sigma$ envelope covers the tested continua but excludes unknown response-shape and physical-model errors. Figures~\ref{fig:gas_indices} and \ref{fig:published_rates} and Table~\ref{tab:newq} use this limit and the December/January ranges; electronic tables retain uncorrected estimates.

\input{generated/figure_published_rates.tex}

\citet{PaekCNComa2026} resolved an extended inbound CN coma at $r_h<3$ au and increasing $Q$(CN) relative to $Af\rho$, supporting a gas contribution to blue coma growth. Their CN-sensitive 400-nm band samples different emission from C$_2$ in our broadband $g$ images.

In July 2025, JCMT gave $3\sigma$ limits of $Q$(HCN)$<1.7\times10^{24}$ s$^{-1}$ at $r_h=4.01$--3.97 au and $Q$(CO)$<1.1\times10^{27}$ s$^{-1}$ at 3.94--3.84 au \citep{Hinkle2026JCMT}. Including later observations gave $Q$(HCN)$\propto r_h^{-n}$ with $n=12.7^{+6.9}_{-2.5}$, describing parent-molecule activation before our daughter-species detections. The epochs preclude a simultaneous HCN/CN ratio, and this gas-production index has a different physical meaning from the dust $Af\rho(0)$ index.

\subsubsection{Gas-production ratios}
Table~\ref{tab:gasratios} and Figure~\ref{fig:gas_ratios} compare C$_3$, CN, C$_2$, and Ni. The common gray scale cancels in same-spectrum ratios. Without cross-band statistical covariance for C$_3$, we remove that scale term and report the maximum first-order standard deviation allowed by the marginal errors: $\sigma_x/|y|+|x|\sigma_y/y^2$ for $x/y$. Continuum-choice and response-shape errors are excluded; December's continuum range is separate. C$_2$/CN and Ni I/CN propagate band errors without the shared scale. We divide upper-limit numerators by detected denominators and omit ratios with undetected denominators. Ratios retain the distinct C$_3$, CN/C$_2$, and atomic source models.

$Q$(C$_2$)/$Q$(CN) rises from $0.257\pm0.014$ on 2025 September 14/15 to $0.481\pm0.080$ on November 26. LRIS Ni I/CN is $0.0336\pm0.0012$ for coma and $0.0311\pm0.0014$ for coma--tail. Its Ni rates, $(2.18$--$2.48)\times10^{23}$ atoms s$^{-1}$, are comparable to the August 21 value $10^{23.30\pm0.07}$ \citep{Rahatgaonkar2025}, subject to epoch, fluorescence, and source-model differences. C$_3$/CN is $0.327\pm0.024$, $0.214\pm0.009$, and $0.296\pm0.054$ in August, September, and November. December's 0.422--0.493 range divides the two solar-continuum C$_3$ estimates by nominal CN production, excluding CN uncertainty. LRIS gives $0.713\pm0.042$ for coma and $0.823\pm0.050$ for coma--tail. Continuum and source-model dependence limit compositional interpretation.

The January 21 ATel reports $Q$(CN)$=(4.54\pm0.05)\times10^{24}$, $Q$(C$_2$)$=(5.85\pm0.05)\times10^{24}$, and $Q$(C$_3$)$=(5.79\pm0.06)\times10^{23}$ s$^{-1}$ \citep{ATel17841}. Our CN limit, $4.48\times10^{24}$ s$^{-1}$, is close to their value; our C$_2$ limit, $8.47\times10^{24}$ s$^{-1}$, includes theirs. Published ATel ratios are plotted separately. The September ATel identifies emission without rates \citep{ATel17503}.

On August 24, \citet{Hoogendam2025KCWI} measured emission-profile $e$-folding radii of $593.7\pm14.8$ km for Ni and $841.0\pm15.4$ km for CN; the Ni coma is more concentrated. Post-perihelion profiles are more extended \citep{Hoogendam2026Post}. These empirical radii describe the spatial dependence of Ni I/CN and its aperture correction; they are distinct from Haser survival lengths.

\FloatBarrier
\input{generated/table_gas_ratios.tex}
\input{generated/figure_gas_ratios.tex}

\FloatBarrier
\subsection{Infrared results}\label{sec:nir}
\subsubsection{MOSFIRE coma profiles and aperture measurements}\label{sec:mosfire_profiles}
Elliptical Gaussian fits within 3 arcsec give July MOSFIRE comet minor-axis FWHMs of 1.31, 1.27, and 1.79 arcsec in $J,H,K_s$; geometric-mean widths are 1.42, 1.36, and 2.03 arcsec. Comparison sources fall within 1.5 arcsec of a 2MASS position, have A/B photometric quality without relevant artifact flags, and aperture fluxes within one magnitude of the comet. Their minor-axis FWHMs are 0.58--0.61 arcsec in $J$ (two sources), 0.50 arcsec in $H$ (one), and 1.44 arcsec in $K_s$ (one). Trailing and structured residuals limit their use as PSF estimates.

\input{generated/figure_nir_images.tex}

February minor-axis FWHMs are 1.87 and 1.79 arcsec in $J$ and $K_s$, with geometric means of 1.93 and 1.94 arcsec. Two similarly bright $J$ stars have widths of 1.15--1.42 arcsec, supporting an extended comet core. No $K_s$ star passes the same selection. Tracking and background residuals affect these profile comparisons independently of the color calibration.

Additional February apertures of 1.590 and 0.430 arcsec provide instrumental-count checks. The smaller aperture matches the physical area of the GNIRS rectangle ($0.68\times2$ arcsec at $\Delta=1.905$ au), with a circular radius of 909 km. The primary F2 extraction ($1.43\times14.4$ arcsec at $\Delta=2.562$ au) instead requires an equal-area circular radius of approximately 4760 km, or 2.25 arcsec in February. The 0.430-arcsec aperture requires a cross-filter PSF correction within the approximately 1.8-arcsec core, and coma asymmetry changes the weighting between rectangles and circles. Figure~\ref{fig:nir_spectral_colors} uses calibrated MOSFIRE radii of 1.4384--3.5960 arcsec, with aperture and epoch differences retained in the spectroscopic comparison.

\subsubsection{Near-infrared spectra, colors, and variability}
\input{generated/figure_nir_spectra.tex}
We restore the solar spectrum $F_\odot(\lambda)$ to reflectance $R(\lambda)$ for synthetic photometry, assuming solar-equivalent continua for the comparison stars. The photon-weighted signal through transmission $T_b(\lambda)$ is
\begin{equation}
I_b\propto\int R(\lambda)F_\odot(\lambda)T_b(\lambda)\lambda\,d\lambda.
\end{equation}
We integrate valid samples against the appropriate AB or Vega reference. Incomplete coverage makes broadband colors model-dependent.

The July sidereal transfer gives $Y-J=0.466$ mag at 2.16-arcsec radius. The three bridge-star estimates (0.413, 0.520, and 0.466 mag) have 0.053-mag sample scatter. Maximum shifts with aperture, background, and anchor spectrum are 0.082, 0.069, and 0.078 mag; their quadrature sum with the full bridge scatter gives the 0.143-mag sensitivity scale in Table~\ref{tab:nirpop} and Figures~\ref{fig:nir_spectral_colors} and \ref{fig:nir_movis}. Correlations, spatial and transparency transfer, tracking-mode PSFs, contamination, and anchor mismatch remain incompletely quantified. A 25--35-pixel $J$-anchor background gives 0.501 mag. The nominal color is 0.227 mag redder than the native-filter solar value of 0.238 mag, without establishing a significant difference.

In the same aperture, July colors are $J-H=0.239\pm0.052$, $H-K_s=0.454\pm0.150$, and $J-K_s=0.693\pm0.202$ mag. Radii of 1.4384 and 2.1576 arcsec (8 and 12 pixels; 12 nominal) give colors of 0.187--0.239, 0.304--0.454, and 0.491--0.693 mag, respectively. At either radius, adopted errors combine the absolute aperture difference with the formal flux error in quadrature. The latter is only 0.003--0.004 mag, from $\sigma_m=(2.5/\ln10)\sigma_F/F$ and independent statistical errors between bands. The plotted empirical errors measure aperture sensitivity, with unestablished Gaussian $1\sigma$ coverage and unconstrained common contamination and calibration bias. Adding single-band calibration sensitivities would require their covariance. Larger-aperture measurements remain in the electronic table. Stellar residuals and differing physical apertures limit a July--February color-evolution test.

Table~\ref{tab:partialcolors} and Figure~\ref{fig:nir_spectral_colors} give Vega colors over the accepted 2MASS passband intervals. F2 JH gives $J-H=0.241\pm0.005$ mag from spectral scatter; the 0.198--0.274-mag range across correlated extractions dominates its uncertainty. The trimmed HK spectrum lacks $H-K_s$ and $J-K_s$ coverage and was obtained on a different date, preventing direct broadband combination with JH.

GNIRS partial colors $(J-H,H-K_s,J-K_s)$ are $(0.270,-0.050,0.219)$ mag on December 1 and $(0.277,-0.067,0.210)$ mag on December 2. Uncertainty scales including half the block separation are $(0.027,0.020,0.008)$ and $(0.053,0.026,0.027)$ mag. We combine relative band fluxes with common exposure weights, preserving $J-H+H-K_s=J-K_s$. Coverage differs slightly between blocks and samples only part of each passband.

February MOSFIRE gives $J-K_s=0.303\pm0.051$ mag and $R_{K_s}/R_J=0.952\pm0.045$ at 2.1576-arcsec radius, consistent with a near-solar continuum. The error combines measurement, reference-star scatter, aperture, and sky terms under the assumed solar infrared reference spectra; shared spectral mismatch remains unquantified. The color changes from 0.305 to 0.280 mag over radii of 1.4384--3.5960 arcsec.

GNIRS $J-H$ differs by 0.054 and 0.106 mag between blocks on December 1 and 2, exceeding the 0.007-mag difference between nightly means. Similar $H-K_s$ offsets exceed local scatter errors. Calibration, chromatic slit losses, and extraction differences prevent resolving physical variability. Table~\ref{tab:partialcolors} and Figure~\ref{fig:nir_spectral_colors} compare the correlated extractions: F2 extraction radii of 20, 40, 60, and 80 pixels (3.6, 7.2, 10.8, and 14.4 arcsec at 0.18 arcsec pixel$^{-1}$\footnote{\url{https://gemini-iraf-flamingos-2-cookbook.readthedocs.io/en/latest/F2_overview.html}}), and GNIRS 1-, 2-, 3-arcsec and native extractions. Symbols distinguish these choices, with filled squares for primary spectral and February imaging measurements and open symbols for July.

\FloatBarrier
\input{generated/table_partialcolors.tex}
\input{generated/figure_nir_spectral_colors.tex}

\subsubsection{Comparison with solar colors and small-body populations}
Matched spectral intervals and Vega references give solar $J-H=0.324$ mag for F2 and 0.331/0.326 mag for December 1/2 GNIRS. Comet excesses are $-0.082$, $-0.061$, and $-0.049$ mag, respectively; GNIRS excesses are $-0.094/-0.111$ mag in $H-K_s$ and $-0.155/-0.159$ mag in $J-K_s$. These intervals show a neutral to weakly blue continuum within comparison-star and response uncertainties. Full 2MASS solar colors are $(J-H,H-K_s,J-K_s)=(0.332,0.043,0.374)$ mag; the interval-matched values account for incomplete coverage.

Table~\ref{tab:nirpop} gives unweighted medians and sample standard deviations from the object-averaged MBOSS catalog \citep{Hainaut2012}.\footnote{Published 2012 CDS catalog J/A+A/546/A115, Table 2: \url{https://cdsarc.cds.unistra.fr/viz-bin/cat/J/A+A/546/A115}.} Each object contributes once per available color. We divide gray and red Centaurs at $B-R=(B-V)+(V-R)=1.6$ mag and combine resonant, classical, scattered, and detached TNOs, including large icy objects. We derive $J-K$ for objects with both $J-H$ and $H-K$, whose epoch sets may differ. Sample sizes vary by color, and dispersions include measurement noise and population diversity.

MBOSS $J-H$ medians are 0.39 mag for both Centaur groups, 0.40 mag for TNOs, and 0.45 mag for Jupiter Trojans; $H-K$ and $J-K$ medians span $-0.06$--0.16 and 0.33--0.59 mag. July MOSFIRE $H-K_s$ is redder than these medians, December GNIRS lies near the red-Centaur value, and February MOSFIRE $J-K_s$ falls within the TNO dispersion. Filter variants in MBOSS and partial passbands in our spectra limit these comparisons. Its only comet nucleus with both $J-H$ and $H-K$ is 49P/Arend--Rigaux.

For active comae, \citet{JewittCometColors1988} summarizes weighted mean $J-H$ of $0.46\pm0.07$ and $0.41\pm0.06$ mag and $H-K$ of $0.10\pm0.08$ and $0.14\pm0.05$ mag for the 14- and 9-comet samples of \citet{JewittCometColors1986,JewittCometColors1988}. The errors include systematic contributions. P/Giclas lacks $H-K$ in the first sample, and Bradfield is excluded from the second $H-K$ mean because of possible thermal contamination. \citet{HannerCometInfrared1984} contributes measurements to the 1986 sample. C/2005 E2 (McNaught) has $J-H=0.48\pm0.01$, $H-K=0.01\pm0.01$, and $J-K=0.49\pm0.01$ mag on 2005 September 26--27 \citep[their Table 1]{PicazzioMcNaughtNIR2010}. Table~\ref{tab:nirpop} also retains the active- and inactive-comet compilations of \citet{DeLeon2020Borisov}. C-, P-, and D-type asteroid colors overlap in 2MASS \citep{Sykes2000,GravHolman2004}. Comparisons between free coma grains and surface regoliths depend on grain size, porosity, phase angle, gas, and ice.

\begin{table}[!htbp]\centering\small
\caption{NIR colors and contextual population comparisons}\label{tab:nirpop}
\setlength{\tabcolsep}{4pt}
\begin{tabular}{lrrrr}\hline
Sample & $Y-J$ & $J-H$ & $H-K_s$ & $J-K_s$\\\hline
3I MOSFIRE, July 11, $1.44^{\prime\prime}$ & --- & $0.187\pm0.052$ & $0.304\pm0.150$ & $0.491\pm0.202$\\
3I MOSFIRE, July 11, $2.16^{\prime\prime}$ & $0.466\pm0.143^{\ast}$ & $0.239\pm0.052$ & $0.454\pm0.150$ & $0.693\pm0.202$\\
3I MOSFIRE, February 6 & --- & --- & --- & $0.303\pm0.051$\\
Sun, native MOSFIRE synthetic & 0.238 & 0.300 & 0.056 & 0.356\\
Sun, UKIRT (Willmer 2018) & 0.27 & --- & --- & ---\\
3I F2, September 5/6 & --- & $0.241\pm0.005$ & --- & ---\\
3I GNIRS, December 1 & --- & $0.270\pm0.027$ & $-0.050\pm0.020$ & $0.219\pm0.008$\\
3I GNIRS, December 2 & --- & $0.277\pm0.053$ & $-0.067\pm0.026$ & $0.210\pm0.027$\\
Sun, full 2MASS synthetic & --- & 0.332 & 0.043 & 0.374\\\hline
\multicolumn{5}{l}{MBOSS (2012): median $\pm$ sample dispersion [number of objects]}\\
Sample & $Y-J$ & $J-H$ & $H-K$ & $J-K$\\\hline
Gray Centaurs & --- & $0.39\pm0.09\ [11]$ & $0.16\pm0.09\ [9]$ & $0.55\pm0.12\ [9]$\\
Red Centaurs & --- & $0.39\pm0.06\ [6]$ & $-0.06\pm0.06\ [6]$ & $0.33\pm0.07\ [6]$\\
TNOs & --- & $0.40\pm0.21\ [119]$ & $0.05\pm0.11\ [39]$ & $0.43\pm0.19\ [39]$\\
Jupiter Trojans & --- & $0.45\pm0.06\ [12]$ & $0.13\pm0.04\ [12]$ & $0.59\pm0.09\ [12]$\\
49P/Arend--Rigaux nucleus & --- & $0.36\ [1]$ & $0.14\ [1]$ & $0.50\ [1]$\\\hline
\multicolumn{5}{l}{Comet comae: weighted mean $\pm$ published uncertainty [number of comets]}\\
1986 coma sample & --- & $0.46\pm0.07\ [14]$ & $0.10\pm0.08\ [13]$ & ---\\
1988 coma sample & --- & $0.41\pm0.06\ [9]$ & $0.14\pm0.05\ [8]$ & ---\\\hline
\multicolumn{5}{l}{Individual active comet: published colors and errors}\\
C/2005 E2 (McNaught) & --- & $0.48\pm0.01$ & $0.01\pm0.01$ & $0.49\pm0.01$\\\hline
\multicolumn{5}{l}{Other comet compilations: $J-H$ median [reported observations]}\\
Active short-period comets & --- & $0.42\ [12]$ & --- & ---\\
Inactive short-period comets & --- & $0.41\ [2]$ & --- & ---\\
Active long-period comets & --- & $0.48\ [3]$ & --- & ---\\
Inactive long-period comets & --- & $0.37\ [1]$ & --- & ---\\\hline
\end{tabular}
\par\smallskip\parbox{.96\textwidth}{\footnotesize Colors are in magnitudes. F2/GNIRS use accepted passband intervals; MOSFIRE uses approximate native-filter Vega calibration. The asterisk marks the July $Y-J$ sensitivity scale; July $JHK_s$ errors include aperture sensitivity. These empirical scales have unestablished Gaussian coverage and exclude common contamination or calibration bias. February errors exclude shared reference-star mismatch. MBOSS uses heterogeneous $K$ filters without transformation to $K_s$, with $J-K$ derived for paired objects and errors describing sample dispersion. For 49P, catalog measurement errors are 0.141 mag in each of $J-H$ and $H-K$; missing shared-band covariance prevents a $J-K$ error estimate. MBOSS has no $Y$ or long-period-comet NIR entries. The 1986 and 1988 coma means \citep{JewittCometColors1986,JewittCometColors1988} are quoted from \citet{JewittCometColors1988}, with errors on weighted means including systematics. Different $J-H$ and $H-K$ samples and weights prevent deriving a mean $J-K$. McNaught colors and errors follow Table 1 of \citet{PicazzioMcNaughtNIR2010}; the averaging aperture is unspecified. The final four rows give medians and observation counts from \citet{DeLeon2020Borisov}, drawing on \citet{HannerCometInfrared1984,Popescu2016,Hainaut2012,Sykes2000,PicazzioMcNaughtNIR2010}; no dispersions are supplied. Dashes denote unavailable colors. Solar colors use Vega references, with UKIRT $(Y-J)_\odot=0.27$ mag from \citet{Willmer2018}. Filter systems and scattering states limit taxonomic comparisons.}
\end{table}

Figure~\ref{fig:nir_movis} compares solar-relative colors with VISTA asteroids \citep{Popescu2016,Popescu2018}: July 11 MOSFIRE and December 1/2 GNIRS in $J-H$ versus $H-K_s$, and July MOSFIRE in $Y-J$ versus $J-K_s$. Both panels use 1886 classified objects with all three catalog colors ($Y-J,J-K_s,H-K_s$) finite and $0<\sigma\leq0.10$ mag; $J-H=(J-K_s)-(H-K_s)$. VISTA solar colors are $(Y-J,J-H,H-K_s,J-K_s)_\odot=(0.219,0.254,0.082,0.336)$ mag \citep{Popescu2018}. Crosses mark mean synthetic colors of A, B, C, D, K, L, O, Q, R, S, T, V, and X templates, with one object each for O and R.

The left panel adds the MBOSS medians, 49P nucleus, and Jewitt--Meech coma samples, subtracting solar $(J-H,H-K)$ of $(0.23,0.06)$ mag for MBOSS \citep{HainautMBOSSColors2002}, $(0.32,0.07)$ mag for \citet{JewittCometColors1986}, and $(0.31,0.06)$ mag for \citet{JewittCometColors1988}. Bars show MBOSS sample dispersions, Jewitt--Meech errors on means, and 49P measurement errors. Group summaries may sample different objects in each color. McNaught lacks matched solar colors and appears only in Table~\ref{tab:nirpop}.

July points use the 2.16-arcsec aperture and empirical uncertainties, including the 0.143-mag $Y-J$ scale. Relative to matched solar colors, July $H-K_s$ is redder than December, all three epochs have slightly blue $J-H$, and July $Y-J$ and $J-K_s$ are redder than solar. Contamination, calibration transfer, and differing extractions limit evidence for intrinsic evolution. Solar subtraction leaves differences among the MOSFIRE, partial GNIRS, VISTA, and historical $K/K_s$ responses. Marginal error bars omit shared-$H$ covariance on the left and shared-$J$ covariance on the right.

\input{generated/figure_nir_movis.tex}
\FloatBarrier

\subsubsection{Water-ice models and absorption constraints}
Absorption depth is $D=1-\langle R\rangle/R_c$ relative to continuum $R_c$. Atmospheric gaps near 1.4 and 1.9 $\mu$m make the ice-band depths continuum-sensitive, so we also fit pure-ice templates.

\citet{Yang2020Ice} found no ice bands in IRTF/SpeX and Gemini/GNIRS spectra of 2I/Borisov, limiting ice to approximately 10\% of the coma cross section under their mixture model. For \obj, \citet{Yang2025} modeled July IRTF/SpeX 2.0-$\mu$m absorption with approximately 70\% Tagish Lake material and 30\% 10-$\mu$m ice grains; the 1.5-$\mu$m band was undetected. Their order-of-magnitude areal fraction differs from our fitted ice contribution to scattered light at 1.20 $\mu$m.

Following \citet{Yang2020Ice}, we compute 150 K crystalline-ice spectra for radii $a=1,10,100~\mu$m from the optical constants of \citet{Mastrapa2008,Mastrapa2020PDS}. The Hapke equivalent-slab model includes Fresnel surface terms and a mean absorption path equal to the grain diameter. We also calculate homogeneous-sphere Mie efficiencies and single-scattering albedo $w=Q_{\rm sca}/Q_{\rm ext}$ \citep{Protopapa2018Ice}. Both use an isotropic geometric-albedo mapping and 4.7-nm FWHM smoothing, omitting porosity, dirty-ice mixtures, and a measured ice phase function. Figure~\ref{fig:nir_spectra} shows 10\% mixtures with continua fitted over 1.15--1.30 and 1.70--1.78 $\mu$m, normalized at 1.20 $\mu$m.

We fit a linear or quadratic continuum times a mixture of smooth and normalized ice spectra, replacing the amorphous-carbon component of \citet{Yang2020Ice} with a smooth continuum. The nonnegative fraction measures ice-scattered light at 1.20 $\mu$m; mass or area fractions require additional grain assumptions. Profiling over continuum coefficients gives $\Delta\chi^2=9$ ceilings after scaling scatter errors by continuum residuals. The ranges span both scattering models and continuum orders, excluding unknown smooth calibration errors. Because the ice fraction can approach zero, nondetection leaves grain radius unconstrained.

Figure~\ref{fig:nir_spectra} combines F2 settings and nightly GNIRS blocks. The F2 cutoff removes usable 2-$\mu$m and $K_s$ coverage; retained JH constrains the short-wavelength ice-band shape with strong continuum dependence. Neither instrument yields absorption robust to the tested continuum and calibration choices.

For 10-$\mu$m grains, ceilings span 15--36\% of the 1.20-$\mu$m scattered light in F2 JH, 0.6--2.7\% for GNIRS December 1, and 0.7--2.5\% for December 2 (Table~\ref{tab:ice}). Unquantified telluric and smooth response errors, together with different coverage and slit extractions, prevent an ice-depletion inference.

\FloatBarrier
\input{generated/table_ice.tex}

\section{Discussion and Conclusions}\label{sec:discussion}
\subsection{Activity, dust dynamics, and surface evolution}
The evolution from a concentrated coma and compact sunward feature to an extended tail is consistent with accumulation and radiation-pressure sorting of dust released over many dates \citep{JewittLuu2025Development}. Slowly moving grains can persist as production weakens. The synchrones and syndynes allow a range of release ages and grain sizes, while our cadence cannot resolve the periodic angle modulation of the high-latitude jet reported by \citet{SerraRicart2025Jet}. Similar continuum colors can accompany different spatial dust distributions, as also seen in C/2020 V2 \citep{Voitko2026V2}.

Our red-band activity measurements extend the baseline of \citet{Graykowski2026} to 7.12 au. The steeper outbound decline persists when a separate Liverpool normalization is allowed, although the fitted indices change from $-2.10\pm0.14$ and $-3.34\pm0.43$ to $-1.64\pm0.11$ and $-3.10\pm0.18$ inbound and outbound, respectively. Delayed release, changing active areas, and retained grains can produce different branch normalizations. The perihelion gap prevents us from identifying when these changes occurred or whether a larger scattering area preceded the observed maximum on 2025 December 19, approximately 51 days after perihelion. The seven-day profile peaks near $t=+52$ days.

For fixed grain properties, the cross-section derivative measures dust replenishment minus escape and destruction. The aperture maximum occurs when these terms balance, so long crossing times can delay it relative to peak nucleus production. Thermal lag or changing illumination could also sustain outbound release. The subsequent fading shows that losses exceed replenishment, but its linear extrapolation cannot establish a unique cessation date. Phase-function, relative-calibration, and sampling uncertainties further affect the maximum's amplitude.

At fixed $\beta$, our perpendicular epoch medians exceed the July \textit{HST} value by factors of 1.9--2.6; our axial sunward estimates exceed the corresponding \textit{HST} estimate by factors of 1.4--2.1 \citep{Jewitt2025HST}. The sunward coefficients, 70.3--105.3 m s$^{-1}$, bracket Tianwen-1's 100 m s$^{-1}$ \citep{Ren2026Tianwen}; at its inferred $\beta=10^{-3}$--$10^{-2}$, our speeds are 2.2--10.5 m s$^{-1}$, overlapping its 3--10 m s$^{-1}$ estimate. The approximately 5-m s$^{-1}$ speed adopted for 100-$\mu$m grains by \citet{JewittLuu2025Development} uses the \textit{HST} normalization with a different size conversion. Our illustrative radial-tail projection converts \citet{Liu2026DustIce}'s perpendicular coefficient to $16.5\pm1.3$ m s$^{-1}$, within our 14.5--19.5-m s$^{-1}$ range; their sunward boundary extends beyond our half-maximum definition. Faster sunward motion is consistent with preferential dayside ejection, although differences in epoch, seeing, grain age, and width definition limit tests of temporal acceleration.

The transverse and sunward models give dust losses of $2.7\times10^9$ and $1.0\times10^{10}$ kg for 1-$\mu$m--1-mm grains over the observed baseline. With the 1.3-km nucleus radius of \citet{Hui2026}, these correspond to a global dust-equivalent shell approximately 0.3--1 m thick. Comparing these budgets with \citet{Belyakov2026Dust} requires accounting for the different grain-size and optical-property weighting of scattered and thermal emission, as well as velocity, albedo, and maximum grain size. Concentrated activity could excavate more deeply than the global equivalent; the active area, dust-to-ice ratio, and initial mantle thickness remain unknown.

Cosmic-ray processing has been proposed to explain \obj's volatile and continuum properties \citep{Maggiolo2025GCR}. Delayed methane-release models admit different combinations of initial dust content and irradiation exposure \citep{Rawat2026Methane}. Our erosion estimates alone cannot distinguish these histories without an independent constraint on processing depth.

\subsection{Continuum colors, gas, and ice}
The three GMOS epochs with matched $gri$ photometry have consistent colors despite substantial morphological evolution. Changes in spatial grain distribution therefore need not produce a measurable change in aperture-averaged color. Grain size, phase function, and gas emission all affect broadband colors. In particular, the C$_2$ Swan bands at 486--518.5 nm, including the 516.5-nm bandhead, contribute to $g$ and can broaden its coma or reduce dust-jet contrast. Features shared by $g$ and $r$ support a dust contribution.

C$_2$ production approximately doubles between 2025 September 14--15 and November 26 (Table~\ref{tab:newq}), supporting a growing gas contribution across perihelion. The November spectrum precedes the largest measured red-band cross section by 23 days, leaving the timing of peak C$_2$ production unresolved. Liverpool's $g-r=0.370\pm0.036$ mag on 2025 December 19 is bluer than the GMOS mean of 0.732 mag, indicating enhanced relative $g$ brightness. The December--February C$_2$ limits allow declining production after November. Non-simultaneous spectra, other gas bands, dust color, and calibration differences prevent a unique interpretation of the December color change; $r$ can also contain gas emission. Quantifying the gas fraction requires a spatial gas model integrated through the imaging aperture and filter response. Our long-slit data support a qualitative comparison and identify gas contamination as a limitation of $g$-band $Af\rho$.

The rise in C$_2$/CN is less sensitive to a shared flux-normalization error than either production rate alone. Relative spectral response, molecular lifetimes and distributions, illumination, and distributed grain sources still affect the ratio, while late C$_3$ measurements depend strongly on continuum subtraction. Inbound CN with carbon-chain upper limits \citep{Lazzarin2026CN} and prominent outbound CN, C$_3$, and C$_2$ with normal-comet mixing ratios \citep{Kawakita2026Spectra} document stronger carbon-chain emission outbound, as also reported by \citet{Ahuja2026}. Their water-normalized abundances require conversion for comparison with our CN-normalized ratios. Spatial profiles \citep{Mura2026Profiles} test the CN and metal scale lengths used in aperture corrections. LRIS Ni I provides a refractory-species measurement, but Ni I/CN requires distinct fluorescence and source models before interpretation as an elemental abundance.

The neutral to weakly blue NIR spectra agree qualitatively with the turnover near 1.1 $\mu$m reported by \citet{Medler2026} for a continuum resembling 2I/Borisov. Curved grain reflectance can produce this behavior. February MOSFIRE colors are near-solar under the reference-star assumption; July $Y-J$ remains sensitive to stellar calibration and contamination. December nightly spectral differences fall below the block repeatability scale. Residual stars, unequal apertures, and passband differences limit color-evolution tests, and comparisons with inactive bodies must account for scattering by free coma grains rather than regoliths.

For 10-$\mu$m pure-ice grains, GNIRS limits their contribution to roughly 1--3\% of the scattered light at 1.20 $\mu$m under the tested continuum and scattering models; unknown smooth calibration errors are excluded. The early ice spectra of \citet{Yang2025} and the outer-coma dirty-ice interpretation of \citet{Liu2026DustIce} sample different epochs and regions. An icy population that contributes little to our inner-coma continuum remains allowed. MIRI silicate spectra constrain grain composition through thermal emission \citep{Belyakov2026Dust}.

\subsection{Comparison with 1I, 2I, and the interstellar population}
The sustained, resolved activity of \obj\ resembles 2I/Borisov more closely than 1I/`Oumuamua. Yet 1I showed nongravitational acceleration without a detected dust coma \citep{Micheli2018Oumuamua}, and 2I had unusually CO-rich gas \citep{Bodewits2020BorisovCO,Cordiner2020BorisovCO}. Similar optical colors across these objects can accompany different volatile inventories, grain properties, and source geometries. The 1.3-km radius inferred for 3I \citep{Hui2026} exceeds 2I's 0.5-km \textit{HST} upper limit under their respective albedo assumptions \citep{Jewitt2020BorisovNucleus}; unknown active areas limit comparisons of production per unit surface area. Comae also complicate the rotation measurements obtainable from 1I's large-amplitude, tumbling lightcurve \citep{Meech2017,Drahus2018Tumbling}.

Three detections provide too small and selected a sample to establish the relative abundance of these properties. Active comae increase the discovery volume, while size, albedo, trajectory, cadence, perihelion distance, and observing coverage influence the detected sample. Following each encounter helps distinguish intrinsic differences among parent bodies from their responses to solar heating.

Our Keck and Gemini observations show changing morphology and gas production, including an increasing C$_2$/CN ratio, alongside stable optical colors. Image-based dust speeds imply losses of $10^9$--$10^{10}$ kg, equivalent to a global layer approximately 0.3--1 m thick under the adopted models. The NIR spectra place model-dependent limits on the ice contribution to scattered light, with no robust absorption detection.

\begin{acknowledgments}
The Liverpool Telescope is operated on the island of La Palma by Liverpool John Moores University in the Spanish Observatorio del Roque de los Muchachos of the Instituto de Astrofisica de Canarias with financial support from the UK Science and Technology Facilities Council, based on observations at the PATAG proposal AZ22A01 (PI: F. Romanov).

We thank Josh Walawender, Ian Wong, and Matthew Belyakov for their assistance with the MOSFIRE observations, reduction of the Gemini data, and observation planning.

D.F. conducted this research at the Jet Propulsion Laboratory, California Institute of Technology, under a contract with the National Aeronautics and Space Administration (80NM0018D0004).

This work has been supported by the Czech Science Foundation through grant 25-16789S (J. Hanu\v{s}).

This work was supported by a NASA Keck PI Data Award, administered by the NASA Exoplanet Science Institute. Data presented herein were obtained at the W. M. Keck Observatory from telescope time allocated to the National Aeronautics and Space Administration through the agency's scientific partnership with the California Institute of Technology and the University of California.

Some of the data presented herein were obtained at Keck Observatory, which is a private 501(c)3 non-profit organization operated as a scientific partnership among the California Institute of Technology, the University of California, and the National Aeronautics and Space Administration. The Observatory was made possible by the generous financial support of the W. M. Keck Foundation. The raw Keck Observatory data presented herein, once public, can be accessed from the Keck Observatory Archive (KOA), which is operated by the W. M. Keck Observatory and the NASA Exoplanet Science Institute (NExScI), under contract with the National Aeronautics and Space Administration.

The authors wish to recognize and acknowledge the very significant cultural role and reverence that the summit of Mauna Kea has always had within the indigenous Hawaiian community. We are most fortunate to have the opportunity to conduct observations from this mountain.

We thank Prof. Karen Meech, Dr. Elena Sabbi, the Gemini Observatory team, and the Shadow the Scientists (StS) team for supporting these observations and public observing sessions. Dr. Sabbi was PI of GS-2025B-DD-103, which provided the campaign's first Gemini South DDT observations. StS is made possible through the Creating Equity in STEAM (CrEST) program at the University of California, Santa Cruz, and support from the Heising-Simons Foundation.

Based on observations obtained at the international Gemini Observatory, a program of NSF NOIRLab, which is managed by the Association of Universities for Research in Astronomy (AURA) under a cooperative agreement with the U.S. National Science Foundation on behalf of the Gemini Observatory partnership: the U.S. National Science Foundation (United States), National Research Council (Canada), Agencia Nacional de Investigaci\'{o}n y Desarrollo (Chile), Ministerio de Ciencia, Tecnolog\'{i}a e Innovaci\'{o}n (Argentina), Minist\'{e}rio da Ci\^{e}ncia, Tecnologia, Inova\c{c}\~{o}es e Comunica\c{c}\~{o}es (Brazil), and Korea Astronomy and Space Science Institute (Republic of Korea).

This research benefited from B. T. Bolin's participation in the 2026 workshop ``Exploring Planetary Systems in the Era of Time-domain Astronomy'', hosted by the Institute for Astronomy at the University of Hawai'i and supported by award 2106927 from the National Science Foundation's Astronomy \& Astrophysics Research Grants program.

We used ChatGPT (OpenAI) to assist with manuscript editing, development of analysis code and figure preparation. The authors are responsible for the scientific results and the final manuscript.
\end{acknowledgments}
\facilities{Keck:I (LRIS, MOSFIRE), Gemini:North (GMOS, GNIRS), Gemini:South (GMOS, Flamingos-2), Liverpool:2m (IO:O)}
\software{LPipe \citep{PerleyLPipe2019}, Astropy, NumPy, SciPy, Matplotlib, Photutils, sbpy \citep{Mommert2019}, JPL Horizons}

\FloatBarrier

\bibliographystyle{aasjournalv7.1}
\bibliography{references/verified_references}
\end{document}

%% file: authors.tex
\author[0000-0002-4950-6323]{Bryce T. Bolin}
\affiliation{Eureka Scientific, Inc., 2452 Delmer Street, Oakland, CA 94602, USA}
\email[show]{bbolin@eurekasci.com}
\author[0000-0001-5703-2108]{Jeffrey Cooke}
\affiliation{Centre for Astrophysics and Supercomputing, Swinburne University of Technology, Hawthorn, VIC 3122, Australia}
\affiliation{ARC Centre of Excellence for Gravitational Wave Discovery, Swinburne University of Technology, Hawthorn, VIC 3122, Australia}
\email{jeffreycooke@swin.edu.au}
\author[0000-0002-1428-7036]{Brian C. Lemaux}
\affiliation{International Gemini Observatory/NSF NOIRLab, 670 N. A`ohoku Place, Hilo, HI 96720, USA}
\affiliation{Department of Physics and Astronomy, University of California, Davis, 1 Shields Avenue, Davis, CA 95616, USA}
\email{}
\author[0000-0001-7895-8209]{Marco Micheli}
\affiliation{ESA NEO Coordination Centre, Largo Galileo Galilei 1, I-00044 Frascati (RM), Italy}
\email{}
\author[0000-0003-0774-884X]{Davide Farnocchia}
\affiliation{Jet Propulsion Laboratory, California Institute of Technology, 4800 Oak Grove Drive, Pasadena, CA 91109, USA}
\email{}
\author[0009-0000-8709-5273]{Laura-May Abron}
\affiliation{Griffith Observatory, 2800 East Observatory Road, Los Angeles, CA 90027, USA}
\email{}
\author[0000-0002-2250-8687]{Leo Y. Alcorn}
\affiliation{W. M. Keck Observatory, 65-1120 Mamalahoa Highway, Kamuela, HI 96743, USA}
\email{}
\author[0000-0002-8262-2924]{Michael W. Coughlin}
\affiliation{School of Physics and Astronomy, University of Minnesota, 116 Church Street S.E., Minneapolis, MN 55455, USA}
\email{}
\author[0000-0002-4223-103X]{Christoffer Fremling}
\affiliation{Caltech Optical Observatories, California Institute of Technology, 1200 East California Boulevard, Pasadena, CA 91125, USA}
\email{}
\author[0000-0002-2934-3723]{Josef Hanu\v{s}}
\affiliation{Charles University, Faculty of Mathematics and Physics, Astronomical Institute, V Hole\v{s}ovi\v{c}k\'{a}ch 2, 180\,00 Prague, Czech Republic}
\email{}
\author[0000-0003-3953-9532]{Willem B. Hoogendam}
\affiliation{Institute for Astronomy, University of Hawai`i, 2680 Woodlawn Drive, Honolulu, HI 96822, USA}
\email{}
\author[0000-0002-7053-5495]{Carl Ingebretsen}
\affiliation{William H. Miller III Department of Physics and Astronomy, Johns Hopkins University, 3400 N. Charles Street, Baltimore, MD 21218, USA}
\email{}
\author[0000-0001-7285-373X]{Oleksandra Ivanova}
\affiliation{Astronomical Institute of the Slovak Academy of Sciences, P.O. Box 18, 059 60 Tatransk\'{a} Lomnica, Slovak Republic}
\affiliation{Main Astronomical Observatory of the National Academy of Sciences of Ukraine, 27 Akademika Zabolotnoho Street, UA-03143 Kyiv, Ukraine}
\email{}
\author[0000-0002-6230-0151]{David O. Jones}
\affiliation{Institute for Astronomy, University of Hawai`i, 640 N. A`ohoku Place, Hilo, HI 96720, USA}
\email{dojones@hawaii.edu}
\author{Piera King}
\affiliation{International Gemini Observatory/NSF NOIRLab}
\email{}
\author[0000-0002-9321-3202]{Ludmilla Kolokolova}
\affiliation{Department of Astronomy, University of Maryland, 4296 Stadium Drive, Room 1113, College Park, MD 20742-2421, USA}
\email{}
\author[0000-0002-6013-9384]{Keith S. Noll}
\affiliation{Planetary Science Institute, 1700 East Fort Lowell, Suite 106, Tucson, AZ 85719, USA}
\email{}
\author[0000-0002-5268-7735]{Filipp D. Romanov}
\affiliation{American Association of Variable Star Observers (AAVSO), 1770 Massachusetts Avenue, Suite 121, Cambridge, MA 02140, USA}
\email{filipp.romanov.27.04.1997@gmail.com}
\author[0000-0003-4631-1149]{Benjamin J. Shappee}
\affiliation{Institute for Astronomy, University of Hawai`i, 2680 Woodlawn Drive, Honolulu, HI 96822, USA}
\email{}
\author[0000-0003-1949-4621]{Karleyne Silva}
\affiliation{International Gemini Observatory/NSF NOIRLab, Casilla 603, La Serena, Chile}
\email{}
\author{Matt Wahl}
\affiliation{W. M. Keck Observatory, 65-1120 Mamalahoa Highway, Kamuela, HI 96743, USA}
\email{}
\author[0009-0006-8529-3229]{Pablo Candia}
\affiliation{International Gemini Observatory/NSF NOIRLab, Casilla 603, La Serena, Chile}
\email{}
\author[0000-0002-9020-5004]{Kristin Chiboucas}
\affiliation{International Gemini Observatory/NSF NOIRLab, 670 N. A`ohoku Place, Hilo, HI 96720, USA}
\email{}
\author[0000-0002-7272-9234]{Eleazar R. Carrasco}
\affiliation{International Gemini Observatory/NSF NOIRLab, Casilla 603, La Serena, Chile}
\email{}
\author[0000-0002-4628-3726]{Veronica Firpo}
\affiliation{International Gemini Observatory/NSF NOIRLab, Casilla 603, La Serena, Chile}
\email{}
\author[0000-0003-2530-3000]{Jeong-Eun Heo}
\affiliation{International Gemini Observatory/NSF NOIRLab, Casilla 603, La Serena, Chile}
\email{}
\author{Jameeka Marshall}
\affiliation{Department of Astronomy and Astrophysics, University of California, Santa Cruz, 1156 High Street, Santa Cruz, CA 95064, USA}
\email{}
\author{Emily Peavy}
\affiliation{International Gemini Observatory/NSF NOIRLab, 670 N. A`ohoku Place, Hilo, HI 96720, USA}
\email{}
\author{Erin Figura}
\affiliation{International Gemini Observatory/NSF NOIRLab, 670 N. A`ohoku Place, Hilo, HI 96720, USA}
\email{}
\author[0000-0001-7523-140X]{Denise Hung}
\affiliation{International Gemini Observatory/NSF NOIRLab, 670 N. A`ohoku Place, Hilo, HI 96720, USA}
\email{}
\author[0000-0001-9003-0999]{Ariel Graykowski}
\affiliation{SETI Institute, Carl Sagan Center, 339 Bernardo Avenue, Suite 200, Mountain View, CA 94043, USA}
\email{}
\author[0000-0003-4603-556X]{Teo Mo\v{c}nik}
\affiliation{International Gemini Observatory/NSF NOIRLab, 670 N. A`ohoku Place, Hilo, HI 96720, USA}
\email{}
\author{Matias Rodriguez}
\affiliation{International Gemini Observatory/NSF NOIRLab, 670 N. A`ohoku Place, Hilo, HI 96720, USA}
\email{}
\author[0000-0003-0773-1888]{David J. Tholen}
\affiliation{Institute for Astronomy, University of Hawai`i, 2680 Woodlawn Drive, Honolulu, HI 96822, USA}
\email{}
\author[0000-0002-5396-946X]{Peter Vere\v{s}}
\affiliation{Minor Planet Center, Center for Astrophysics | Harvard \& Smithsonian, 60 Garden Street, MS-15, Cambridge, MA 02138, USA}
\email{}
\author{Leila Alamos}
\affiliation{International Gemini Observatory/NSF NOIRLab, 670 N. A`ohoku Place, Hilo, HI 96720, USA}
\email{}
\author[0000-0002-3370-187X]{Eunchong Kim}
\affiliation{International Gemini Observatory/NSF NOIRLab, 670 N. A`ohoku Place, Hilo, HI 96720, USA}
\email{}
\author[0009-0006-0758-2113]{Jen Miller}
\affiliation{International Gemini Observatory/NSF NOIRLab, 670 N. A`ohoku Place, Hilo, HI 96720, USA}
\email{}
\author[0000-0003-0408-9850]{Percy Gomez}
\affiliation{W. M. Keck Observatory, 65-1120 Mamalahoa Highway, Kamuela, HI 96743, USA}
\email{}
\author{Heather Hershley}
\affiliation{W. M. Keck Observatory, 65-1120 Mamalahoa Highway, Kamuela, HI 96743, USA}
\email{}
\author[0000-0001-9589-3793]{Michael Lundquist}
\affiliation{W. M. Keck Observatory, 65-1120 Mamalahoa Highway, Kamuela, HI 96743, USA}
\email{}
\author{Rita Morris}
\affiliation{W. M. Keck Observatory, 65-1120 Mamalahoa Highway, Kamuela, HI 96743, USA}
\email{}
\author[0000-0002-2536-1633]{Hyewon Suh}
\affiliation{International Gemini Observatory/NSF NOIRLab, 670 N. A`ohoku Place, Hilo, HI 96720, USA}
\email{hyewon.suh@noirlab.edu}
\author[0000-0002-6822-2254]{Emanuele Paolo Farina}
\affiliation{International Gemini Observatory/NSF NOIRLab, 670 N. A`ohoku Place, Hilo, HI 96720, USA}
\email{emanuele.farina@noirlab.edu}
\author[0000-0003-1724-2885]{Ryan Ridden}
\affiliation{School of Physical and Chemical Sciences, University of Canterbury, Christchurch, New Zealand}
\email{ryan.ridden@canterbury.ac.nz}
\author[0000-0002-3569-1038]{Mark Suhr}
\affiliation{Centre for Astrophysics and Supercomputing, Swinburne University of Technology, Hawthorn, VIC 3122, Australia}
\email{msuhr@swin.edu.au}

%% file: observation_table.tex
\begingroup\scriptsize\setlength{\tabcolsep}{2pt}
\begin{longtable}{@{}lllrrrrrrr@{}}
\caption{Individual imaging filters and spectroscopic settings.}\label{tab:observations}\\
\toprule
UTC date/time & Facility/instrument & Filter/grating & $t_{\rm exp}$ & $X$ & $r_h$ & $\Delta$ & $\alpha$ & $\delta_E$ & $t-T_p$\\\midrule\endfirsthead
\multicolumn{10}{l}{Table~\ref{tab:observations} continued}\\\toprule
UTC date/time & Facility/instrument & Filter/grating & $t_{\rm exp}$ & $X$ & $r_h$ & $\Delta$ & $\alpha$ & $\delta_E$ & $t-T_p$\\\midrule\endhead
\bottomrule\endfoot
2025-07-11 07:10 & K1/MOSFIRE & Y$^\ast$ & 611 & 1.62 & 4.172 & 3.217 & 5.4 & -0.85 & -110.2\\
2025-07-11 07:09 & K1/MOSFIRE & J & 707 & 1.59 & 4.172 & 3.217 & 5.4 & -0.85 & -110.2\\
2025-07-11 07:18 & K1/MOSFIRE & H & 786 & 1.55 & 4.172 & 3.217 & 5.4 & -0.85 & -110.2\\
2025-07-11 07:27 & K1/MOSFIRE & Ks & 786 & 1.50 & 4.172 & 3.217 & 5.4 & -0.85 & -110.2\\
2025-08-27 23:58 & GS/GMOS-S & g & 60 & 1.17 & 2.637 & 2.589 & 22.3 & 0.40 & -62.5\\
2025-08-27 23:56 & GS/GMOS-S & r & 60 & 1.17 & 2.637 & 2.589 & 22.3 & 0.40 & -62.5\\
2025-08-27 23:54 & GS/GMOS-S & i & 60 & 1.16 & 2.637 & 2.589 & 22.3 & 0.40 & -62.5\\
2025-08-28 00:02 & GS/GMOS-S & u & 240 & 1.18 & 2.637 & 2.589 & 22.3 & 0.40 & -62.5\\
2025-08-28 00:23 & GS/GMOS-S & B480+/540 & 900 & 1.24 & 2.636 & 2.589 & 22.3 & 0.40 & -62.5\\
2025-08-28 00:42 & GS/GMOS-S & B480+/550 & 900 & 1.32 & 2.636 & 2.589 & 22.3 & 0.40 & -62.5\\
2025-09-03 23:30$^\dagger$ & GS/F2 & HK grism & 1020 & 1.26 & 2.426 & 2.567 & 23.1 & 0.62 & -55.5\\
2025-09-05 23:59$^\dagger$ & GS/F2 & JH grism & 340 & 1.41 & 2.367 & 2.562 & 23.2 & 0.68 & -53.5\\
2025-09-05 23:39 & GS/GMOS-S & u & 100 & 1.30 & 2.367 & 2.562 & 23.2 & 0.68 & -53.5\\
2025-09-05 23:37 & GS/GMOS-S & g & 100 & 1.30 & 2.367 & 2.562 & 23.2 & 0.68 & -53.5\\
2025-09-05 23:36 & GS/GMOS-S & r & 100 & 1.29 & 2.367 & 2.562 & 23.2 & 0.68 & -53.5\\
2025-09-05 23:35 & GS/GMOS-S & i & 100 & 1.29 & 2.367 & 2.562 & 23.2 & 0.68 & -53.5\\
2025-09-05 23:34 & GS/GMOS-S & Z & 100 & 1.28 & 2.367 & 2.562 & 23.2 & 0.68 & -53.5\\
2025-09-06 23:32 & GS/GMOS-S & g & 25 & 1.30 & 2.338 & 2.559 & 23.2 & 0.71 & -52.5\\
2025-09-06 23:30 & GS/GMOS-S & r & 25 & 1.29 & 2.338 & 2.559 & 23.2 & 0.71 & -52.5\\
2025-09-06 23:27 & GS/GMOS-S & i & 25 & 1.28 & 2.338 & 2.559 & 23.2 & 0.71 & -52.5\\
2025-09-06 23:24 & GS/GMOS-S & Z & 25 & 1.27 & 2.338 & 2.559 & 23.2 & 0.71 & -52.5\\
2025-09-08 23:35 & GS/GMOS-S & u & 100 & 1.37 & 2.280 & 2.555 & 23.2 & 0.77 & -50.5\\
2025-09-08 23:34 & GS/GMOS-S & g & 100 & 1.36 & 2.280 & 2.555 & 23.2 & 0.77 & -50.5\\
2025-09-08 23:33 & GS/GMOS-S & r & 100 & 1.36 & 2.280 & 2.555 & 23.2 & 0.77 & -50.5\\
2025-09-08 23:32 & GS/GMOS-S & i & 100 & 1.35 & 2.280 & 2.555 & 23.2 & 0.77 & -50.5\\
2025-09-08 23:31 & GS/GMOS-S & Z & 100 & 1.35 & 2.280 & 2.555 & 23.2 & 0.77 & -50.5\\
2025-09-14 23:38 & GS/GMOS-S & g & 50 & 1.62 & 2.111 & 2.542 & 22.6 & 0.95 & -44.5\\
2025-09-14 23:36 & GS/GMOS-S & r & 60 & 1.60 & 2.111 & 2.542 & 22.6 & 0.95 & -44.5\\
2025-09-15 00:04 & GS/GMOS-S & B480+/540 & 1200 & 1.86 & 2.110 & 2.542 & 22.6 & 0.95 & -44.5\\
2025-09-15 00:28 & GS/GMOS-S & B480+/550 & 1200 & 2.18 & 2.110 & 2.542 & 22.6 & 0.95 & -44.5\\
2025-11-26 15:46 & GN/GMOS-N & g & 36 & 1.35 & 1.702 & 1.962 & 30.2 & 2.38 & 28.2\\
2025-11-26 15:47 & GN/GMOS-N & r & 24 & 1.35 & 1.702 & 1.962 & 30.2 & 2.38 & 28.2\\
2025-11-26 15:48 & GN/GMOS-N & i & 14 & 1.34 & 1.702 & 1.962 & 30.2 & 2.38 & 28.2\\
2025-11-26 15:45 & GN/GMOS-N & Z & 16 & 1.35 & 1.702 & 1.962 & 30.2 & 2.38 & 28.2\\
2025-11-26 14:33 & GN/GMOS-N & B480+/540 & 60 & 1.91 & 1.700 & 1.963 & 30.2 & 2.38 & 28.1\\
2025-11-26 14:39 & GN/GMOS-N & B480+/550 & 60 & 1.83 & 1.701 & 1.962 & 30.2 & 2.38 & 28.1\\
2025-12-01 14:28$^\dagger$ & GN/GNIRS & 32/mm, B1 & 1708 & 1.46 & 1.815 & 1.905 & 30.6 & 2.40 & 33.1\\
2025-12-01 15:12$^\dagger$ & GN/GNIRS & 32/mm, B2 & 1220 & 1.28 & 1.816 & 1.905 & 30.6 & 2.40 & 33.1\\
2025-12-02 14:28$^\dagger$ & GN/GNIRS & 32/mm, B1 & 1464 & 1.43 & 1.840 & 1.895 & 30.6 & 2.40 & 34.1\\
2025-12-02 15:03$^\dagger$ & GN/GNIRS & 32/mm, B2 & 1464 & 1.27 & 1.840 & 1.895 & 30.6 & 2.40 & 34.1\\
2025-12-19 12:08 & GN/GMOS-N & g & 10 & 1.46 & 2.295 & 1.798 & 24.1 & 2.16 & 51.0\\
2025-12-22 12:02 & GN/GMOS-N & g & 40 & 1.34 & 2.382 & 1.802 & 22.1 & 2.07 & 54.0\\
2025-12-22 12:03 & GN/GMOS-N & r & 40 & 1.34 & 2.382 & 1.802 & 22.1 & 2.07 & 54.0\\
2025-12-22 12:04 & GN/GMOS-N & i & 40 & 1.33 & 2.382 & 1.802 & 22.1 & 2.07 & 54.0\\
2025-12-22 12:05 & GN/GMOS-N & Z & 40 & 1.33 & 2.382 & 1.802 & 22.1 & 2.07 & 54.0\\
2025-12-22 12:29 & GN/GMOS-N & B480+/540 & 180 & 1.23 & 2.383 & 1.802 & 22.1 & 2.07 & 54.0\\
2025-12-22 12:38 & GN/GMOS-N & B480+/550 & 180 & 1.20 & 2.383 & 1.802 & 22.1 & 2.07 & 54.0\\
2026-01-20 09:53 & GN/GMOS-N & g & 50 & 1.03 & 3.279 & 2.297 & 1.4 & 0.78 & 82.9\\
2026-01-20 09:55 & GN/GMOS-N & r & 50 & 1.02 & 3.279 & 2.297 & 1.4 & 0.78 & 82.9\\
2026-01-20 09:57 & GN/GMOS-N & i & 50 & 1.02 & 3.279 & 2.298 & 1.4 & 0.78 & 82.9\\
2026-01-20 10:01 & GN/GMOS-N & Z & 100 & 1.02 & 3.279 & 2.298 & 1.4 & 0.78 & 82.9\\
2026-01-21 08:03 & GN/GMOS-N & B480+/540 & 450 & 1.25 & 3.309 & 2.326 & 1.0 & 0.74 & 83.8\\
2026-01-21 08:17 & GN/GMOS-N & B480+/550 & 450 & 1.20 & 3.309 & 2.326 & 1.0 & 0.74 & 83.9\\
2026-02-05 05:49 & GN/GMOS-N & g & 50 & 1.40 & 3.794 & 2.868 & 5.9 & 0.17 & 98.7\\
2026-02-05 05:51 & GN/GMOS-N & r & 50 & 1.39 & 3.794 & 2.868 & 5.9 & 0.17 & 98.7\\
2026-02-05 05:53 & GN/GMOS-N & i & 50 & 1.38 & 3.794 & 2.868 & 5.9 & 0.17 & 98.8\\
2026-02-05 05:56 & GN/GMOS-N & Z & 100 & 1.36 & 3.794 & 2.868 & 5.9 & 0.17 & 98.8\\
2026-02-06 -- & K1/MOSFIRE & J & 52 & 1.08 & 3.833 & 2.917 & 6.2 & 0.14 & 99.9\\
2026-02-06 -- & K1/MOSFIRE & Ks & 140 & 1.07 & 3.833 & 2.917 & 6.2 & 0.14 & 99.9\\
2026-02-07 07:42 & GN/GMOS-N & g & 50 & 1.04 & 3.862 & 2.954 & 6.5 & 0.11 & 100.8\\
2026-02-07 07:44 & GN/GMOS-N & r & 50 & 1.03 & 3.862 & 2.954 & 6.5 & 0.11 & 100.8\\
2026-02-07 07:46 & GN/GMOS-N & i & 50 & 1.03 & 3.863 & 2.954 & 6.5 & 0.11 & 100.8\\
2026-02-07 07:49 & GN/GMOS-N & Z & 100 & 1.03 & 3.863 & 2.954 & 6.5 & 0.11 & 100.8\\
2026-02-17 09:59 & GN/GMOS-N & g & 72 & 1.16 & 4.196 & 3.397 & 8.8 & -0.16 & 110.9\\
2026-02-17 10:07 & GN/GMOS-N & r & 144 & 1.19 & 4.196 & 3.398 & 8.8 & -0.16 & 110.9\\
2026-02-17 10:04 & GN/GMOS-N & i & 72 & 1.18 & 4.196 & 3.398 & 8.8 & -0.16 & 110.9\\
2026-02-17 10:08 & GN/GMOS-N & Z & 144 & 1.19 & 4.196 & 3.398 & 8.8 & -0.16 & 110.9\\
2026-02-17 10:31 & GN/GMOS-N & B480+/540 & 500 & 1.27 & 4.197 & 3.399 & 8.8 & -0.16 & 110.9\\
2026-02-17 10:45 & GN/GMOS-N & B480+/550 & 500 & 1.33 & 4.197 & 3.399 & 8.8 & -0.16 & 111.0\\
2026-03-27 08:26 & GN/GMOS-N & g & 120 & 1.56 & 5.465 & 5.293 & 10.5 & -0.62 & 148.9\\
2026-03-27 08:29 & GN/GMOS-N & r & 120 & 1.59 & 5.465 & 5.293 & 10.5 & -0.62 & 148.9\\
2026-03-27 08:32 & GN/GMOS-N & i & 120 & 1.61 & 5.465 & 5.293 & 10.5 & -0.62 & 148.9\\
2026-03-27 08:37 & GN/GMOS-N & Z & 240 & 1.65 & 5.466 & 5.294 & 10.5 & -0.62 & 148.9\\
2026-03-27 07:30 & GN/GMOS-N & B480+/540 & 1800 & 1.26 & 5.464 & 5.291 & 10.5 & -0.62 & 148.8\\
2026-03-27 08:06 & GN/GMOS-N & B480+/550 & 1800 & 1.43 & 5.465 & 5.292 & 10.5 & -0.62 & 148.8\\
2026-05-15 06:16 & K1/LRIS & G & 1380 & -- & 7.118 & 7.761 & 6.0 & -0.62 & 197.8\\
2026-05-15 06:16 & K1/LRIS & R & 900 & -- & 7.118 & 7.761 & 6.0 & -0.62 & 197.8\\
2026-05-15 06:35 & K1/LRIS & I & 720 & -- & 7.119 & 7.762 & 6.0 & -0.62 & 197.8\\
2025-09-01 00:00$^\ddagger$ & K1/LRIS & 400/3400 & -- & 2.00 & 2.515 & 2.575 & 22.8 & 0.52 & -58.5\\
2025-09-01 00:00$^\ddagger$ & K1/LRIS & 400/8500 & -- & 2.10 & 2.515 & 2.575 & 22.8 & 0.52 & -58.5\\
\end{longtable}
\noindent\footnotesize Times are UTC exposure-weighted midpoints, except $^\dagger$ (input start time) and $^\ddagger$ (confirmed September 1 date; 00:00 supplies geometry because the midpoint is unavailable). $^\ast$July $Y$ is an uncalibrated stack with incomplete target coverage; airmass comes from its reference exposure and 611 s is the total input time, with nonuniform coverage. Exposure times are in seconds, distances in au, angles in degrees, and $t-T_p$ in days. $\delta_E$ is the signed orbital-plane viewing angle \citep{Bolin2021LD2}. GS/GN and K1 denote Gemini South/North and Keck I. GMOS B480+ uses G5327 (South) or G5309 (North) with 1-arcsec slits. GNIRS uses 32/mmSB\_G5533, orders 3--7, and a 0.68-arcsec slit; F2 uses JH\_G5801 or HK\_G5802 and an 8-pixel (1.43-arcsec) slit. LRIS uses 400/3400 blue and 400/8500 red dispersers. Reduced headers verify neither the adopted 0.7-arcsec LRIS slit nor binning; its 70-pixel extraction spans 9.45 arcsec assuming unbinned 0.135-arcsec pixels. The electronic exposure log identifies correlated calibrated alternatives, counted once.
\endgroup

%% file: generated/figure_optical_images.tex
\begingroup\renewcommand{\theHfigure}{\arabic{figure}.optical_images.1}
\begin{figure}[!htbp]\centering
\includegraphics[page=1,width=\textwidth,height=.68\textheight,keepaspectratio]{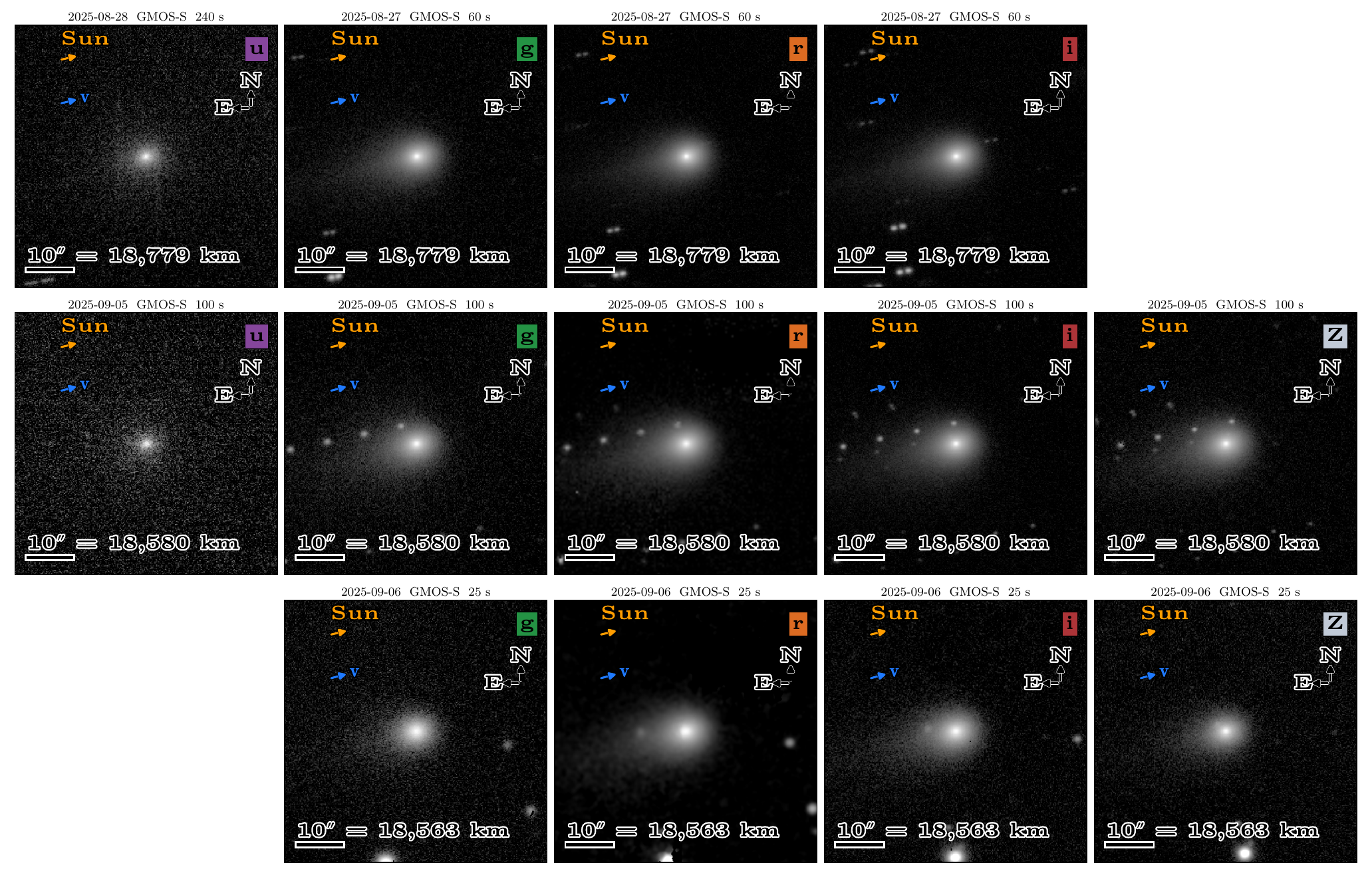}
\caption{Gemini South GMOS imaging in $u,g,r,i,Z$ columns. August 28 $u$ shares the August 27 $gri$ row and carries its own date. Panels give date, summed integration, compass, scale, Sun direction, and positive heliocentric-velocity direction. North is up and east left; absent filters are blank. September 5, 8, and 14 $r$ use 0.48-arcsec valid-pixel bins with bilinear rendering. September 6 $r$ uses a mask-normalized 0.94-arcsec-FWHM Gaussian average to suppress mask holes, reducing display resolution. Photometry and profiles use unsmoothed data. Exposure coverage is spatially nonuniform.}\label{fig:optical_images}
\end{figure}\endgroup

\begingroup\renewcommand{\theHfigure}{\arabic{figure}.optical_images.2}
\addtocounter{figure}{-1}
\begin{figure}[!htbp]\centering
\includegraphics[page=2,width=\textwidth,height=.68\textheight,keepaspectratio]{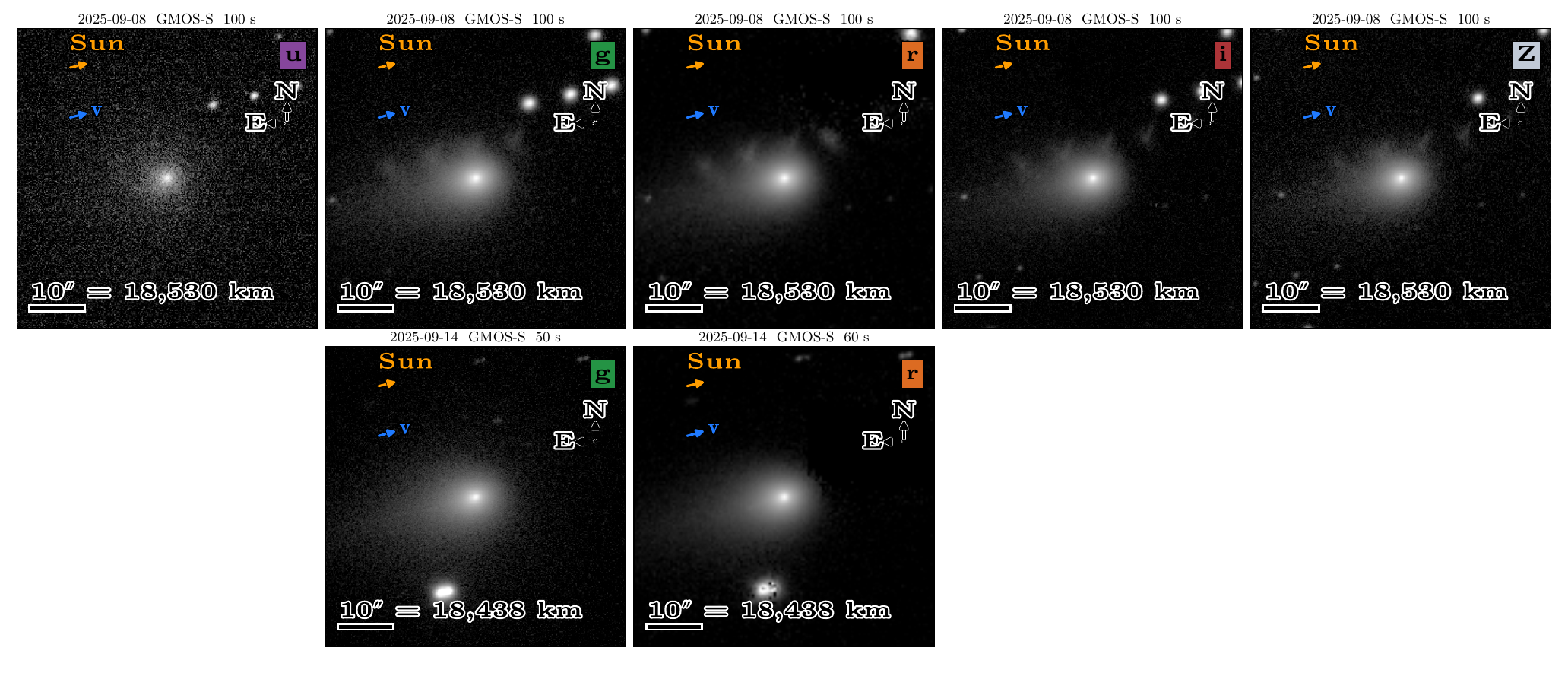}
\caption{Continued.}\label{fig:optical_images_p2}
\end{figure}\endgroup

%% file: generated/figure_optical_north.tex
\begingroup\renewcommand{\theHfigure}{\arabic{figure}.optical_north.1}
\begin{figure}[!htbp]\centering
\includegraphics[page=1,width=\textwidth,height=.68\textheight,keepaspectratio]{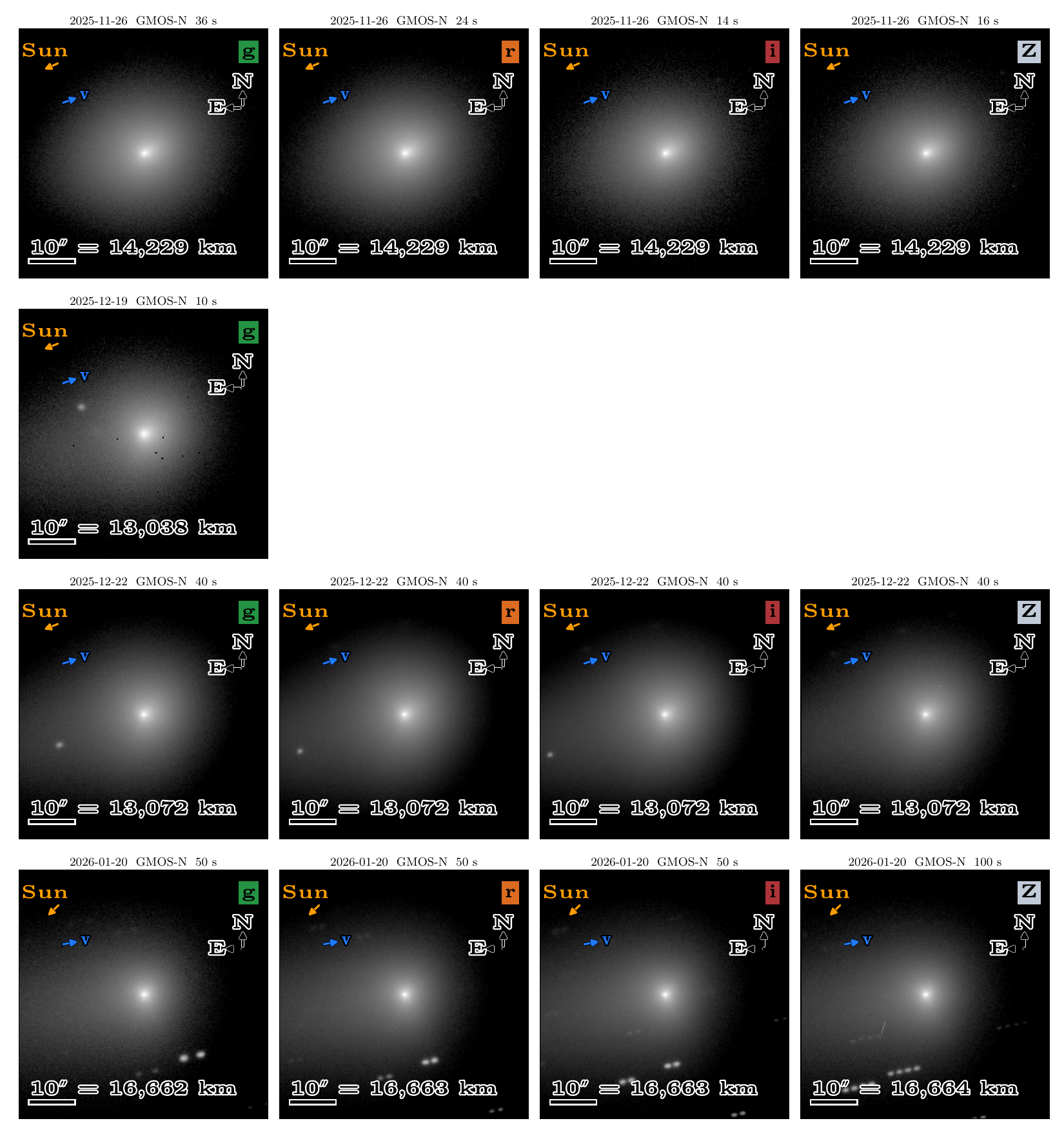}
\caption{Gemini North GMOS and Keck LRIS imaging in $g,r/R,i,Z$ columns, following Figure~\ref{fig:optical_images}. Absent filters are blank and the LRIS row is centered. White circles locate the comet in March GMOS and May LRIS images; masked pixels are black.}\label{fig:optical_north}
\end{figure}\endgroup

\begingroup\renewcommand{\theHfigure}{\arabic{figure}.optical_north.2}
\addtocounter{figure}{-1}
\begin{figure}[!htbp]\centering
\includegraphics[page=2,width=\textwidth,height=.68\textheight,keepaspectratio]{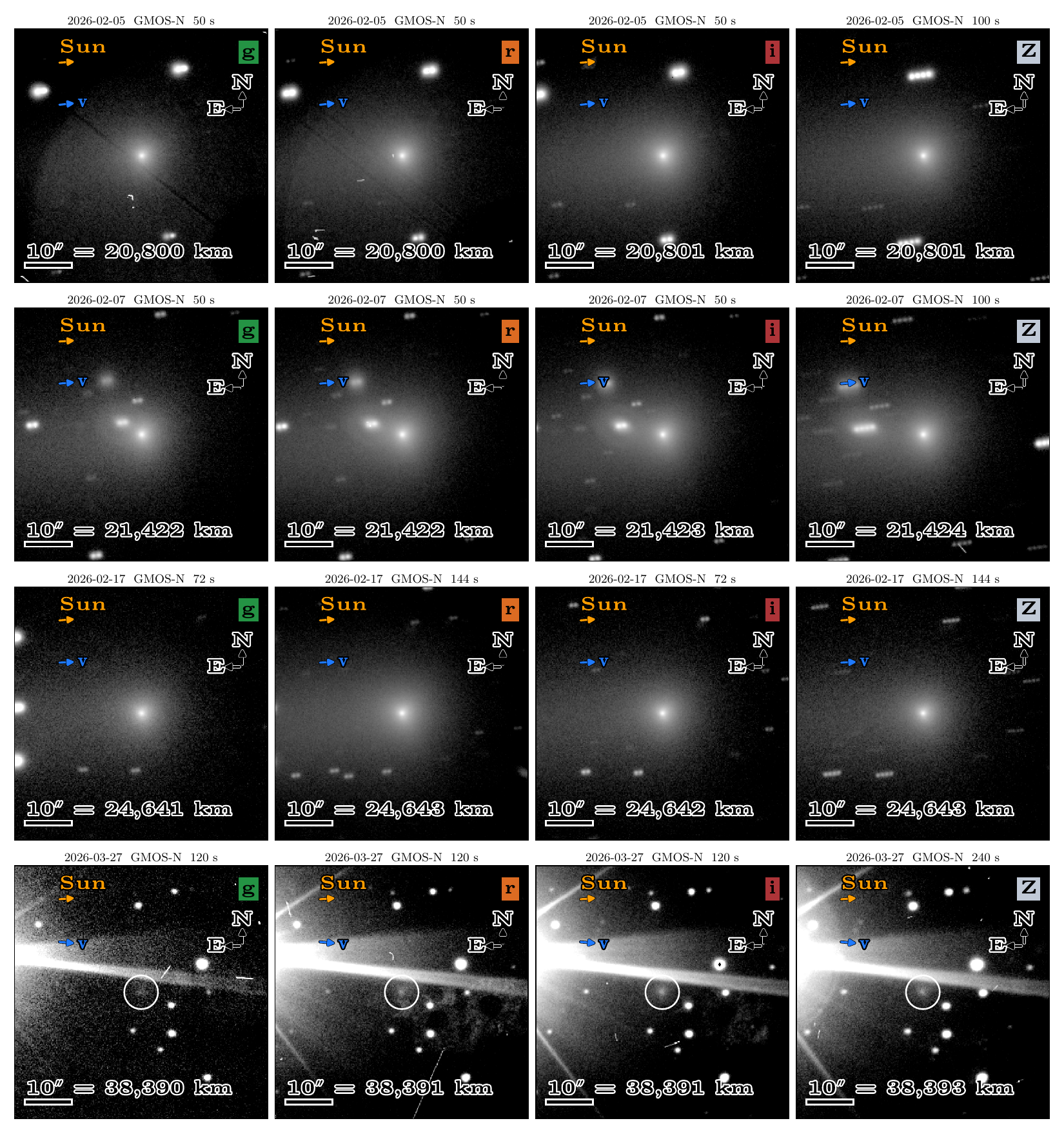}
\caption{Continued.}\label{fig:optical_north_p2}
\end{figure}\endgroup

\begingroup\renewcommand{\theHfigure}{\arabic{figure}.optical_north.3}
\addtocounter{figure}{-1}
\begin{figure}[!htbp]\centering
\includegraphics[page=3,width=\textwidth,height=.68\textheight,keepaspectratio]{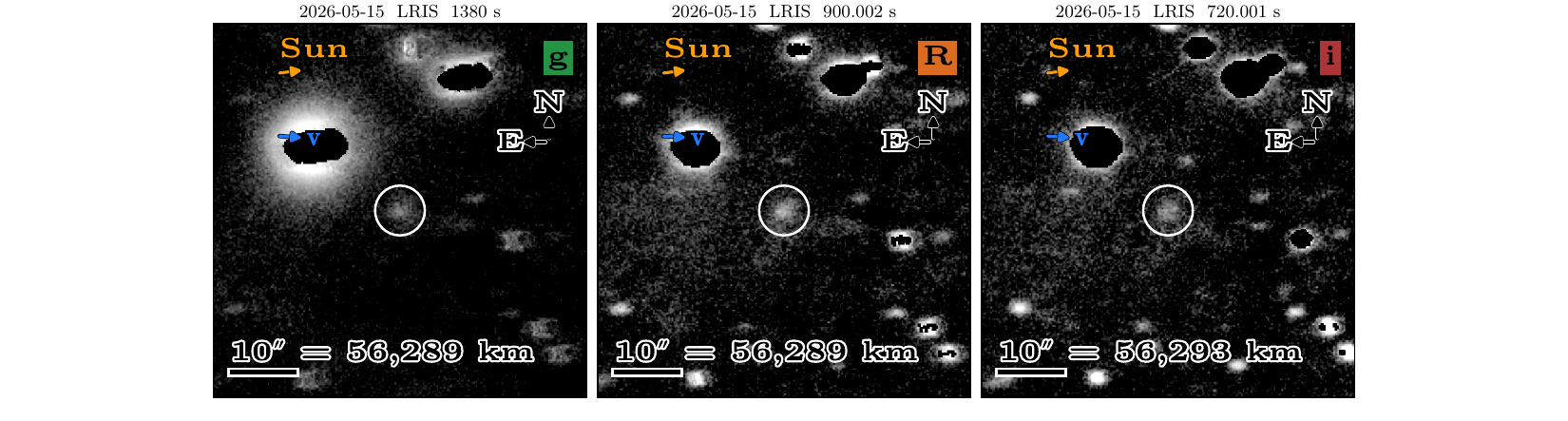}
\caption{Continued.}\label{fig:optical_north_p3}
\end{figure}\endgroup

%% file: photometry_table.tex
\begingroup\small\setlength{\tabcolsep}{4pt}
\begin{longtable}{lllrrrr}
\caption{Photometry and Halley--Marcus phase-corrected activity in 10,000-km-radius apertures.}\label{tab:conditionalphot}\\
\toprule
UTC date & Instrument & Band & $t-T_p$ (d) & $\theta_\rho$ (arcsec) & Magnitude & $A(0)f\rho$ (cm)\\\midrule\endfirsthead
\multicolumn{7}{l}{Table~\ref{tab:conditionalphot} continued}\\\toprule
UTC date & Instrument & Band & $t-T_p$ (d) & $\theta_\rho$ (arcsec) & Magnitude & $A(0)f\rho$ (cm)\\\midrule\endhead
\bottomrule\endfoot
2025-08-27 & GMOS-S & g & -62.5 & 5.33 & $16.512\pm0.036$ & $552\pm18$\\
2025-08-27 & GMOS-S & i & -62.5 & 5.33 & $15.495\pm0.030$ & $826\pm23$\\
2025-08-27 & GMOS-S & r & -62.5 & 5.33 & $15.758\pm0.030$ & $725\pm20$\\
2025-08-28 & GMOS-S & u & -62.5 & 5.33 & $17.694\pm0.067$ & $610\pm38$\\
2025-09-05 & GMOS-S & z & -53.5 & 5.38 & $14.984\pm0.033$ & $1045\pm31$\\
2025-09-05 & GMOS-S & g & -53.5 & 5.38 & $16.102\pm0.040$ & $649\pm24$\\
2025-09-05 & GMOS-S & i & -53.5 & 5.38 & $15.109\pm0.031$ & $949\pm27$\\
2025-09-08 & GMOS-S & z & -50.5 & 5.40 & $14.831\pm0.033$ & $1111\pm33$\\
2025-09-08 & GMOS-S & g & -50.5 & 5.40 & $15.913\pm0.040$ & $712\pm26$\\
2025-09-08 & GMOS-S & i & -50.5 & 5.40 & $14.956\pm0.030$ & $1010\pm28$\\
2025-09-08 & GMOS-S & u & -50.5 & 5.40 & $16.575\pm0.091$ & $1271\pm106$\\
2025-09-14 & GMOS-S & r$^{M}$ & -44.5 & 5.42 & $14.775\pm0.208$ & $1115\pm213$\\
2025-11-26 & GMOS-N & z$^{c}$ & 28.2 & 7.03 & $12.026\pm0.033$ & $5552\pm167$\\
2025-11-26 & GMOS-N & r$^{M}$ & 28.2 & 7.03 & $12.523\pm0.202$ & $3998\pm745$\\
2025-12-22 & GMOS-N & r$^{c}$ & 54.0 & 7.65 & $13.583\pm0.030$ & $2114\pm59$\\
2026-01-20 & GMOS-N & z & 82.9 & 6.00 & $14.663\pm0.033$ & $1090\pm33$\\
2026-01-20 & GMOS-N & r & 82.9 & 6.00 & $15.011\pm0.030$ & $900\pm25$\\
2026-02-05 & GMOS-N & z & 98.8 & 4.81 & $16.057\pm0.033$ & $749\pm23$\\
2026-02-07 & GMOS-N & z & 100.8 & 4.67 & $16.222\pm0.032$ & $724\pm21$\\
2026-02-07 & GMOS-N & g & 100.8 & 4.67 & $17.176\pm0.036$ & $522\pm17$\\
2026-02-07 & GMOS-N & i & 100.8 & 4.67 & $16.296\pm0.030$ & $688\pm19$\\
2026-02-07 & GMOS-N & r & 100.8 & 4.67 & $16.512\pm0.030$ & $631\pm17$\\
2026-02-17 & GMOS-N & z & 110.9 & 4.06 & $16.955\pm0.032$ & $625\pm18$\\
2026-02-17 & GMOS-N & g & 110.9 & 4.06 & $18.079\pm0.036$ & $386\pm13$\\
2026-02-17 & GMOS-N & i & 110.9 & 4.06 & $17.068\pm0.030$ & $573\pm16$\\
2026-02-17 & GMOS-N & r & 110.9 & 4.06 & $17.302\pm0.030$ & $517\pm14$\\
2026-05-15 & LRIS & r & 197.8 & 1.78 & $22.024\pm0.116$ & $91\pm10$\\
2026-05-15 & LRIS & g & 197.8 & 1.78 & $22.876\pm0.140$ & $63\pm8$\\
2026-05-15 & LRIS & i & 197.8 & 1.78 & $21.753\pm0.197$ & $104\pm19$\\
2025-07-17 & Liverpool$^L$ & r & -103.5 & 4.51 & $17.119\pm0.030$ & $428\pm12$\\
2025-08-02 & Liverpool$^L$ & g & -87.6 & 4.94 & $17.225\pm0.271$ & $458\pm114$\\
2025-08-02 & Liverpool$^L$ & r & -87.6 & 4.94 & $16.595\pm0.020$ & $536\pm10$\\
2025-08-13 & Liverpool$^L$ & g & -76.6 & 5.16 & $16.629\pm0.020$ & $657\pm12$\\
2025-08-13 & Liverpool$^L$ & r & -76.6 & 5.16 & $16.109\pm0.010$ & $694\pm6$\\
2025-08-14 & Liverpool$^L$ & g & -75.6 & 5.18 & $16.685\pm0.091$ & $612\pm51$\\
2025-08-14 & Liverpool$^L$ & r & -75.6 & 5.18 & $16.055\pm0.060$ & $717\pm39$\\
2025-08-17 & Liverpool$^L$ & g & -72.6 & 5.22 & $16.479\pm0.030$ & $699\pm19$\\
2025-08-17 & Liverpool$^L$ & r & -72.6 & 5.22 & $15.909\pm0.010$ & $773\pm7$\\
2025-08-27 & Liverpool$^L$ & g & -62.6 & 5.32 & $16.054\pm0.020$ & $845\pm16$\\
2025-08-27 & Liverpool$^L$ & r & -62.6 & 5.32 & $15.553\pm0.080$ & $877\pm65$\\
2025-09-03 & Liverpool$^L$ & g & -55.6 & 5.36 & $15.837\pm0.010$ & $877\pm8$\\
2025-09-03 & Liverpool$^L$ & r & -55.6 & 5.36 & $15.277\pm0.010$ & $961\pm9$\\
2025-12-06 & Liverpool$^L$ & g & 37.7 & 7.41 & $12.734\pm0.030$ & $5806\pm161$\\
2025-12-07 & Liverpool$^L$ & g & 38.7 & 7.45 & $13.038\pm0.020$ & $4456\pm82$\\
2025-12-19 & Liverpool$^L$ & g & 50.7 & 7.66 & $12.803\pm0.030$ & $6426\pm177$\\
2025-12-19 & Liverpool$^L$ & r & 50.7 & 7.66 & $12.433\pm0.020$ & $5910\pm109$\\
2026-01-11 & Liverpool$^L$ & g & 73.6 & 6.73 & $14.099\pm0.030$ & $2617\pm73$\\
2026-01-11 & Liverpool$^L$ & r & 73.6 & 6.73 & $13.729\pm0.020$ & $2410\pm44$\\
2026-01-20 & Liverpool$^L$ & g & 82.6 & 6.02 & $14.785\pm0.030$ & $1684\pm47$\\
2026-01-20 & Liverpool$^L$ & r & 82.6 & 6.02 & $14.205\pm0.030$ & $1880\pm52$\\
2026-01-28 & Liverpool$^L$ & g & 91.4 & 5.34 & $15.905\pm0.030$ & $962\pm26$\\
2026-01-28 & Liverpool$^L$ & r & 91.4 & 5.34 & $15.255\pm0.020$ & $1146\pm21$\\
2026-02-08 & Liverpool$^L$ & g & 102.5 & 4.55 & $16.605\pm0.050$ & $969\pm45$\\
2026-02-08 & Liverpool$^L$ & r & 102.5 & 4.55 & $16.075\pm0.030$ & $1034\pm29$\\
2026-02-21 & Liverpool$^L$ & g & 115.4 & 3.82 & $17.506\pm0.080$ & $809\pm60$\\
2026-02-21 & Liverpool$^L$ & r & 115.4 & 3.82 & $17.196\pm0.040$ & $704\pm26$\\
\end{longtable}
\noindent\footnotesize Magnitudes are approximate SDSS AB equivalents; native values and transformations are retained electronically. Tonry et al. (2012) stellar transformations leave cometary emission-band differences uncorrected. Without simultaneous $g-r$, PS1 $0.60\pm0.15$ supplies the passband correction. LRIS $R$ uses measured $g-R$. $^c$: small missing-area correction. $^M$: MPC-scaled SDSS-like $r$, with a 0.20-mag systematic floor, excluded from colors. $^L$: published Liverpool Sloan photometry, adjusted by less than 0.008 mag to the common aperture, without further passband or fitted zero-point shifts. Gas can contribute to $g$-band activity.
\endgroup

%% file: generated/figure_color_color.tex
\begingroup\renewcommand{\theHfigure}{\arabic{figure}.color_color.1}
\begin{figure}[!htbp]\centering
\includegraphics[page=1,width=\textwidth,height=.68\textheight,keepaspectratio]{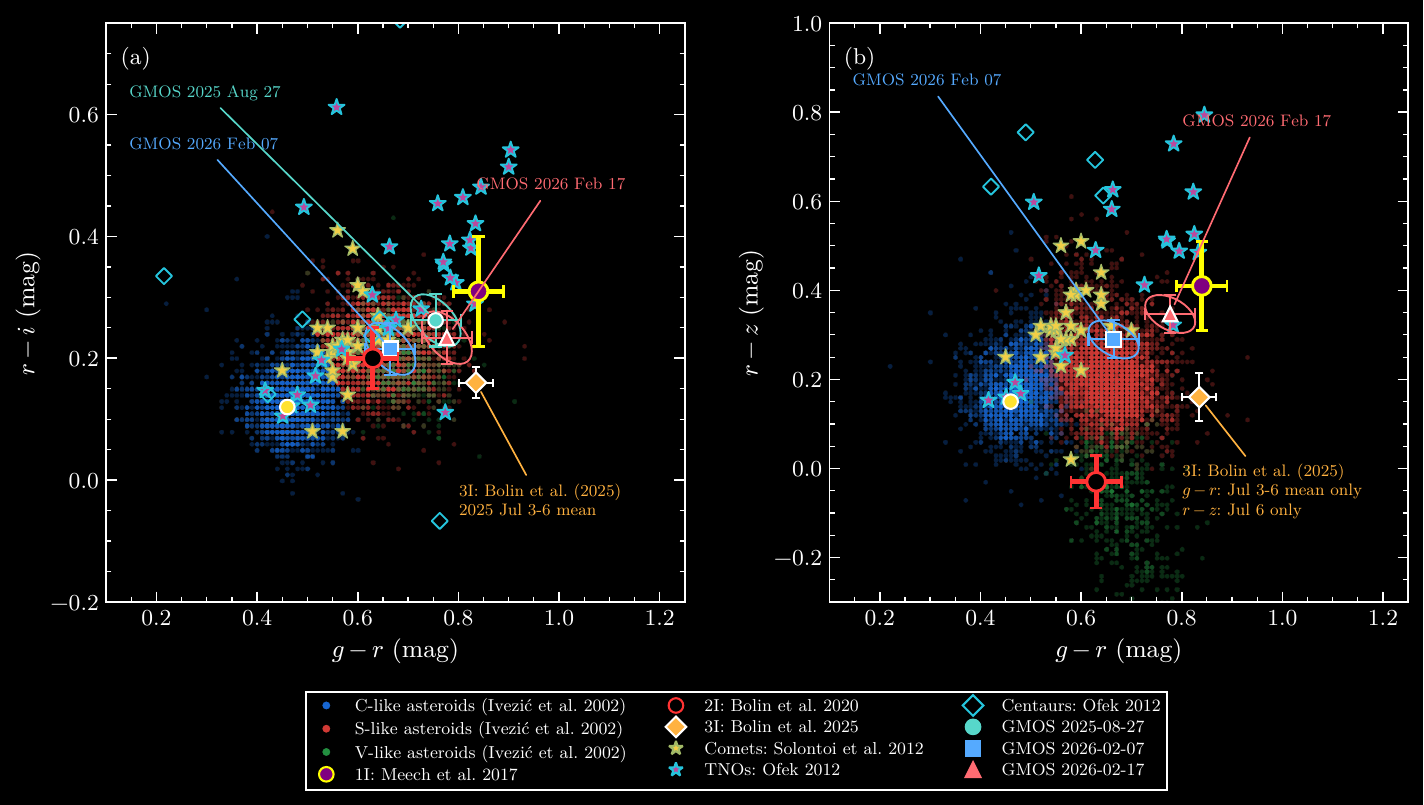}
\caption{Optical colors: $g-r$ versus $r-i$ (left) and $r-z$ (right). Blue/red/green asteroid groups follow \citet{Ivezic2002}; comet and outer-Solar-System samples follow \citet{Solontoi2012} and \citet{Ofek2012}. The 1I point is purple with yellow errors; 2I has a black core with red outline and errors, using the values in \citet{Bolin2025Discovery}. Left: arithmetic means of the July 3 and 6 discovery colors, $g-r=0.835\pm0.034$ and $r-i=0.160\pm0.025$ mag, assuming independent epoch errors and excluding shared calibration uncertainty. Right: the same July 3--6 $g-r$ mean paired with July 6-only $r-z=0.160\pm0.054$ mag. Dated GMOS points use the approximate stellar PS1-to-SDSS transformation \citep{Tonry2012}; ellipses include shared-band covariance. Three epochs have matched calibrated $gri$ colors, two also with $Z$.}\label{fig:color_color}
\end{figure}\endgroup

%% file: generated/figure_colors_time.tex
\begingroup\renewcommand{\theHfigure}{\arabic{figure}.colors_time.1}
\begin{figure}[!htbp]\centering
\includegraphics[page=1,width=\textwidth,height=.68\textheight,keepaspectratio]{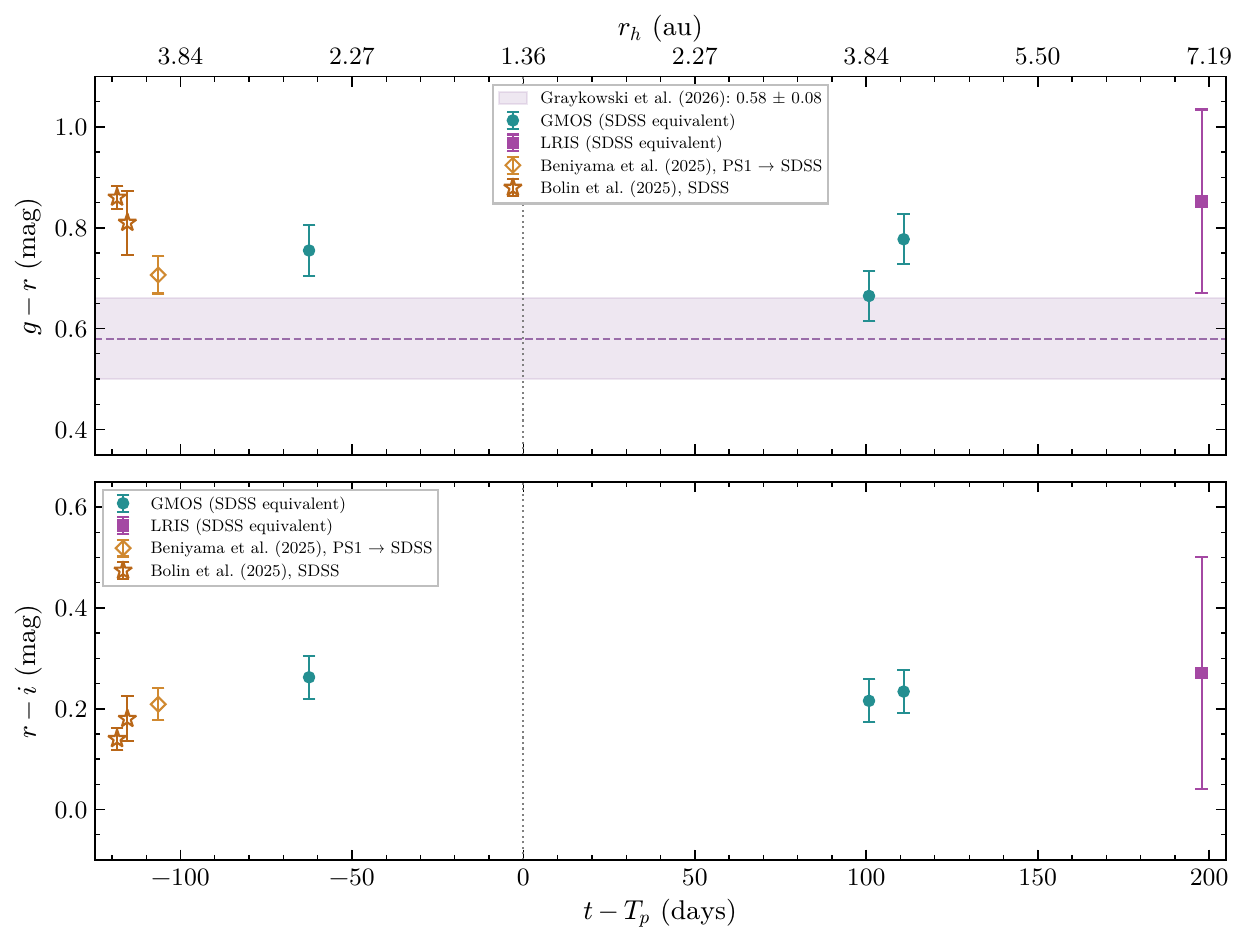}
\caption{Colors versus $t-T_p$, using $T_p=\mathrm{JD}\,2460977.995263$ TDB from JPL Horizons solution 54 \citep{Giorgini1996}, converted to UTC. GMOS/LRIS colors are approximate SDSS equivalents; P200/APO retain published SDSS values. July PS1 colors \citep{Beniyama2025} use the same Tonry transformation. The purple band spans the observing interval of \citet{Graykowski2026}, with weighted mean $g-r=0.58$ mag and reported scatter 0.08 mag; no $r-i$ mean is available. All matched epochs and errors are shown.}\label{fig:colors_time}
\end{figure}\endgroup

%% file: generated/figure_activity.tex
\begingroup\renewcommand{\theHfigure}{\arabic{figure}.activity.1}
\begin{figure}[!htbp]\centering
\includegraphics[page=1,width=\textwidth,height=.68\textheight,keepaspectratio]{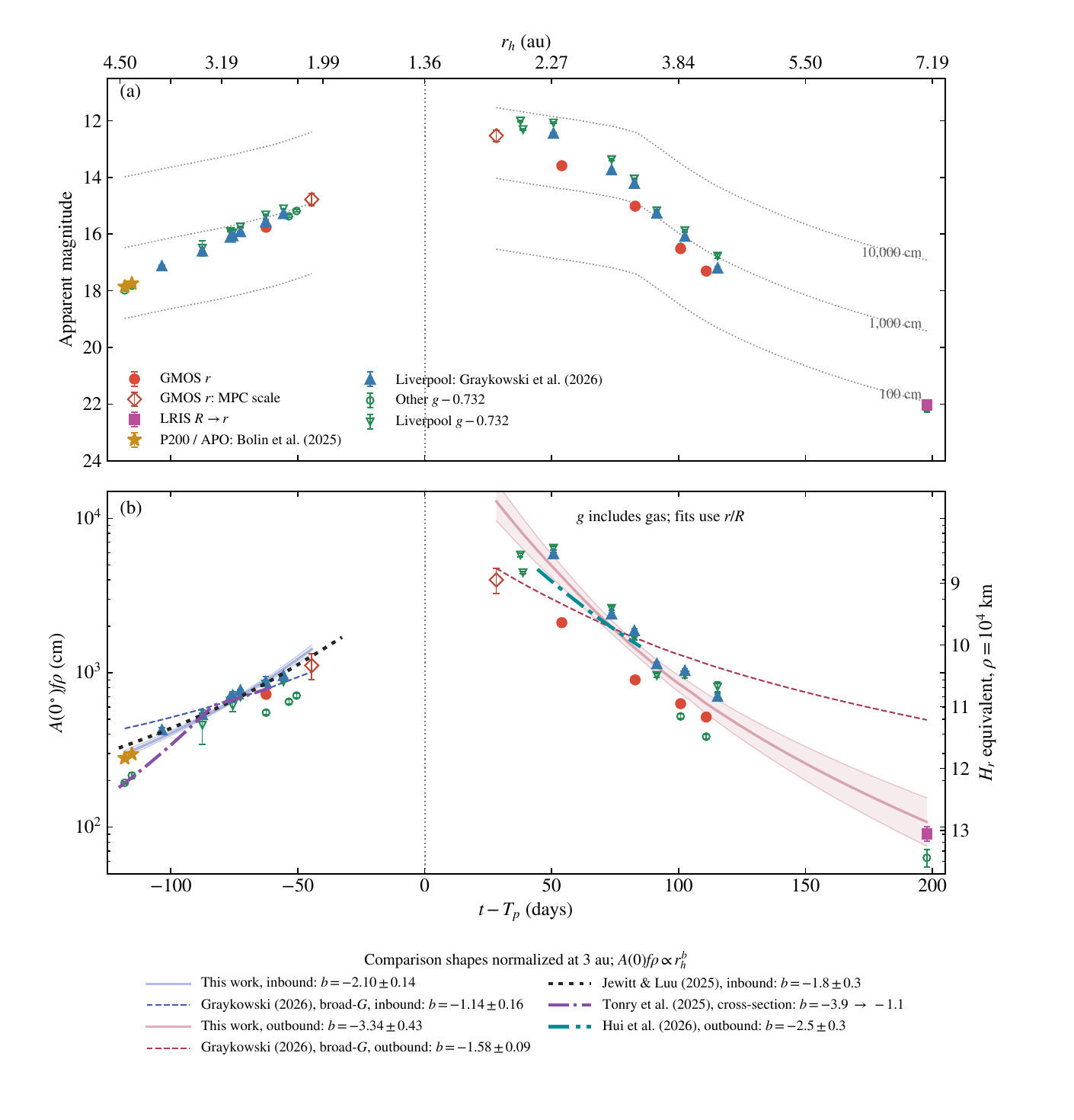}
\caption{Brightness and Halley--Marcus phase-corrected activity versus time from perihelion; the upper axis gives heliocentric distance. Top: approximate SDSS $r$ and $g-0.732$ in a 10,000-km-radius aperture, with constant-$\afrho$ curves. Blue triangles mark Liverpool $r$ (13 epochs), and open green inverted triangles mark shifted $g$ (14 epochs) \citep{Graykowski2026}; open diamonds mark MPC-scaled points. Liverpool magnitudes retain their published zero points, with the small aperture adjustment described in the text. Bottom: red-band fits exclude gas-bearing $g$. Solid curves and shading show our fits and formal $1\sigma$ errors on the mean; the residual scatter is larger. Comparison slopes are $b=-1.14\pm0.16$ inbound and $-1.58\pm0.09$ outbound for broad $G$ \citep[blue/red dashed]{Graykowski2026}; $-1.8\pm0.3$ \citep[black dotted]{JewittLuu2025Development}; $-3.9$ to $-1.1$ across 3.3 au \citep[purple dot-dashed]{Tonry3IATLASPhotometry2025}; and $-2.5\pm0.3$ outbound \citep[teal dash-dot]{Hui2026}. These laws are normalized to our corresponding branch at 3 au and cover the published intervals, including Hui's December--January \textit{HST} data in 200--1000-km-radius apertures. Aperture and phase-law differences limit comparisons to curve shape. Shading applies only to our fits. The right axis gives geometry-normalized $r$ brightness; dotted vertical lines mark perihelion. Some errors are smaller than the symbols.}\label{fig:activity}
\end{figure}\endgroup

%% file: generated/figure_crosssection.tex
\begingroup\renewcommand{\theHfigure}{\arabic{figure}.crosssection.1}
\begin{figure}[!htbp]\centering
\includegraphics[page=1,width=\textwidth,height=.68\textheight,keepaspectratio]{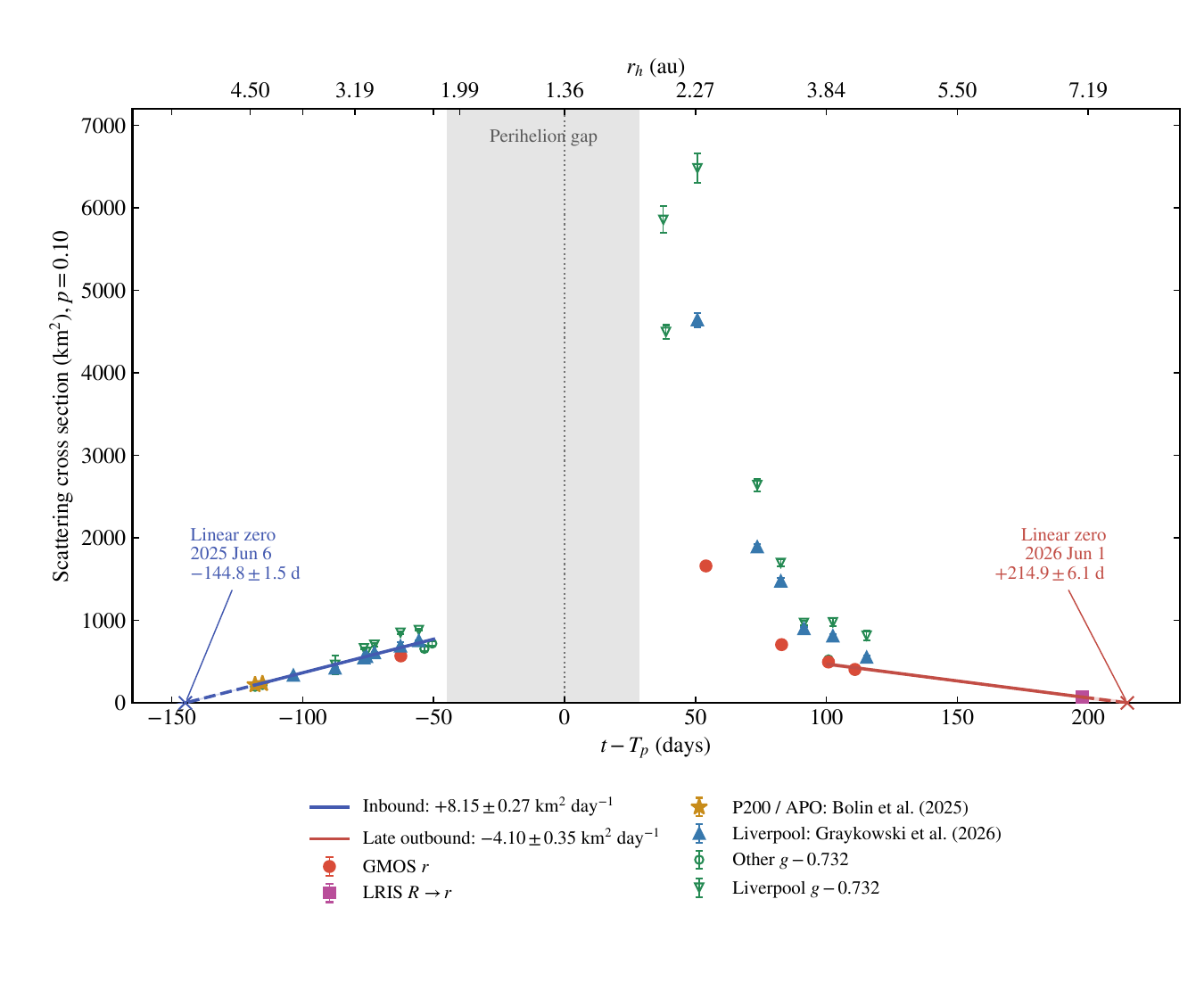}
\caption{Scattering cross section in a 10,000-km-radius aperture, for albedo 0.10, the Halley--Marcus phase law, and approximate SDSS photometry. Triangles show Liverpool data \citep{Graykowski2026}; stars show discovery photometry \citep{Bolin2025Discovery}. The two MPC-normalized GMOS $r$ estimates are omitted from the display and retained in the numerical analysis. Liverpool values use the common aperture and phase correction, with no fitted normalization offsets. Shifted $g$ is a gas-bearing $r$ proxy; fits use $r/R$. The dotted vertical line marks perihelion; gray shading marks the observing gap. Blue/red trends cover $-150\leq t-T_p\leq-50$ days inbound and $+100\leq t-T_p\leq+214.9$ days outbound. Bands give formal $1\sigma$ errors on the fitted mean, scaled by residual scatter. The late fit uses February 7, February 17, and May 15 GMOS/LRIS data; dashed segments extrapolate beyond the fitted epochs. The inbound zero, UT 2025 June 6 ($t-T_p=-144.8\pm1.5$ days), follows the TESS and ZTF prediscovery detections \citep{MartinezPalomera2025TESS,Ye2025Prediscovery}. The outbound zero, UT 2026 June 1 ($t-T_p=+214.9\pm6.1$ days), extrapolates the linear decline. Some errors are smaller than the symbols.}\label{fig:crosssection}
\end{figure}\endgroup

%% file: generated/figure_enhancement.tex
\begingroup\renewcommand{\theHfigure}{\arabic{figure}.enhancement.1}
\begin{figure}[!htbp]\centering
\includegraphics[page=1,width=\textwidth,height=.68\textheight,keepaspectratio]{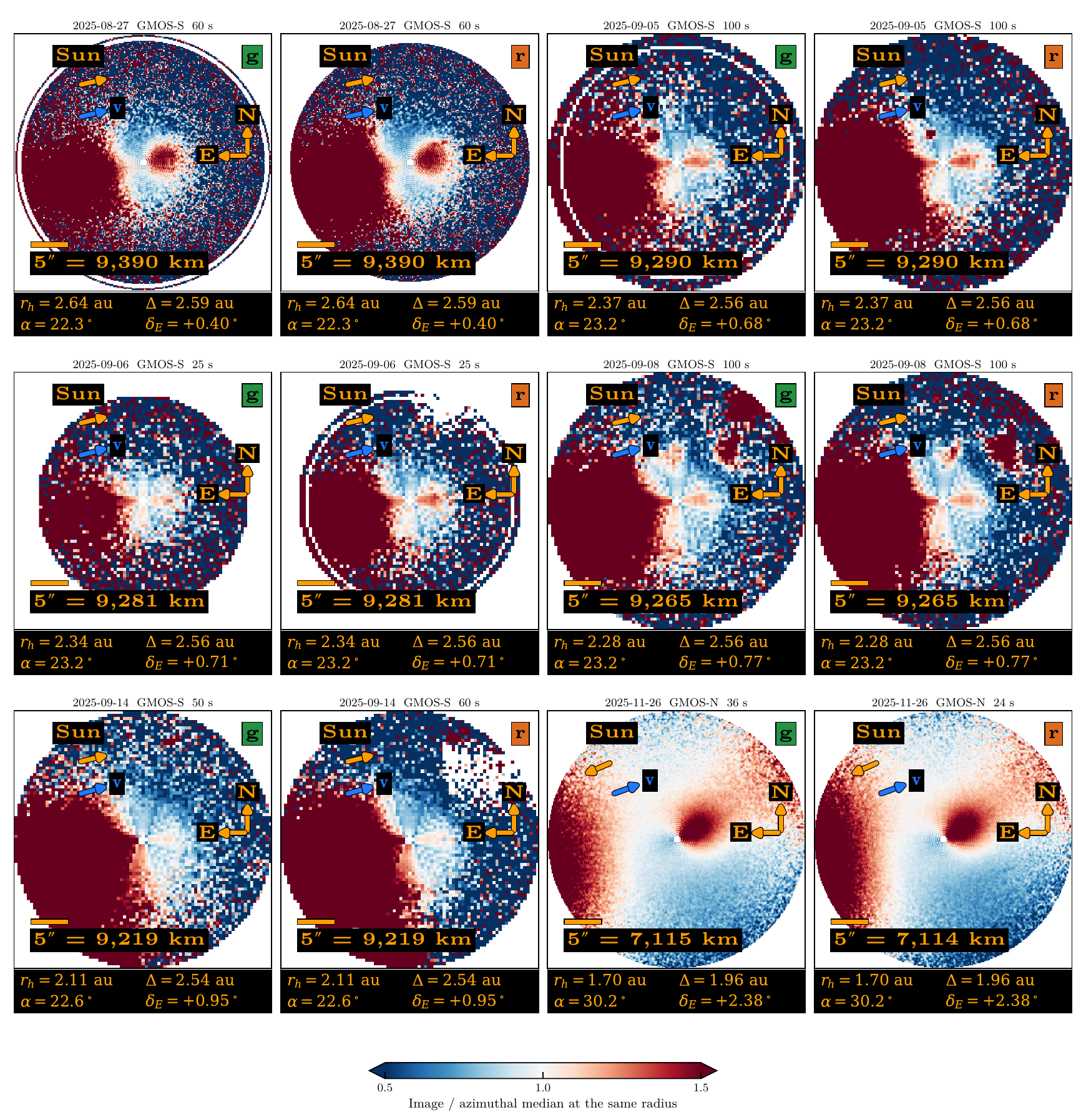}
\caption{Azimuthal-median division of paired GMOS $g/r$ images, with absent bands blank and March excluded. Orange labels give $r_h$, $\Delta$, $\alpha$, and $\delta_E$ (Table~\ref{tab:observations}). Compass/Sun arrows and scale bars are orange; heliocentric-velocity arrows are blue, with black outlines and label backgrounds. North is up and east left. The ratio scale shows deficits in blue, excesses in red, and unity in white; values beyond 0.5--1.5 saturate and masked pixels remain blank. The four September pairs use matched 0.48-arcsec native-area bins and radial annuli; other dates use 0.32-arcsec annuli on a 0.16-arcsec grid. No additional filter is applied. Detector masks, stellar residuals, and centering sensitivity limit jet identification.}\label{fig:enhancement}
\end{figure}\endgroup

\begingroup\renewcommand{\theHfigure}{\arabic{figure}.enhancement.2}
\addtocounter{figure}{-1}
\begin{figure}[!htbp]\centering
\includegraphics[page=2,width=\textwidth,height=.68\textheight,keepaspectratio]{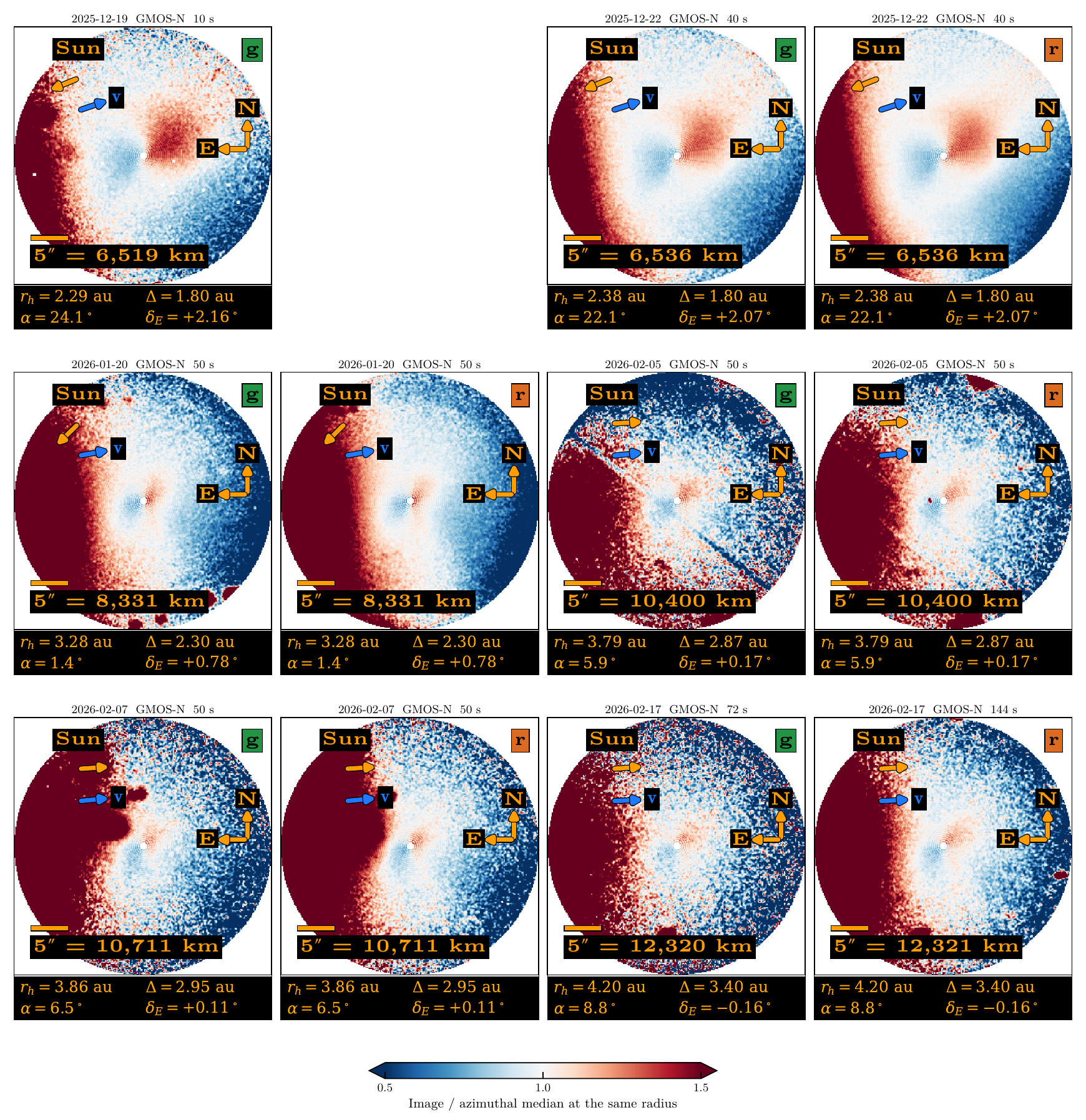}
\caption{Continued.}\label{fig:enhancement_p2}
\end{figure}\endgroup

%% file: generated/figure_dust.tex
\begingroup\renewcommand{\theHfigure}{\arabic{figure}.dust.1}
\begin{figure}[!htbp]\centering
\includegraphics[page=1,width=\textwidth,height=.68\textheight,keepaspectratio]{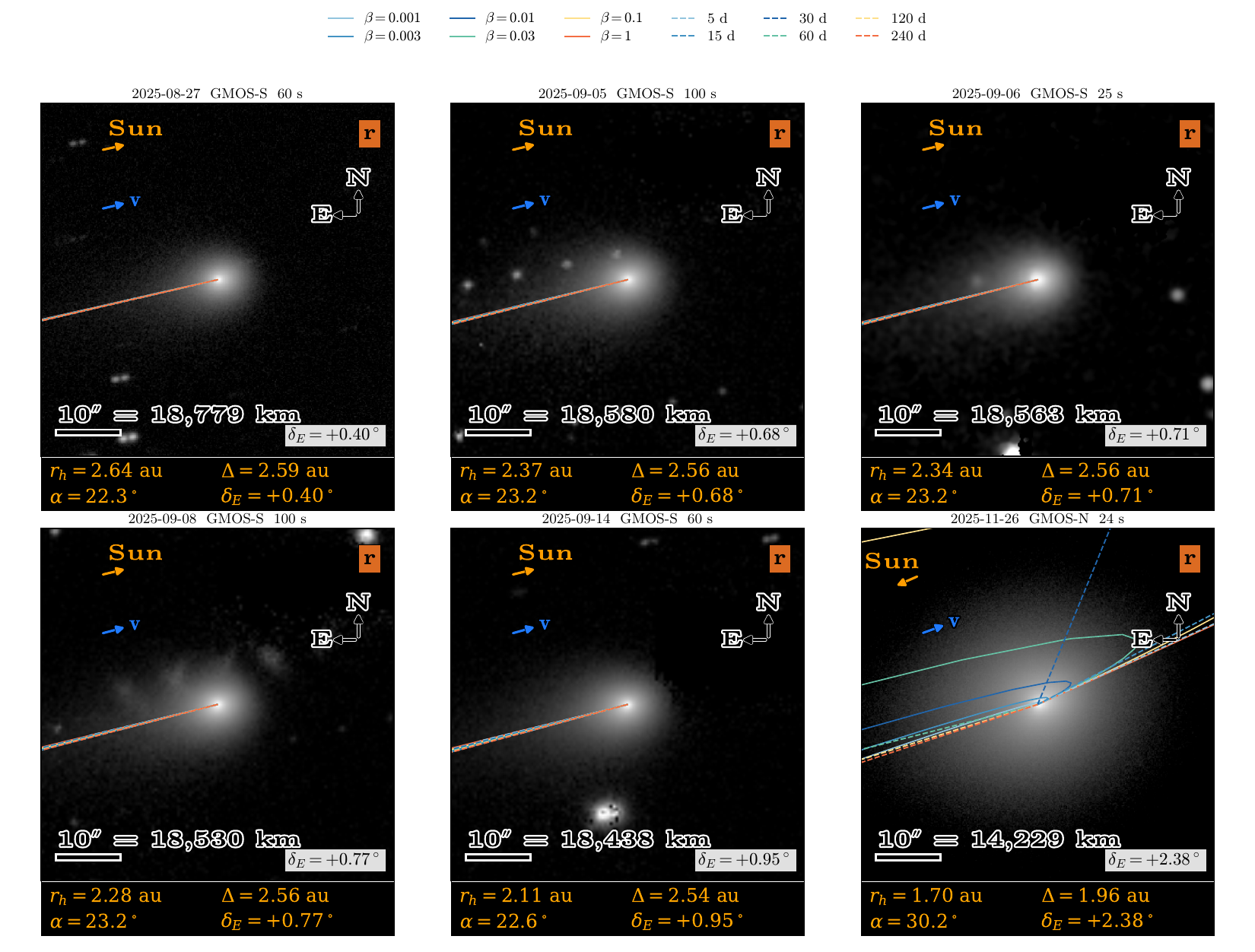}
\caption{Syndynes and synchrones on GMOS $r$ images, excluding March and LRIS. Legends give radiation-pressure parameter $\beta$ and release age; geometry uses imaging midpoints. Sun arrows are orange and heliocentric-velocity arrows blue. Trajectories assume zero ejection speed and include solar gravity and radiation pressure. Overlapping inbound curves limit grain-size and release-time constraints; no size distribution or production history is fitted. September images use the display treatments in Figure~\ref{fig:optical_images}, including the September 6 mask-normalized average. Orange labels give $r_h$, $\Delta$, phase angle $\alpha$, and signed orbital-plane viewing angle $\delta_E$ (Table~\ref{tab:observations}).}\label{fig:dust}
\end{figure}\endgroup

\begingroup\renewcommand{\theHfigure}{\arabic{figure}.dust.2}
\addtocounter{figure}{-1}
\begin{figure}[!htbp]\centering
\includegraphics[page=2,width=\textwidth,height=.68\textheight,keepaspectratio]{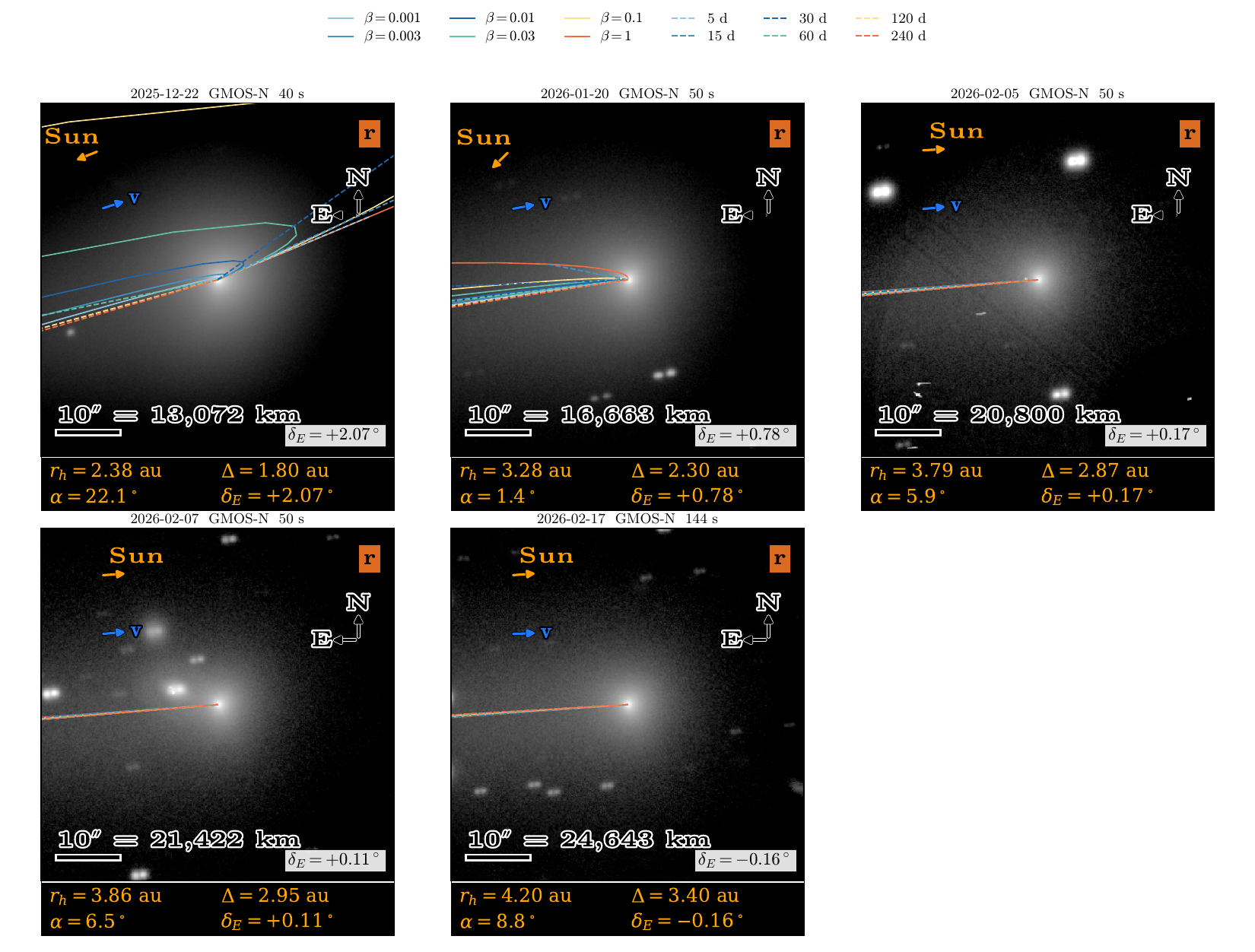}
\caption{Continued.}\label{fig:dust_p2}
\end{figure}\endgroup

%% file: generated/figure_tail_width_examples.tex
\begin{figure}[!htbp]\centering
\includegraphics[width=.91\textwidth]{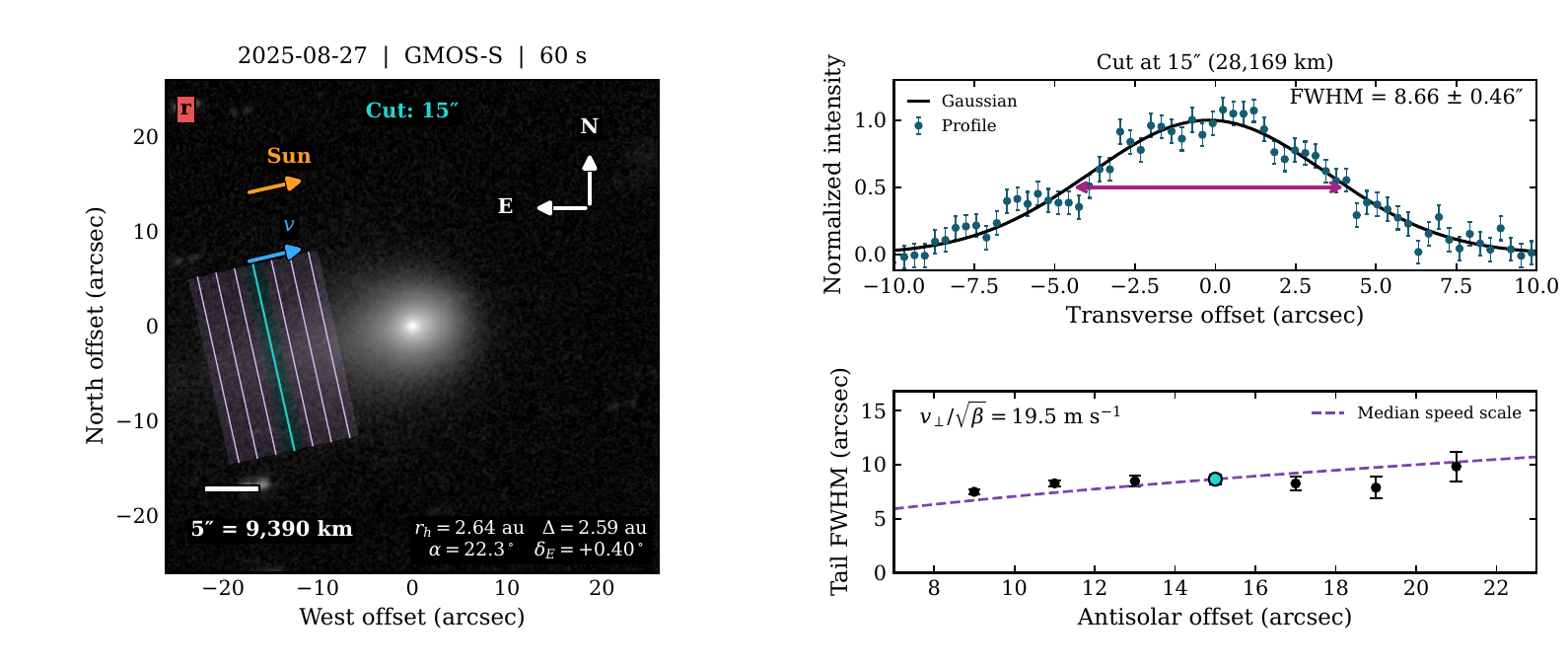}
\caption{GMOS $r$ transverse-width measurement on 2025 August 27. Left: masked image, north up and east left, with accepted 2-arcsec strips; cyan marks the example cut. Sun/heliocentric-velocity arrows are orange/blue. Upper right: profile after linear-background subtraction, normalized to the Gaussian amplitude; magenta marks FWHM. Errors use sky sampling scaled by $\sqrt{\max(1,\chi_\nu^2)}$. Lower right: accepted widths versus projected antisolar distance \citep{Jewitt2019Gault}; the dashed curve uses the epoch-median $U_\perp$. Widths include PSF broadening; formal errors exclude projection, grain-age, and PSF systematics.}\label{fig:tail_width_examples}
\end{figure}

%% file: generated/figure_sunward_width_examples.tex
\begin{figure}[!htbp]\centering
\includegraphics[width=.93\textwidth]{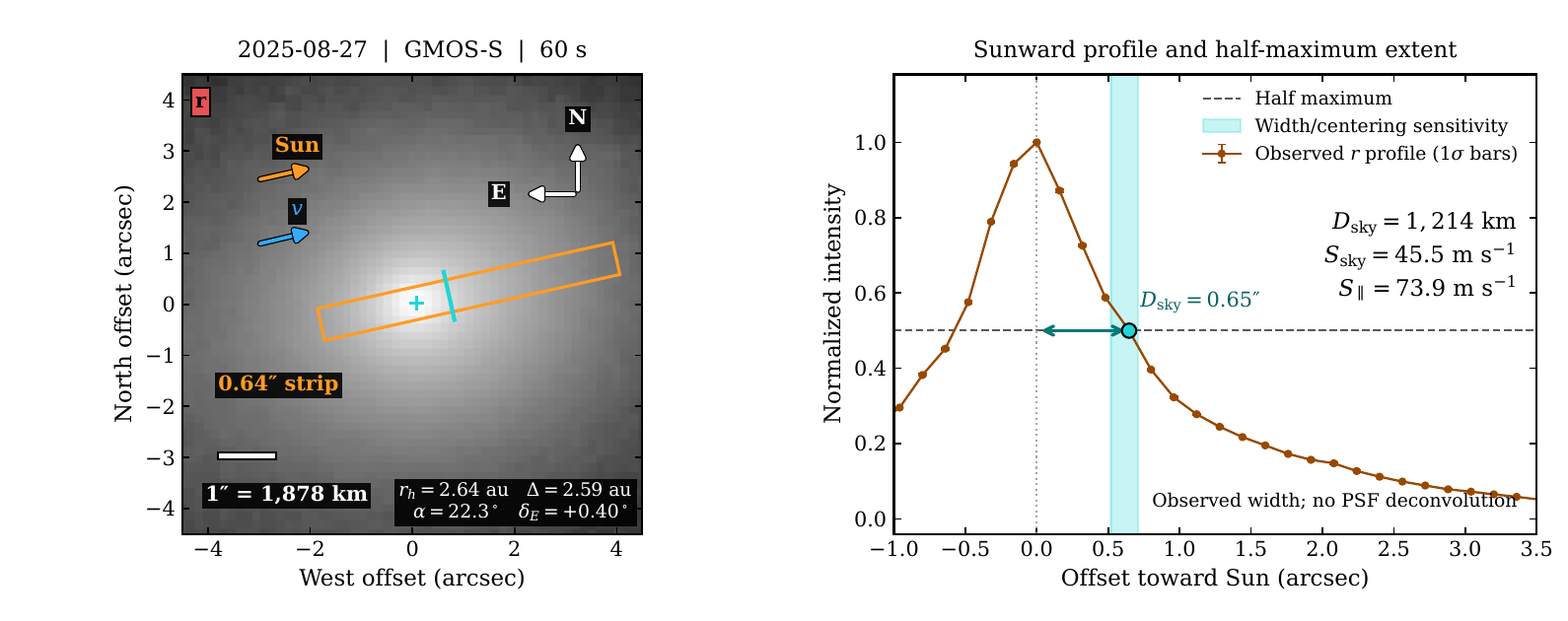}
\caption{GMOS $r$ sunward-width measurement on 2025 August 27 \citep{JewittComaProfiles1987}. Left: north-up, east-left image with the 0.64-arcsec extraction strip (orange), photocenter (cyan cross), and half-maximum location (cyan line); masked pixels are gray. Right: profile normalized to its near-center peak, with the half-maximum crossing in cyan. Statistical $1\sigma$ bars use 10,000 Monte Carlo realizations of the diagonal variance map, recalculating strip medians and peak normalization. Most bars are smaller than symbols; the peak is fixed to unity. Errors exclude interpixel covariance, unrepresented sky errors, seeing, and geometry. Shading spans strip-width and centering sensitivity. $S_{\rm sky}$ and $S_\parallel$ enter $v=S\sqrt{\beta}$ in Equation~\ref{eq:sunward_speed}. The HWHM includes seeing and coma broadening; no sharp turnaround edge is identified.}\label{fig:sunward_width_examples}
\end{figure}

%% file: generated/figure_tail_speed.tex
\begin{figure}[!htbp]\centering
\includegraphics[width=.93\textwidth]{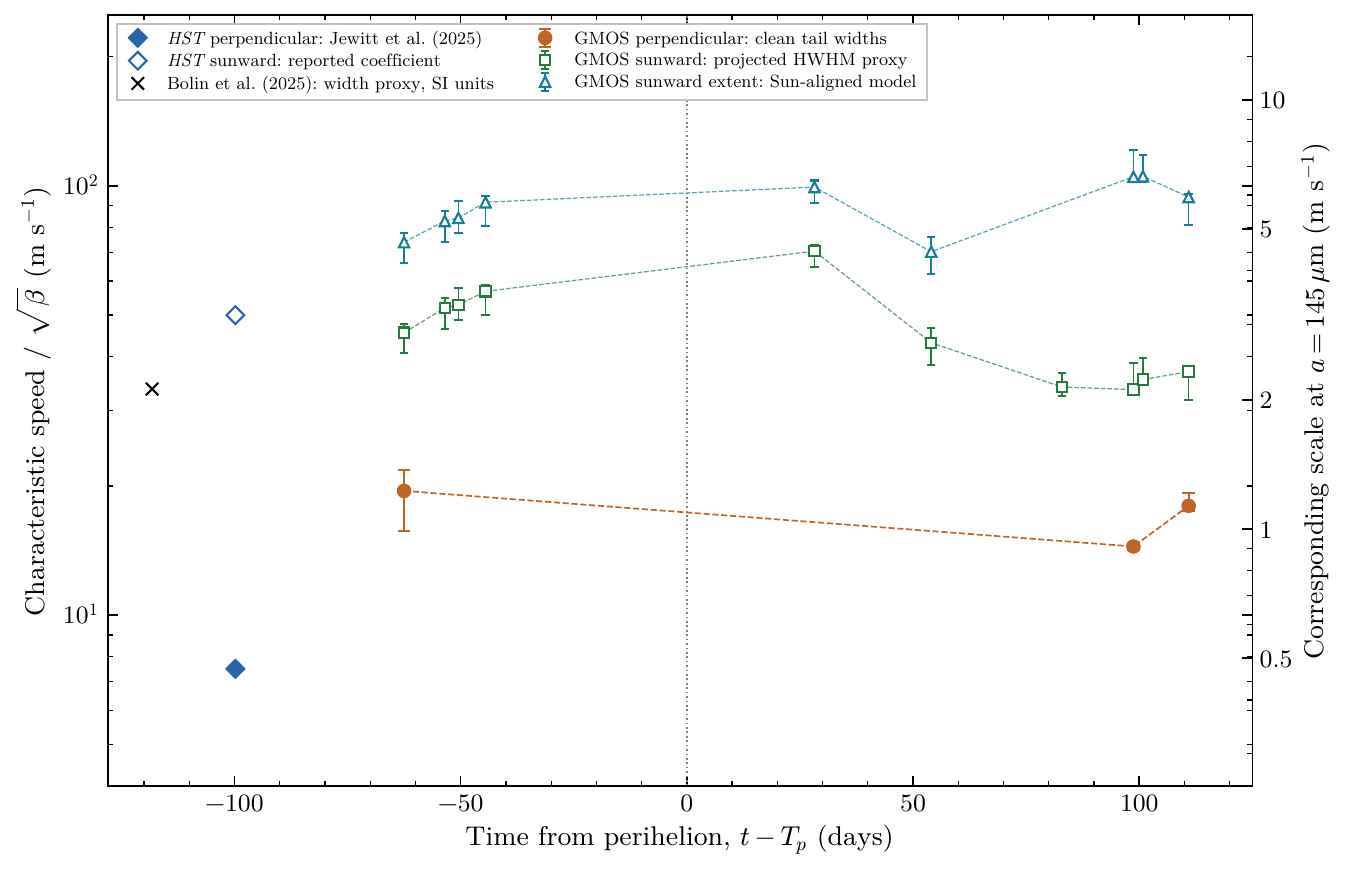}
\caption{Dust-speed constraints from transverse widths and sunward profiles \citep{JewittComaProfiles1987}. Filled orange circles show three artifact-screened perpendicular coefficients, with bars spanning retained cuts. Open green squares show ten projected sunward HWHM coefficients; open teal triangles (``GMOS sunward extent: Sun-aligned model'') show nine estimates that treat HWHM as the axial turnaround distance and correct for phase-angle projection. January's near-opposition geometry permits only a projected estimate. September 6, March, and $u$-only/$g$-only epochs provide no sunward dust constraint. Sunward bars span strip-width and centering sensitivity, excluding unknown PSF corrections; no sharp turnaround edge is identified. Dashed lines show interpolation for the separate mass budgets. Diamonds give \textit{HST} estimates \citep{Jewitt2025HST}; the black cross evaluates the discovery-paper width in SI units \citep{Bolin2025Discovery}. The right axis uses $\beta=0.574/a_{\mu\mathrm{m}}$ at 145 $\mu$m. Perpendicular and axial estimates constrain different velocity components; projected sunward values retain geometric dependence.}\label{fig:tail_speed}
\end{figure}

%% file: generated/figure_dust_mass.tex
\begin{figure}[!htbp]\centering
\includegraphics[width=.95\textwidth,height=.76\textheight,keepaspectratio]{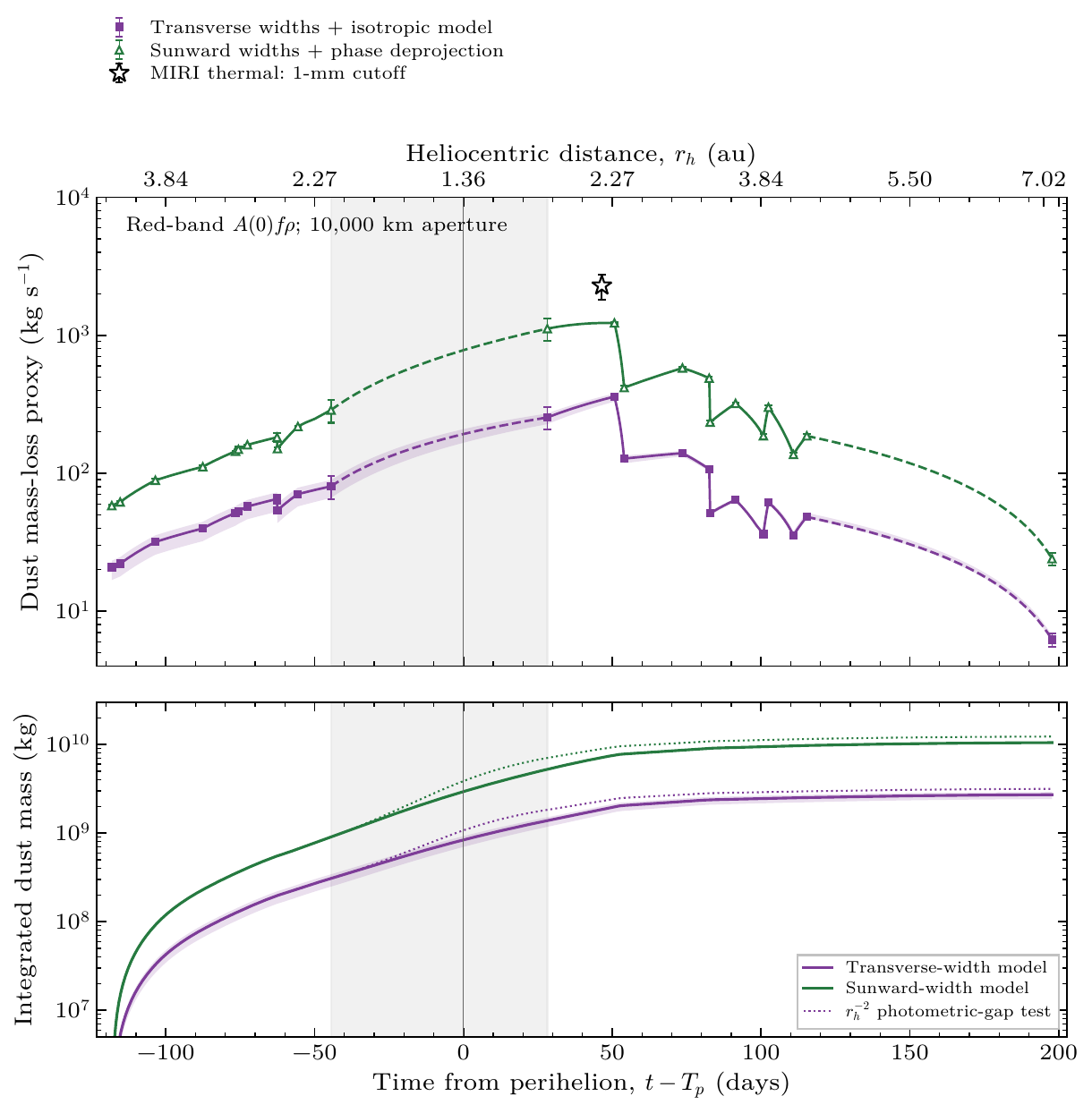}
\caption{Dust mass-loss rates from red-band $A(0)f\rho$ (top) and cumulative mass (bottom), for 1-$\mu$m--1-mm grains, aperture grain-number index $-3$, $\rho_d=1000$ kg m$^{-3}$, and $p_r=0.10$. Purple uses three artifact-screened transverse-width coefficients with an isotropic Gaussian velocity model; shading spans retained-profile sensitivity. Green uses nine sunward HWHM coefficients with a Sun-aligned turnaround model and phase-angle correction, excluding January's nearly end-on geometry. Width coefficients are linearly interpolated and held constant beyond the endpoint epochs; activity is interpolated independently. Errors propagate photometric uncertainty only and may be smaller than symbols. The open star and whiskers show the midpoint and full MIRI thermal-rate range for a 1-mm cutoff \citep{Belyakov2026Dust}. Gray shading marks the perihelion gap; dashed rates bridge gaps longer than 30 days. Dotted cumulative curves test an alternative gap interpolation of $[A(0)f\rho]r_h^2$, divided by daily $r_h^2$. This preserves the bounding observations while allowing a solar-flux-shaped enhancement, without bounding the total mass. The budgets remain conditional on the two velocity models.}\label{fig:dustmass}
\end{figure}

%% file: generated/figure_visible_spectra.tex
\begingroup\renewcommand{\theHfigure}{\arabic{figure}.visible_spectra.1}
\begin{figure}[!htbp]\centering
\includegraphics[page=1,width=\textwidth,height=.68\textheight,keepaspectratio]{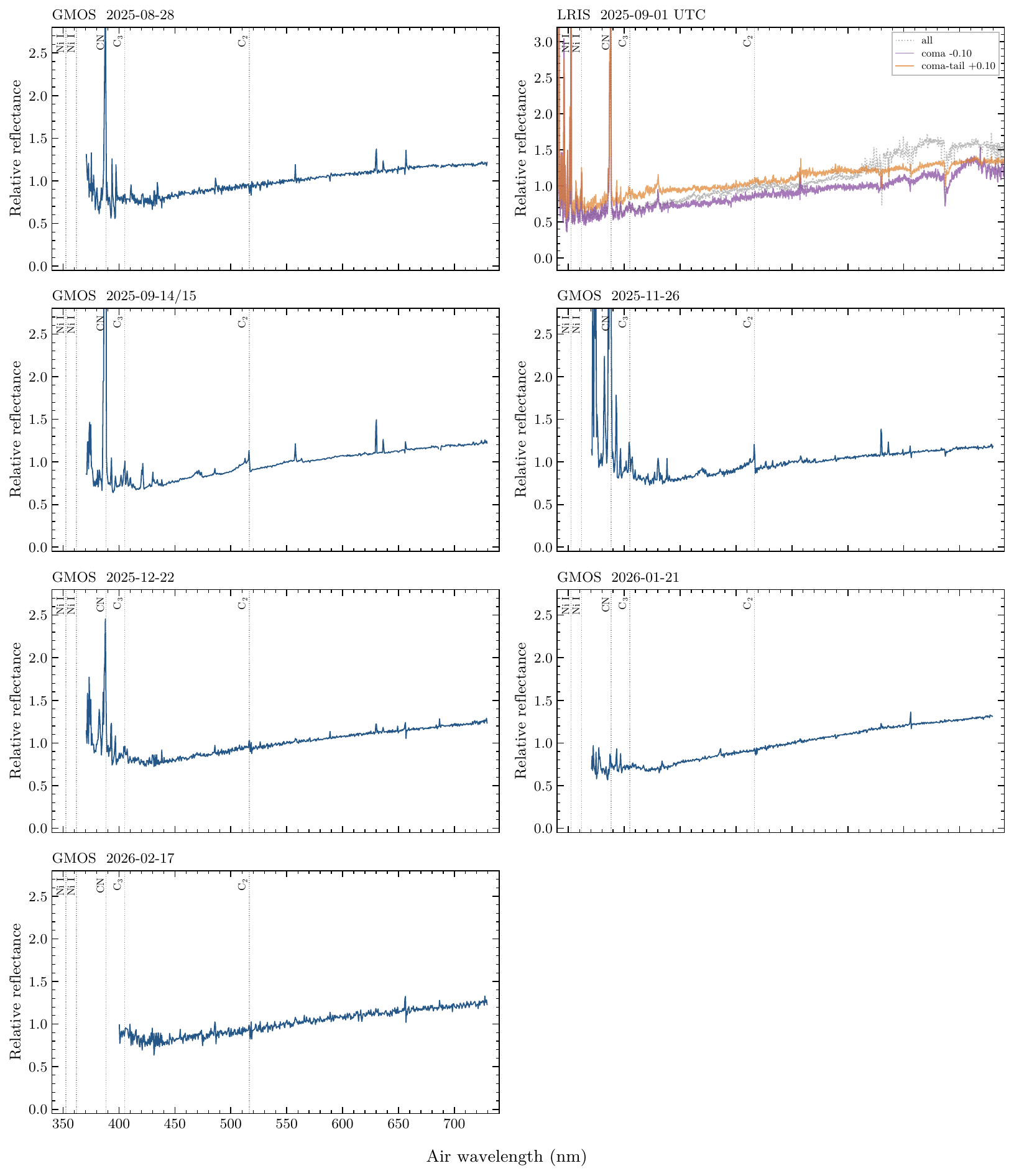}
\caption{Visible reflectance spectra in date order, combining both GMOS central-wavelength settings. Six GMOS epochs and September 1 LRIS are shown; March is excluded. LRIS uses synthetic-solar division and the red-arm continuity correction. Purple/orange curves show coma/coma--tail with display offsets of $-0.10$/$+0.10$; gray shows the unshifted, correlated all-exposure spectrum. Reference lines mark CN, C$_3$, C$_2$, and Ni I. The noisy February 17 region below 400 nm is omitted, retaining C$_3$.}\label{fig:visible_spectra}
\end{figure}\endgroup

%% file: generated/figure_gmos_flux.tex
\begingroup\renewcommand{\theHfigure}{\arabic{figure}.gmos_flux.1}
\begin{figure}[!htbp]\centering
\includegraphics[page=1,width=\textwidth,height=.68\textheight,keepaspectratio]{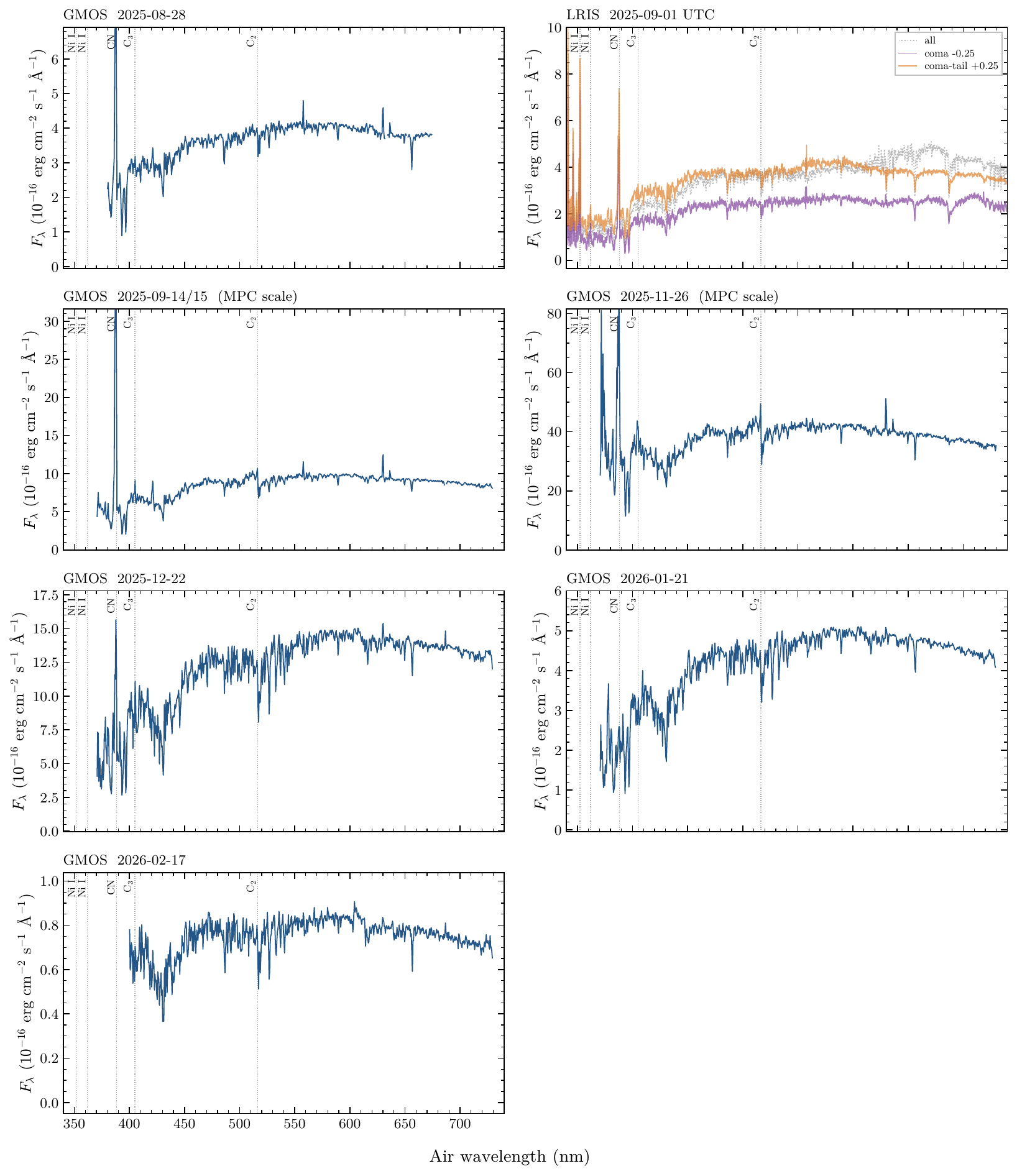}
\caption{Visible flux spectra in date order for six GMOS epochs and LRIS. August uses a measured standard response and imaging; December--February use model responses anchored to imaging. September/November use MPC aperture photometry transferred through GMOS image profiles. LRIS uses G191B2B and a single red-arm continuity factor. Translucent LRIS coma and orange coma--tail curves have offsets of $-0.25$ and $+0.25$ in plotted units ($10^{-16}$ erg cm$^{-2}$ s$^{-1}$ \AA$^{-1}$); gas measurements use unshifted spectra. The noisy February 17 region below 400 nm is omitted, retaining C$_3$.}\label{fig:gmos_flux}
\end{figure}\endgroup

%% file: generated/table_haser_parameters.tex
\begingroup\footnotesize\setlength{\tabcolsep}{3pt}
\begin{longtable}{llrrrrrl}
\caption{Epoch-specific geometry and gas-conversion inputs.}\label{tab:haserparams}\\
\toprule
UTC date & Instrument/extraction & $r_h$ & $\Delta$ & $\dot r_h$ & $g_{\rm CN}/10^{-13}$ & $v_{\rm CN,C_2}$ & $w\times h$\\\midrule\endfirsthead
\multicolumn{8}{l}{Table~\ref{tab:haserparams} continued}\\\toprule
UTC date & Instrument/extraction & $r_h$ & $\Delta$ & $\dot r_h$ & $g_{\rm CN}/10^{-13}$ & $v_{\rm CN,C_2}$ & $w\times h$\\\midrule\endhead
\bottomrule\endfoot
2025-08-28 & GMOS & 2.636 & 2.589 & -52.89 & 0.441 & 0.524 & $1.0\times6.400$\\
2025-09-01 & LRIS coma & 2.515 & 2.575 & -52.05 & 0.506 & 0.536 & $0.7\times9.450$\\
2025-09-01 & LRIS coma-tail & 2.515 & 2.575 & -52.05 & 0.506 & 0.536 & $0.7\times9.450$\\
2025-09-14/15 & GMOS & 2.110 & 2.542 & -47.67 & 0.832 & 0.585 & $1.0\times6.400$\\
2025-11-26 & GMOS & 1.701 & 1.962 & 37.86 & 1.310 & 0.652 & $1.0\times6.456$\\
2025-12-22 & GMOS & 2.383 & 1.802 & 50.92 & 0.607 & 0.551 & $1.0\times6.456$\\
2026-01-21 & GMOS & 3.309 & 2.326 & 55.80 & 0.420 & 0.467 & $1.0\times6.456$\\
2026-02-17 & GMOS & 4.197 & 3.399 & 57.42 & 0.271 & 0.415 & $1.0\times6.456$\\
\end{longtable}
\noindent\footnotesize Distances are in au, $\dot r_h$ and speeds in km s$^{-1}$, dimensions in arcsec, and $g_{\rm CN}$ in erg s$^{-1}$ molecule$^{-1}$. CN/C$_2$ use $v=0.85r_h^{-1/2}$, a 0.8-arcsec Gaussian PSF, GMOS sky transfer, and one-au parent/daughter lengths $(13000,210000)$ and $(22000,66000)$ km. C$_3$ uses $v=0.5$ km s$^{-1}$, a centered rectangle, $g=3.8\times10^{-13}r_h^{-2}$, and lengths $(3100,145000)r_h^2$ km, without PSF/sky correction. LRIS assumes a 0.7-arcsec slit and 9.45-arcsec extraction height for unbinned pixels, with geometry at 2025 September 1 00:00 UTC. December uses $r_h=2.38269281$ au (2.383 au here); the ATel example used 2.368 au.
\endgroup

%% file: generated/table_newq.tex
\begingroup\footnotesize\setlength{\tabcolsep}{3pt}
\begin{longtable}{llrlrr}
\caption{Integrated gas fluxes and production rates.}\label{tab:newq}\\
\toprule
UTC date & Instrument/extraction & $t-T_p$ (d) & Species & $F$ ($10^{-14}$ erg cm$^{-2}$ s$^{-1}$) & $Q$ (s$^{-1}$)\\\midrule\endfirsthead
\multicolumn{6}{l}{Table~\ref{tab:newq} continued}\\\toprule
UTC date & Instrument/extraction & $t-T_p$ (d) & Species & $F$ ($10^{-14}$ erg cm$^{-2}$ s$^{-1}$) & $Q$ (s$^{-1}$)\\\midrule\endhead
\bottomrule\endfoot
2025-08-28 & GMOS & -62.5 & CN & $1.287\pm0.047$ & $(1.85\pm0.07)\times10^{25}$\\
2025-08-28 & GMOS & -62.5 & C$_3$ & $1.868\pm0.111$ & $(6.04\pm0.36)\times10^{24}$\\
2025-08-28 & GMOS & -62.5 & C$_2$ & $<0.244$ & $<3.93\times10^{24}$\\
2025-09-01 & LRIS coma & -58.5 & CN & $0.597\pm0.019$ & $(6.51\pm0.21)\times10^{24}$\\
2025-09-01 & LRIS coma & -58.5 & C$_3$ & $1.560\pm0.043$ & $(4.64\pm0.13)\times10^{24}$\\
2025-09-01 & LRIS coma & -58.5 & C$_2$ & $<0.102$ & $<1.28\times10^{24}$\\
2025-09-01 & LRIS coma & -58.5 & Ni I & $\ldots^{n}$ & $(2.18\pm0.04)\times10^{23}$\\
2025-09-01 & LRIS coma-tail & -58.5 & CN & $0.730\pm0.031$ & $(7.96\pm0.33)\times10^{24}$\\
2025-09-01 & LRIS coma-tail & -58.5 & C$_3$ & $2.201\pm0.042$ & $(6.55\pm0.13)\times10^{24}$\\
2025-09-01 & LRIS coma-tail & -58.5 & C$_2$ & $<0.149$ & $<1.88\times10^{24}$\\
2025-09-01 & LRIS coma-tail & -58.5 & Ni I & $\ldots^{n}$ & $(2.48\pm0.04)\times10^{23}$\\
2025-09-14/15 & GMOS & -44.5 & CN & $6.320\pm1.217$ & $(3.65\pm0.70)\times10^{25}$\\
2025-09-14/15 & GMOS & -44.5 & C$_3$ & $5.100\pm0.985$ & $(7.79\pm1.50)\times10^{24}$\\
2025-09-14/15 & GMOS & -44.5 & C$_2$ & $1.213\pm0.241$ & $(9.37\pm1.86)\times10^{24}$\\
2025-11-26 & GMOS & 28.1 & CN & $14.076\pm3.245$ & $(3.81\pm0.88)\times10^{25}$\\
2025-11-26 & GMOS & 28.1 & C$_3$ & $16.554\pm3.188$ & $(1.13\pm0.22)\times10^{25}$\\
2025-11-26 & GMOS & 28.1 & C$_2$ & $4.992\pm1.053$ & $(1.83\pm0.39)\times10^{25}$\\
2025-12-22 & GMOS & 54.0 & CN & $1.554\pm0.179$ & $(1.36\pm0.16)\times10^{25}$\\
2025-12-22 & GMOS & 54.0 & C$_3$ & $2.803$--$3.277$ & $(5.73$--$6.70)\times10^{24}$\\
2025-12-22 & GMOS & 54.0 & C$_2$ & $<0.977$ & $<1.08\times10^{25}$\\
2026-01-21 & GMOS & 83.9 & CN & $<0.222$ & $<4.48\times10^{24}$\\
2026-01-21 & GMOS & 83.9 & C$_3$ & $0.589$--$0.651$ & $(4.00$--$4.42)\times10^{24}$\\
2026-01-21 & GMOS & 83.9 & C$_2$ & $<0.247$ & $<8.47\times10^{24}$\\
2026-02-17 & GMOS & 111.0 & CN & $<0.175$ & $<9.50\times10^{24}$\\
2026-02-17 & GMOS & 111.0 & C$_3$ & $<0.333$ & $<5.71\times10^{24}$\\
2026-02-17 & GMOS & 111.0 & C$_2$ & $<0.063$ & $<6.15\times10^{24}$\\
\end{longtable}
\noindent\footnotesize Pre-December C$_3$ fluxes use the ATel estimator: mean flux density at 397.5--415 nm minus the pooled median at 379--384 and 393--397 nm, multiplied by 175 \AA. C$_3$ rates use $g=3.8\times10^{-13}r_h^{-2}$ erg s$^{-1}$ molecule$^{-1}$, parent/daughter lengths $(3100,145000)r_h^2$ km, $v=0.5$ km s$^{-1}$, and the extraction rectangle. CN/C$_2$ use the forward PSF/sky model with $v=0.85r_h^{-1/2}$ km s$^{-1}$. Rectangles are $1\times6.400$ arcsec for GMOS-S, $1\times6.456$ arcsec for GMOS-N, and $0.7\times9.45$ arcsec for LRIS. September/November GMOS scales use MPC photometry; December--February use model responses. Errors include noise and quantified gray-scale uncertainty, excluding C$_3$ continuum and chromatic-response systematics. LRIS absolute-scale error is unquantified. December/January C$_3$ ranges span extended/red-only solar-continuum results without statistical coverage. The February limit is the maximum $\max(0,F_j)+3\sigma_j$ over three continua, including noise and gray-scale error. Other limits use $\max(0,\widehat Q)+3\sigma_Q$ or its flux equivalent. Late C$_3$ remains sensitive to continuum/response calibration; the estimator integrates the complex without separating every transition. $^n$Ni I uses a common-amplitude multi-line fit and fixed-speed atomic model. Electronic tables retain signed residuals and the correlated LRIS all-exposure C$_3$ estimate.
\endgroup

%% file: generated/figure_gas_indices.tex
\begingroup\renewcommand{\theHfigure}{\arabic{figure}.gas_indices.1}
\begin{figure}[!htbp]\centering
\includegraphics[page=1,width=\textwidth,height=.68\textheight,keepaspectratio]{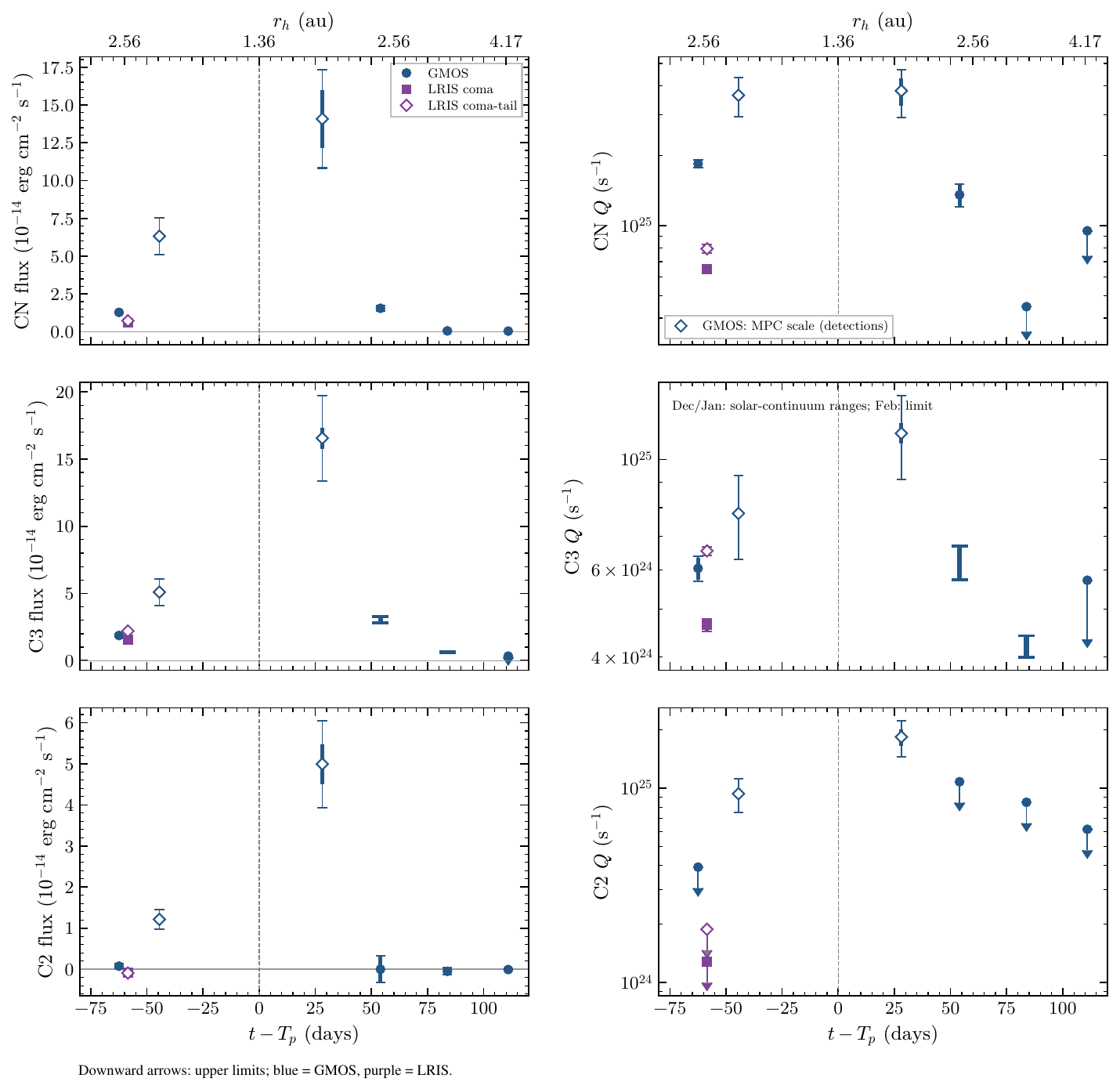}
\caption{Molecular fluxes (left) and production rates (right). C$_3$ uses an extraction-matched rectangular Haser model with $v=0.5$ km s$^{-1}$; pre-December fluxes follow Equation~\ref{eq:c3atel}. December/January brackets span the extended/red-only solar-continuum results, $(5.73$--$6.70)\times10^{24}$ and $(4.00$--$4.42)\times10^{24}$ s$^{-1}$, and represent continuum-choice ranges with no measured midpoint or statistical coverage. The February limit, $Q<5.71\times10^{24}$ s$^{-1}$, is the largest residual plus $3\sigma$ over three continua. It includes noise and gray-scale error, excluding unknown chromatic-response and model errors. CN/C$_2$ use the forward PSF/sky model. GMOS is blue and LRIS purple; squares show LRIS coma and open diamonds coma--tail. Open blue diamonds identify MPC-scaled spectra. Thick/thin bars on other points show statistical/gray-scale-inclusive errors; upper-limit arrows follow instrument colors. Signed fluxes are retained where applicable. March is excluded.}\label{fig:gas_indices}
\end{figure}\endgroup

%% file: generated/table_publishedq.tex
\begingroup\footnotesize\setlength{\tabcolsep}{4pt}
\begin{longtable}{llrlrrl}
\caption{Literature production rates used for comparison.}\label{tab:publishedq}\\
\toprule
UTC date & $t-T_p$ (d) & $r_h$ (au) & Species & $\log Q$ & $\sigma_{\log Q}$ & Source\\\midrule\endfirsthead
\multicolumn{7}{l}{Table~\ref{tab:publishedq} continued}\\\toprule
UTC date & $t-T_p$ (d) & $r_h$ (au) & Species & $\log Q$ & $\sigma_{\log Q}$ & Source\\\midrule\endhead
\bottomrule\endfoot
2025-08-21 & -69.0 & 2.84 & CN & 24.810 & 0.010 & Rahatgaonkar et al. 2025\\
2025-08-21 & -69.0 & 2.84 & NiI & 23.300 & 0.070 & Rahatgaonkar et al. 2025\\
2025-11-26 & 28.0 & 1.70 & C2 & 26.630 & 0.004 & ATel 17561\\
2025-11-26 & 28.0 & 1.70 & C3 & 25.255 & 0.005 & ATel 17561\\
2025-11-26 & 28.0 & 1.70 & CN & 26.859 & 0.013 & ATel 17561\\
2025-12-15 & 47.0 & 2.18 & NiI & 23.890 & 0.030 & Hutsemekers et al. 2026\\
2025-12-21 & 53.0 & 2.35 & NiI & 23.790 & 0.040 & Hutsemekers et al. 2026\\
2025-12-22 & 54.0 & 2.38 & C2 & 26.373 & 0.004 & ATel 17719\\
2025-12-22 & 54.0 & 2.38 & C3 & 24.844 & 0.009 & ATel 17719\\
2025-12-22 & 54.0 & 2.38 & CN & 26.812 & 0.013 & ATel 17719\\
2025-12-26 & 58.0 & 2.50 & NiI & 23.660 & 0.040 & Hutsemekers et al. 2026\\
2026-01-11 & 74.0 & 2.99 & NiI & 23.260 & 0.050 & Hutsemekers et al. 2026\\
2026-01-19 & 82.0 & 3.25 & NiI & 23.060 & 0.060 & Hutsemekers et al. 2026\\
2026-01-21 & 84.0 & 3.31 & C2 & 24.767 & 0.004 & ATel 17841\\
2026-01-21 & 84.0 & 3.31 & C3 & 23.763 & 0.005 & ATel 17841\\
2026-01-21 & 84.0 & 3.31 & CN & 24.657 & 0.005 & ATel 17841\\
2026-01-27 & 90.0 & 3.51 & NiI & 22.880 & 0.070 & Hutsemekers et al. 2026\\
2026-02-07 & 101.0 & 3.87 & NiI & 22.440 & 0.100 & Hutsemekers et al. 2026\\
2025-11-16 & 18.0 & 1.51 & CN & 25.907 & 0.021 & Ahuja et al. 2026\\
2025-11-16 & 18.0 & 1.51 & C2 & 25.845 & 0.015 & Ahuja et al. 2026\\
2025-11-16 & 18.0 & 1.51 & C3 & 25.029 & 0.080 & Ahuja et al. 2026\\
2025-11-17 & 19.0 & 1.52 & CN & 25.954 & 0.018 & Ahuja et al. 2026\\
2025-11-17 & 19.0 & 1.52 & C2 & 25.802 & 0.020 & Ahuja et al. 2026\\
2025-11-17 & 19.0 & 1.52 & C3 & 25.032 & 0.092 & Ahuja et al. 2026\\
2025-11-27 & 29.0 & 1.72 & CN & 25.744 & 0.004 & Ahuja et al. 2026\\
2025-11-27 & 29.0 & 1.72 & C2 & 25.750 & 0.025 & Ahuja et al. 2026\\
2025-11-27 & 29.0 & 1.72 & C3 & 24.832 & 0.049 & Ahuja et al. 2026\\
2025-11-28 & 30.0 & 1.74 & CN & 25.769 & 0.007 & Ahuja et al. 2026\\
2025-11-28 & 30.0 & 1.74 & C2 & 25.744 & 0.040 & Ahuja et al. 2026\\
2025-11-28 & 30.0 & 1.74 & C3 & 24.793 & 0.085 & Ahuja et al. 2026\\
2025-12-04 & 36.0 & 1.89 & CN & 25.614 & 0.021 & Ahuja et al. 2026\\
2025-12-04 & 36.0 & 1.89 & C2 & 25.405 & 0.091 & Ahuja et al. 2026\\
2025-12-04 & 36.0 & 1.89 & C3 & 24.512 & 0.190 & Ahuja et al. 2026\\
2025-12-09 & 41.0 & 2.02 & CN & 25.575 & 0.106 & Ahuja et al. 2026\\
2025-12-09 & 41.0 & 2.02 & C2 & 25.359 & 0.086 & Ahuja et al. 2026\\
2025-12-12 & 44.0 & 2.10 & CN & 25.348 & 0.027 & Ahuja et al. 2026\\
2025-12-12 & 44.0 & 2.10 & C2 & 25.313 & 0.062 & Ahuja et al. 2026\\
2025-12-12 & 44.0 & 2.10 & C3 & 24.638 & 0.240 & Ahuja et al. 2026\\
2025-12-13 & 45.0 & 2.12 & CN & 25.334 & 0.055 & Ahuja et al. 2026\\
2025-12-13 & 45.0 & 2.12 & C2 & 25.336 & 0.095 & Ahuja et al. 2026\\
2025-12-13 & 45.0 & 2.12 & C3 & 24.781 & 0.191 & Ahuja et al. 2026\\
2025-12-17 & 49.0 & 2.24 & CN & 25.232 & 0.079 & Ahuja et al. 2026\\
2025-12-17 & 49.0 & 2.24 & C2 & 25.229 & 0.249 & Ahuja et al. 2026\\
2025-12-17 & 49.0 & 2.24 & C3 & 24.700 & 0.265 & Ahuja et al. 2026\\
2025-12-20 & 52.0 & 2.32 & CN & 25.268 & 0.054 & Ahuja et al. 2026\\
2025-12-20 & 52.0 & 2.32 & C2 & 25.027 & 0.127 & Ahuja et al. 2026\\
2025-12-20 & 52.0 & 2.32 & C3 & 24.481 & 0.391 & Ahuja et al. 2026\\
2026-01-20 & 83.0 & 3.28 & CN & 25.130 & 0.189 & Ahuja et al. 2026\\
2026-01-20 & 83.0 & 3.28 & C2 & 25.004 & 0.393 & Ahuja et al. 2026\\
\end{longtable}
\noindent\footnotesize Rates retain the published fluorescence, outflow speed, scale lengths, apertures, and band coverage, in molecules s$^{-1}$ (atoms s$^{-1}$ for Ni). Geometry and $t-T_p$ use the common orbit and source date. The December 20 value is recalculated from that date, correcting the inconsistent TFP in \citet{Ahuja2026}, Table 3.
\endgroup

%% file: generated/figure_published_rates.tex
\begingroup\renewcommand{\theHfigure}{\arabic{figure}.published_rates.1}
\begin{figure}[!htbp]\centering
\includegraphics[page=1,width=\textwidth,height=.68\textheight,keepaspectratio]{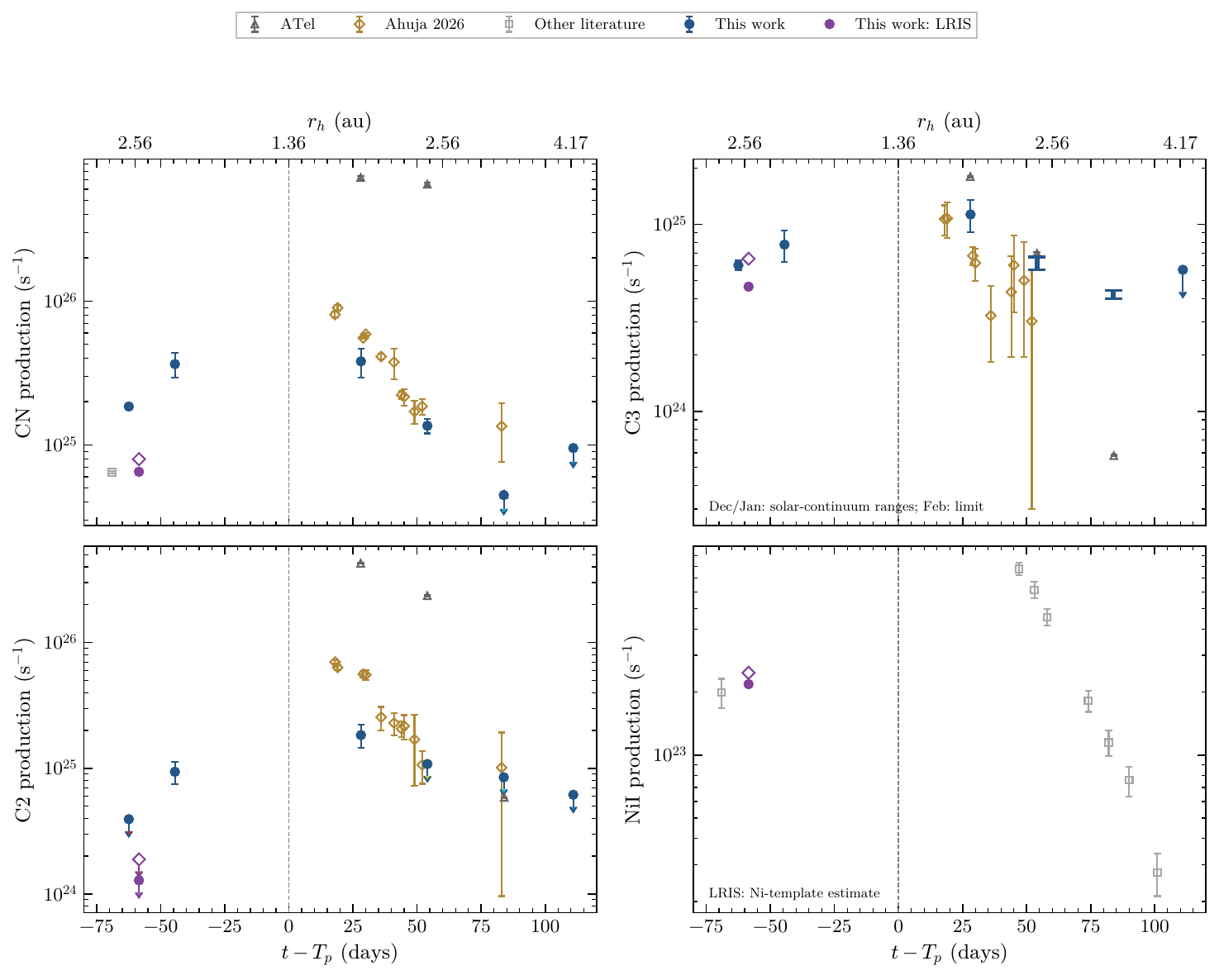}
\caption{Gas-production rates compared with ATels and \citet{Ahuja2026}. December/January C$_3$ brackets span solar-continuum choices, $(5.73$--$6.70)\times10^{24}$ and $(4.00$--$4.42)\times10^{24}$ s$^{-1}$, without statistical coverage. The February limit is $<5.71\times10^{24}$ s$^{-1}$ across three continua (Figure~\ref{fig:gas_indices}). Our C$_3$ rates use the extraction-matched Haser model with $v=0.5$ km s$^{-1}$; pre-December fluxes use the blue-baseline estimator. CN/C$_2$ use the forward PSF/sky model and Ni the atomic model. Purple marks LRIS; open diamonds identify coma--tail. Literature values retain their conventions. Our calibration is independent of those rates; ranges and limits exclude unquantified calibration/model systematics.}\label{fig:published_rates}
\end{figure}\endgroup

%% file: generated/table_gas_ratios.tex
\begingroup\footnotesize\setlength{\tabcolsep}{3pt}
\begin{longtable}{llrlr}
\caption{Gas-production ratios.}\label{tab:gasratios}\\
\toprule
UTC date & Instrument/extraction & $t-T_p$ (d) & Ratio & Value\\\midrule\endfirsthead
\multicolumn{5}{l}{Table~\ref{tab:gasratios} continued}\\\toprule
UTC date & Instrument/extraction & $t-T_p$ (d) & Ratio & Value\\\midrule\endhead
\bottomrule\endfoot
2025-08-28 & GMOS & -62.5 & C$_2$/CN & $<0.213$\\
2025-08-28 & GMOS & -62.5 & C$_3$/CN & $0.327\pm0.024$\\
2025-08-28 & GMOS & -62.5 & C$_3$/C$_2$ & --\\
2025-09-01 & LRIS coma & -58.5 & C$_2$/CN & $<0.198$\\
2025-09-01 & LRIS coma & -58.5 & C$_3$/CN & $0.713\pm0.042$\\
2025-09-01 & LRIS coma & -58.5 & C$_3$/C$_2$ & --\\
2025-09-01 & LRIS coma-tail & -58.5 & C$_2$/CN & $<0.239$\\
2025-09-01 & LRIS coma-tail & -58.5 & C$_3$/CN & $0.823\pm0.050$\\
2025-09-01 & LRIS coma-tail & -58.5 & C$_3$/C$_2$ & --\\
2025-09-14/15 & GMOS & -44.5 & C$_2$/CN & $0.257\pm0.014$\\
2025-09-14/15 & GMOS & -44.5 & C$_3$/CN & $0.214\pm0.009$\\
2025-09-14/15 & GMOS & -44.5 & C$_3$/C$_2$ & $0.832\pm0.062$\\
2025-11-26 & GMOS & 28.1 & C$_2$/CN & $0.481\pm0.080$\\
2025-11-26 & GMOS & 28.1 & C$_3$/CN & $0.296\pm0.054$\\
2025-11-26 & GMOS & 28.1 & C$_3$/C$_2$ & $0.616\pm0.089$\\
2025-12-22 & GMOS & 54.0 & C$_2$/CN & $<0.840$\\
2025-12-22 & GMOS & 54.0 & C$_3$/CN & $0.422$--$0.493$\\
2025-12-22 & GMOS & 54.0 & C$_3$/C$_2$ & --\\
2026-01-21 & GMOS & 83.9 & C$_2$/CN & --\\
2026-01-21 & GMOS & 83.9 & C$_3$/CN & --\\
2026-01-21 & GMOS & 83.9 & C$_3$/C$_2$ & --\\
2026-02-17 & GMOS & 111.0 & C$_2$/CN & --\\
2026-02-17 & GMOS & 111.0 & C$_3$/CN & --\\
2026-02-17 & GMOS & 111.0 & C$_3$/C$_2$ & --\\
2025-09-01 & LRIS coma & -58.5 & Ni I/CN & $0.034\pm0.001$\\
2025-09-01 & LRIS coma-tail & -58.5 & Ni I/CN & $0.031\pm0.001$\\
\end{longtable}
\noindent\footnotesize C$_3$ uses the ATel fixed-speed rectangular model; CN/C$_2$ use the forward PSF/sky model. Shared gray normalization cancels. C$_3$ ratio errors maximize the first-order standard deviation over unknown cross-band covariance: $\sigma_x/|y|+|x|\sigma_y/y^2$ for $x/y$, after removing the shared gray term. They exclude continuum and chromatic-response errors. C$_2$/CN and Ni I/CN errors propagate band uncertainties; undetected denominators are omitted. Species-dependent models preclude a depletion classification. Gas-free continuum controls explain most of the December/January C$_3$ signal; February bounds the tested continua. The December C$_3$/CN range uses nominal CN, excludes its error, and has no statistical coverage.
\endgroup

%% file: generated/figure_gas_ratios.tex
\begingroup\renewcommand{\theHfigure}{\arabic{figure}.gas_ratios.1}
\begin{figure}[!htbp]\centering
\includegraphics[page=1,width=\textwidth,height=.68\textheight,keepaspectratio]{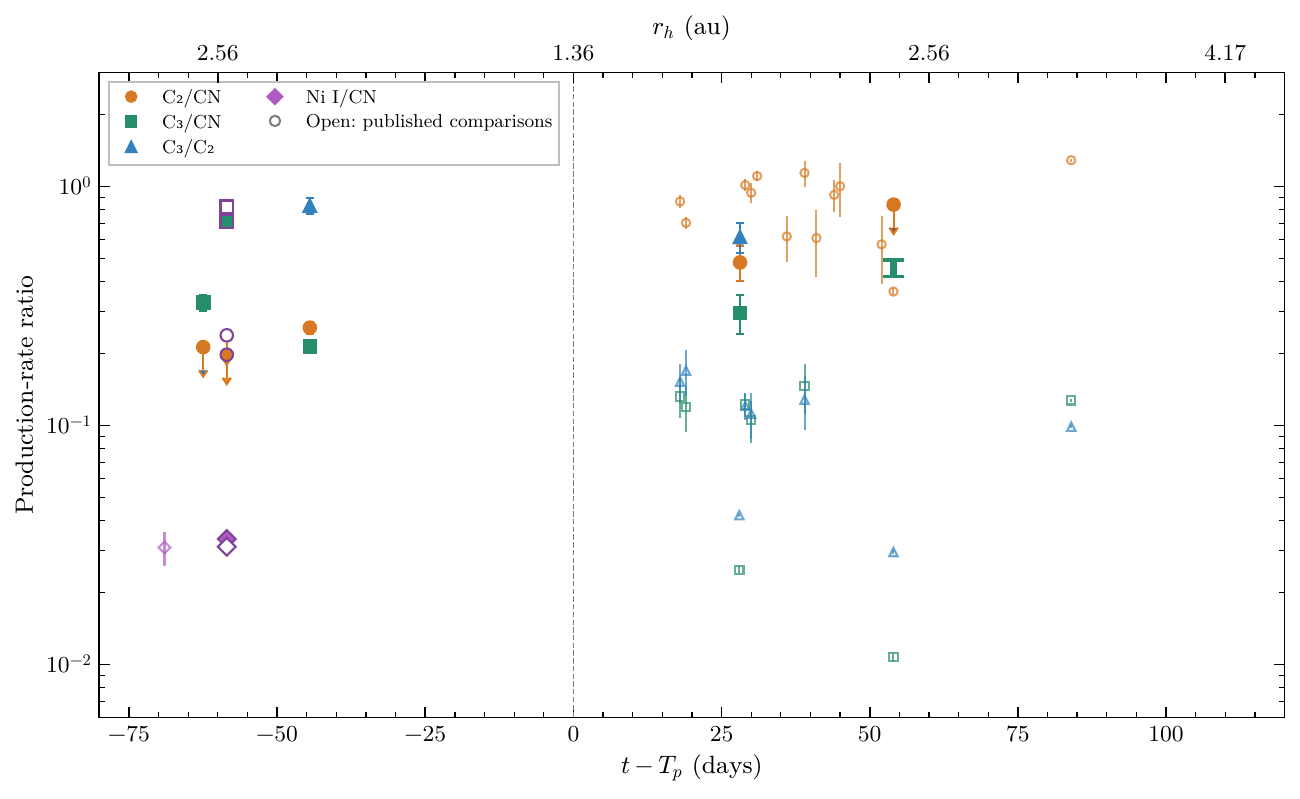}
\caption{Gas-production ratios. The December C$_3$/CN bracket, 0.422--0.493, divides the two-continuum C$_3$ range by nominal CN; it excludes CN uncertainty and has no statistical coverage. Colors and symbols identify species pairs; purple edges mark LRIS and open centers coma--tail. Open comparison points retain published prescriptions. Shared gray normalization cancels. Other C$_3$ bars give the maximum first-order statistical error over unknown cross-band covariance, excluding continuum and chromatic-response systematics. C$_2$/CN and Ni I/CN errors propagate band uncertainties. Arrows mark upper limits; undetected denominators are omitted. Species-dependent source models and December--February C$_3$ continuum sensitivity limit abundance comparisons.}\label{fig:gas_ratios}
\end{figure}\endgroup

%% file: generated/figure_nir_images.tex
\begingroup\renewcommand{\theHfigure}{\arabic{figure}.nir_images.1}
\begin{figure}[!htbp]\centering
\includegraphics[page=1,width=\textwidth,height=.68\textheight,keepaspectratio]{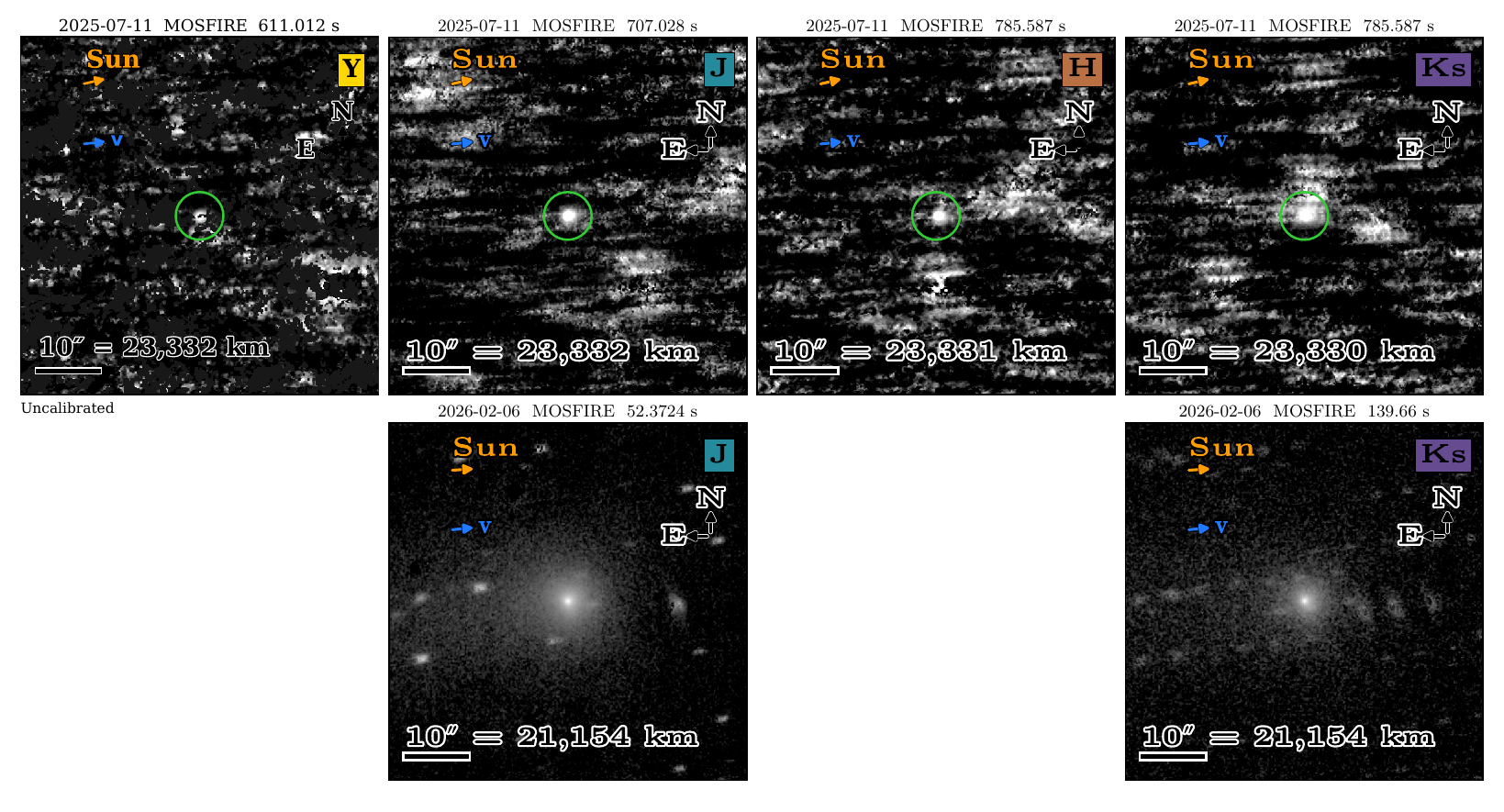}
\caption{MOSFIRE $Y,J,H,K_s$ imaging, north up and east left. Labels give date, summed integration, directions, and scale; Sun/heliocentric-velocity vectors are orange/blue. Each instrumental stack has an independent grayscale stretch. July $Y$ combines 21 exposures totaling 611 s. Dark gray marks nonfinite science/variance values, quality flags, or coverage below three exposures; resampling requires valid contributing pixels. This uncalibrated stack has incomplete inner-coma coverage and nonuniform exposure time. July $Y-J$ uses separate, complete-aperture exposures and sidereal calibration transfer. February $Y/H$ cells are blank; green circles locate the July comet.}\label{fig:nir_images}
\end{figure}\endgroup

%% file: generated/figure_nir_spectra.tex
\begingroup\renewcommand{\theHfigure}{\arabic{figure}.nir_spectra.1}
\begin{figure}[!htbp]\centering
\includegraphics[page=1,width=\textwidth,height=.68\textheight,keepaspectratio]{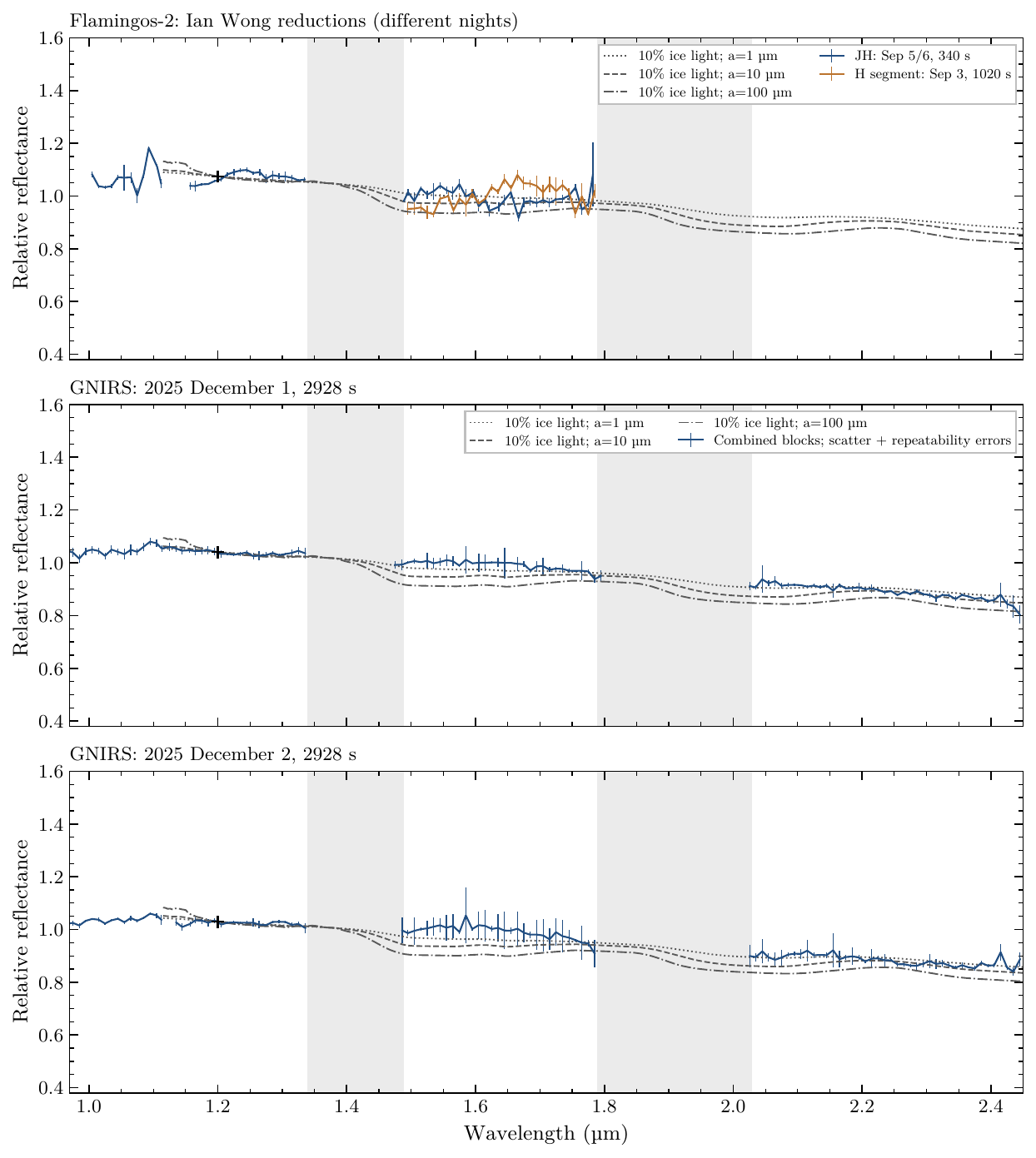}
\caption{F2 and nightly combined GNIRS spectra. The upper panel shows F2 JH alongside the retained H segment of HK from a different night, excluding JH below 1.0 $\mu$m and HK outside 1.45--1.80 $\mu$m. GNIRS blocks are combined separately for December 1 and 2. Errors use local scatter and a block-repeatability floor; gray bands mark excluded atmospheric intervals. Illustrative ice models assign 10\% of the scattered light at 1.20 $\mu$m to grains of radius 1, 10, or 100 $\mu$m. Each model's continuum level and slope are fitted to the spectrum; black crosses mark the 1.20-$\mu$m anchors.}\label{fig:nir_spectra}
\end{figure}\endgroup

%% file: generated/table_partialcolors.tex
\begingroup\small\setlength{\tabcolsep}{4pt}
\begin{longtable}{llrrrr}
\caption{Synthetic NIR colors on accepted portions of the 2MASS passbands.}\label{tab:partialcolors}\\
\toprule
Spectrum & Color & Value (mag) & $\sigma_{\rm scatter}$ & Repeatability floor & Extraction range\\\midrule\endfirsthead
\multicolumn{6}{l}{Table~\ref{tab:partialcolors} continued}\\\toprule
Spectrum & Color & Value (mag) & $\sigma_{\rm scatter}$ & Repeatability floor & Extraction range\\\midrule\endhead
\bottomrule\endfoot
F2 Sep 5/6 & J-H & 0.241 & 0.005 & -- & 0.198--0.274\\
GNIRS Dec 1 & H-Ks & -0.050 & 0.003 & 0.020 & $-0.115$--$-0.005$\\
GNIRS Dec 1 & J-H & 0.270 & 0.002 & 0.027 & 0.184--0.381\\
GNIRS Dec 1 & J-Ks & 0.219 & 0.002 & 0.008 & 0.179--0.266\\
GNIRS Dec 2 & H-Ks & -0.067 & 0.005 & 0.026 & $-0.101$--$-0.027$\\
GNIRS Dec 2 & J-H & 0.277 & 0.004 & 0.053 & 0.189--0.360\\
GNIRS Dec 2 & J-Ks & 0.210 & 0.004 & 0.027 & 0.162--0.265\\
\end{longtable}
\noindent\footnotesize Vega references and solar restoration use the accepted spectral intervals, so these colors differ from full-filter 2MASS/MOSFIRE colors. GNIRS nightly values combine exposure-weighted relative band fluxes; errors use the larger of propagated scatter and half the block color difference. Extraction ranges measure sensitivity across correlated aperture reductions. Standard-star, telluric, chromatic slit-loss, and missing-coverage systematics remain unquantified. F2 HK is restricted to 1.45--1.80 $\mu$m and supplies no $K_s$ color.
\endgroup

%% file: generated/figure_nir_spectral_colors.tex
\begingroup\renewcommand{\theHfigure}{\arabic{figure}.nir_spectral_colors.1}
\begin{figure}[!htbp]\centering
\includegraphics[page=1,width=\textwidth,height=.68\textheight,keepaspectratio]{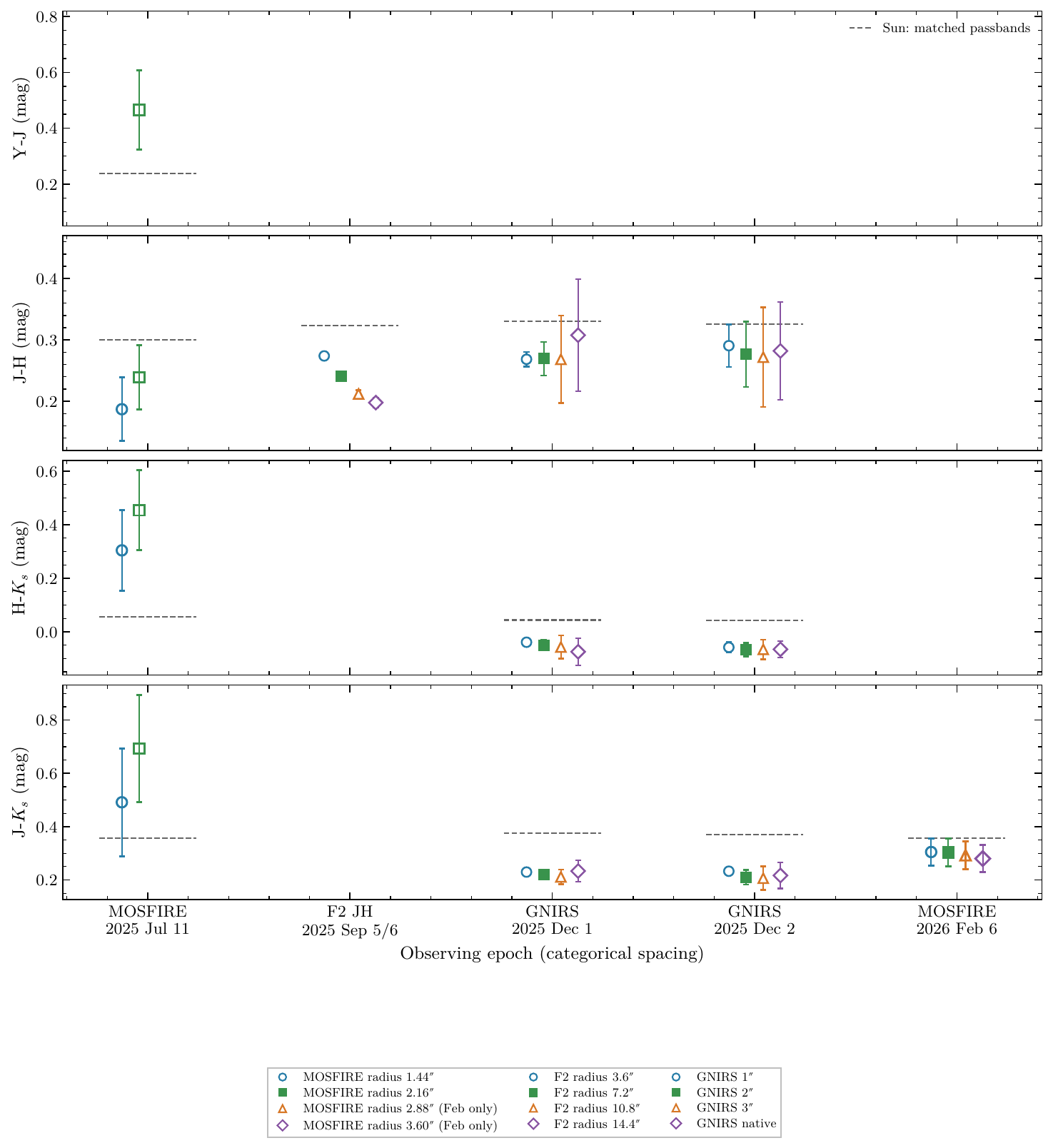}
\caption{NIR imaging and partial-passband spectral colors by aperture. July MOSFIRE $Y-J$ uses a 2.16-arcsec radius and a 0.143-mag sidereal-transfer sensitivity bar without formal $1\sigma$ coverage. F2 extraction radii of 20/40/60/80 pixels span 3.6/7.2/10.8/14.4 arcsec at 0.18 arcsec pixel$^{-1}$ along the slit. Symbols/colors distinguish GNIRS 1/2/3-arcsec/native extractions; filled squares mark primary choices. Nightly GNIRS colors combine block-relative fluxes; errors use the larger of propagated scatter and half the block color difference. F2 errors use local scatter. Extractions are correlated; horizontal offsets separate symbols. F2 cuts preclude $H-K_s$ and $J-K_s$. Gray dashed segments show solar colors over the primary spectral intervals. MOSFIRE $JHK_s$ radii are 1.44/2.16 arcsec in July and 1.44/2.16/2.88/3.60 arcsec in February, with 2.16 arcsec nominal. February errors assume solar infrared reference-star spectra and exclude shared spectral mismatch. July open-symbol bars combine formal flux errors and the full inter-aperture difference in quadrature: 0.052, 0.150, and 0.202 mag for $J-H$, $H-K_s$, and $J-K_s$. These sensitivities lack established Gaussian coverage and exclude shared contamination/calibration biases. Passband and physical-aperture differences limit temporal comparisons. Some errors are smaller than symbols.}\label{fig:nir_spectral_colors}
\end{figure}\endgroup

%% file: generated/figure_nir_movis.tex
\begin{figure}[!htbp]\centering
\includegraphics[width=\textwidth]{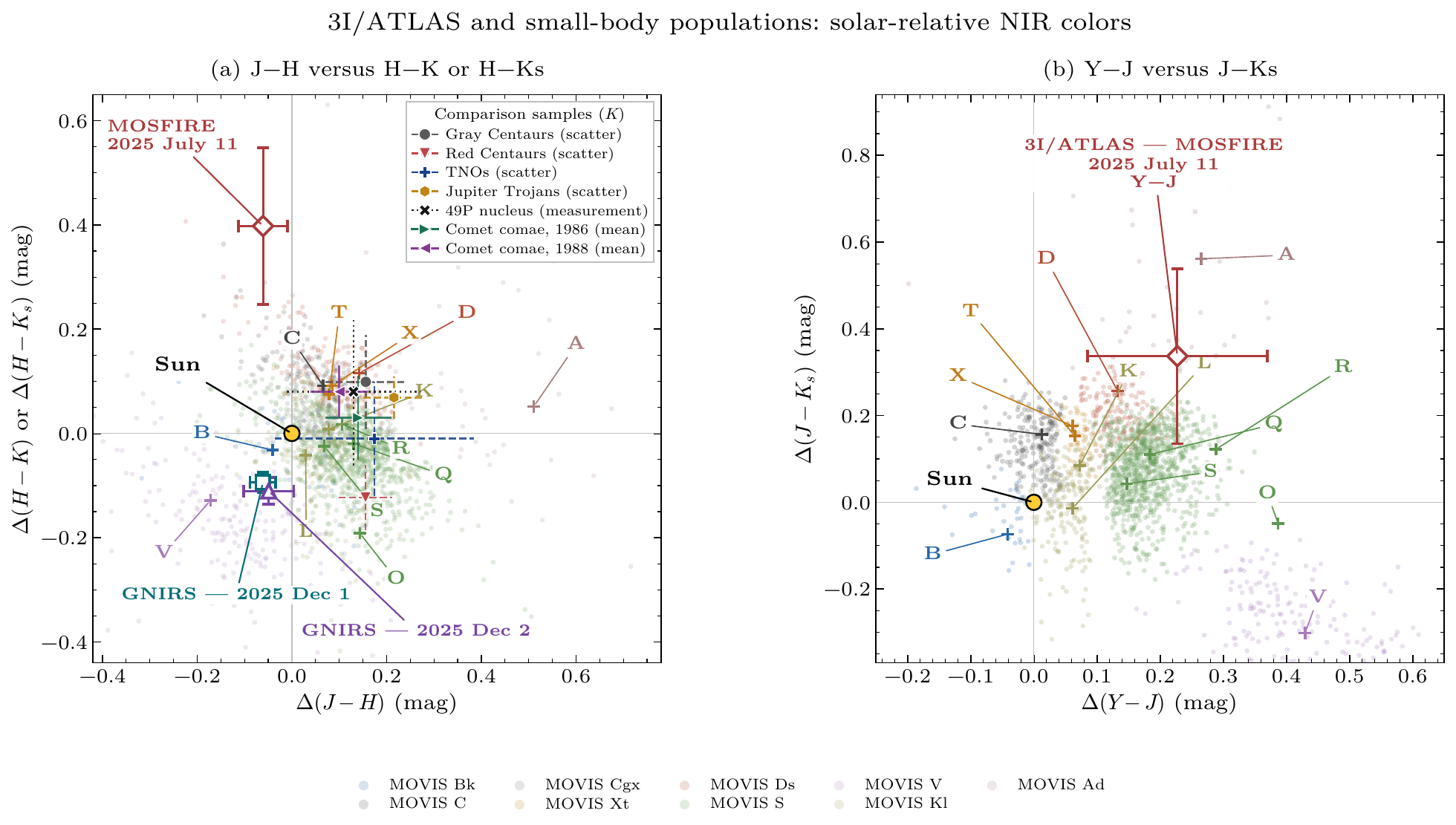}
\caption{Solar-relative NIR colors of \obj\ and comparison populations: (a) $\Delta(J-H)$ versus $\Delta(H-K)$ or $\Delta(H-K_s)$; (b) $\Delta(Y-J)$ versus $\Delta(J-K_s)$. The Sun is at the origin. Faint points show 1886 classified MOVIS asteroids \citep{Popescu2016,Popescu2018}; crosses mark spectral-template means. Open diamonds show July 11 MOSFIRE at 2.16-arcsec radius; open squares/triangles show December 1/2 GNIRS. Filled symbols in (a) show MBOSS gray/red Centaurs, TNOs, Jupiter Trojans, and the 49P nucleus \citep{Hainaut2012}, plus the 1986/1988 comet-coma samples \citep{JewittCometColors1988}. Dashed bars give MBOSS sample dispersions, solid bars comet-sample mean errors, and dotted bars 49P measurement errors. Table~\ref{tab:nirpop} gives per-color counts; group means can use different objects in each color. Each dataset uses its adopted solar colors, including the MBOSS convention \citep{HainautMBOSSColors2002}. Historical $K$, MOSFIRE $K_s$, partial-passband GNIRS, and VISTA colors retain response differences. July bars measure empirical calibration/aperture sensitivity without established Gaussian coverage; GNIRS bars measure scatter/repeatability. Marginal bars omit shared-band covariance. Stellar contamination and calibration uncertainty limit the July--December comparison. Only epochs with both plotted colors are shown.}\label{fig:nir_movis}
\end{figure}

%% file: generated/table_ice.tex
\begingroup\small\setlength{\tabcolsep}{4pt}
\begin{longtable}{lrr}
\caption{Ice-template sensitivity at 1.20 microns.}\label{tab:ice}\\
\toprule
Spectrum & Grain radius ($\mu$m) & Ice-light ceiling range (\%)\\\midrule\endfirsthead
\multicolumn{3}{l}{Table~\ref{tab:ice} continued}\\\toprule
Spectrum & Grain radius ($\mu$m) & Ice-light ceiling range (\%)\\\midrule\endhead
\bottomrule\endfoot
F2 JH & 1 & 35.7--100.0\\
F2 JH & 10 & 15.2--36.0\\
F2 JH & 100 & 9.7--20.8\\
GNIRS 20251201 & 1 & 1.5--7.2\\
GNIRS 20251201 & 10 & 0.6--2.7\\
GNIRS 20251201 & 100 & 0.3--1.4\\
GNIRS 20251202 & 1 & 1.8--7.2\\
GNIRS 20251202 & 10 & 0.7--2.5\\
GNIRS 20251202 & 100 & 0.3--1.2\\
\end{longtable}
\noindent\footnotesize Ceilings give the nonnegative ice fraction of scattered light at 1.20 $\mu$m, profiled over linear/quadratic continua at $\Delta\chi^2=9$. Ranges span Hapke-equivalent-slab and Mie-SSA models for 150 K crystalline ice; broad residuals inflate the scatter. Limits exclude unmeasured smooth response errors. Ice mass remains unconstrained, and grain size is degenerate with abundance.
\endgroup

%% file: references/verified_references.bib
@article{HamuyStandards1992,
 author={Hamuy, M. and Walker, A. R. and Suntzeff, N. B. and Gigoux, P. and Heathcote, S. R. and Phillips, M. M.},
 title={{Southern Spectrophotometric Standards. I.}},
 year={1992}, journal={PASP}, volume={104}, pages={533--552}, doi={10.1086/133028}}

@article{HamuyStandards1994,
 author={Hamuy, M. and Suntzeff, N. B. and Heathcote, S. R. and Walker, A. R. and Gigoux, P. and Phillips, M. M.},
 title={{Southern Spectrophotometric Standards. II.}},
 year={1994}, journal={PASP}, volume={106}, pages={566--589}, doi={10.1086/133417}}

@article{HusserPHOENIX2013,
 author={Husser, T.-O. and Wende-von Berg, S. and Dreizler, S. and Homeier, D. and Reiners, A. and Barman, T. and Hauschildt, P. H.},
 title={{A new extensive library of PHOENIX stellar atmospheres and synthetic spectra}},
 year={2013}, journal={A\&A}, volume={553}, pages={A6}, doi={10.1051/0004-6361/201219058}, eprint={1303.5632}, archivePrefix={arXiv}}

@misc{Denneau2025,
 author={Denneau, L. and Siverd, R. and Tonry, J. and others},
 title={{3I/ATLAS = C/2025 N1 (ATLAS)}}, year={2025},
 howpublished={Minor Planet Electronic Circular 2025-N12},
 doi={10.48377/MPEC/2025-N12}, url={https://minorplanetcenter.net/mpec/K25/K25N12.html}}

@article{Bolin2018,
 author={Bolin, Bryce T. and Weaver, Harold A. and Fernandez, Yanga R. and others},
 title={{APO Time Resolved Color Photometry of Highly-Elongated Interstellar Object 1I/'Oumuamua}},
 year={2018}, journal={ApJL}, volume={852}, pages={L2}, doi={10.3847/2041-8213/aaa0c9}, eprint={1711.04927}, archivePrefix={arXiv}}

@article{Bolin2020Borisov,
 author={Bolin, Bryce T. and Lisse, Carey M. and Kasliwal, Mansi M. and others},
 title={{Characterization of the Nucleus, Morphology and Activity of Interstellar Comet 2I/Borisov by Optical and Near-Infrared GROWTH, Apache Point, IRTF, ZTF and Keck Observations}},
 year={2020}, journal={AJ}, volume={160}, pages={26}, doi={10.3847/1538-3881/ab9305}, eprint={1910.14004}, archivePrefix={arXiv}}

@article{BolinLisse2020,
 author={Bolin, Bryce T. and Lisse, Carey M.},
 title={{Constraints on the Spin-Pole Orientation, Jet Morphology and Rotation of Interstellar Comet 2I/Borisov with Deep \textit{HST} Imaging}},
 year={2020}, journal={MNRAS}, volume={497}, pages={4031--4041}, doi={10.1093/mnras/staa2192}, eprint={1912.07386}, archivePrefix={arXiv}}

@article{Ye2020Borisov,
 author={Ye, Quanzhi and Kelley, Michael S. P. and Bolin, Bryce T. and others},
 title={{Pre-discovery Activity of New Interstellar Comet 2I/Borisov Beyond 5 AU}},
 year={2020}, journal={AJ}, volume={159}, pages={77}, doi={10.3847/1538-3881/ab659b}, eprint={1911.05902}, archivePrefix={arXiv}}

@article{Bolin2021LD2,
 author={Bolin, Bryce T. and Fernandez, Yanga R. and Lisse, Carey M. and others},
 title={{Initial Characterization of Active Transitioning Centaur, P/2019 LD2 (ATLAS), Using Hubble, Spitzer, ZTF, Keck, Apache Point Observatory, and GROWTH Visible and Infrared Imaging and Spectroscopy}},
 year={2021}, journal={AJ}, volume={161}, pages={116}, doi={10.3847/1538-3881/abd94b}, eprint={2011.03782}, archivePrefix={arXiv}}

@article{Bolin2025Discovery,
 author={Bolin, Bryce T. and Belyakov, Matthew and Fremling, Christoffer and others},
 title={{Interstellar comet 3I/ATLAS: discovery and physical description}},
 year={2025}, journal={MNRAS Letters}, volume={542}, pages={L139--L143}, doi={10.1093/mnrasl/slaf078}, eprint={2507.05252}, archivePrefix={arXiv}}

@article{Belyakov2025Optical,
 author={Belyakov, Matthew and Fremling, Christoffer and Graham, Matthew J. and others},
 title={{Palomar and Apache Point Spectrophotometry of Interstellar Comet 3I/ATLAS}},
 year={2025}, journal={arXiv e-prints}, eprint={2507.11720}, archivePrefix={arXiv}, url={https://arxiv.org/abs/2507.11720}}

@article{Rahatgaonkar2025,
 author={Rahatgaonkar, Rohan and Carvajal, Juan Pablo and Puzia, Thomas H. and others},
 title={{Very Large Telescope Observations of Interstellar Comet 3I/ATLAS. II. From Quiescence to Glow: Dramatic Rise of Ni I Emission and Incipient CN Outgassing at Large Heliocentric Distances}},
 year={2025}, journal={ApJL}, volume={995}, pages={L34}, doi={10.3847/2041-8213/ae1cbc}, eprint={2508.18382}, archivePrefix={arXiv}}

@article{Belyakov2026Volatiles,
 author={Belyakov, Matthew and Wong, Ian and Bolin, Bryce T. and Davis, M. Ryleigh and Bromley, Steven J. and Lisse, Carey M. and Brown, Michael E.},
 title={{The Volatile Inventory of 3I/ATLAS as Seen with \textit{JWST}/MIRI}},
 year={2026}, journal={ApJL}, volume={1001}, pages={L11}, doi={10.3847/2041-8213/ae5700}, eprint={2601.22034}, archivePrefix={arXiv}}

@article{Belyakov2026Dust,
 author={Belyakov, Matthew and Wong, Ian and Lisse, Carey M. and Davis, M. Ryleigh and Bolin, Bryce T. and Martin, Audrey and Pontoppidan, Klaus M. and Blake, Geoffrey A. and Chen, Christine and Brown, Michael E.},
 title={{\textit{JWST}/MIRI Finds ISM-like Amorphous Silicates in Interstellar Comet 3I/ATLAS}},
 year={2026}, journal={ApJL}, volume={1009}, pages={L41}, doi={10.3847/2041-8213/aea84e}, eprint={2606.27535}, archivePrefix={arXiv}, url={https://authors.library.caltech.edu/records/8fjmd-gka22}}

@article{Graykowski2026,
 author={Graykowski, Ariel and Bolin, Bryce and Abron, Laura-May and others},
 title={{Multiband Color Monitoring of 3I/ATLAS through Ground-Based Relay Observations}},
 year={2026}, journal={arXiv e-prints}, eprint={2609.22021}, archivePrefix={arXiv}, url={https://arxiv.org/abs/2609.22021}}

@article{Moreno2017,
 author={Moreno, Fernando and Pozuelos, Francisco and Novakovic, Bojan and Licandro, Javier and Cabrera-Lavers, Antonio and Bolin, Bryce and others},
 title={{The Splitting of Double-component Active Asteroid P/2016 J1 (PANSTARRS)}},
 year={2017}, journal={ApJL}, volume={837}, pages={L3}, doi={10.3847/2041-8213/aa6036}, eprint={1702.03665}, archivePrefix={arXiv}}

@article{Hui2026,
 author={Hui, Man-To and Jewitt, David and Mutchler, Max J. and Agarwal, Jessica and Kim, Yoonyoung},
 title={{Nucleus and Postperihelion Activity of Interstellar Object 3I/ATLAS Observed by the Hubble Space Telescope}},
 year={2026}, journal={ApJL}, volume={999}, pages={L37}, doi={10.3847/2041-8213/ae471c}, eprint={2601.21569}, archivePrefix={arXiv}}

@article{Yang2025,
 author={Yang, Bin and Meech, Karen J. and Connelley, Michael and Zhao, Ruining and Keane, Jacqueline V.},
 title={{Spectroscopic Characterization of Interstellar Object 3I/ATLAS: Water Ice in the Coma}},
 year={2025}, journal={ApJL}, volume={992}, pages={L9}, doi={10.3847/2041-8213/ae08a7}, eprint={2507.14916}, archivePrefix={arXiv}}

@article{Medler2026,
 author={Medler, Kyle and Hoogendam, Willem B. and Ashall, Christopher and Yang, Bin and Wray, James J. and Shappee, Benjamin J. and Meech, Karen J. and Tucker, Michael A. and Auchettl, Katie and Desai, Dhvanil D. and Hinkle, Jason T. and Hoffman, Andrew M. and Huber, Mark E. and Jones, David O. and Zhao, Ruining},
 title={{Contemporaneous Optical and Near-infrared Observations of the Interstellar Comet 3I/ATLAS Pre- and Postperihelion}},
 year={2026}, journal={\aj}, volume={172}, pages={25}, doi={10.3847/1538-3881/ae6b71}, eprint={2602.23586}, archivePrefix={arXiv}}

@article{Hutsemekers2026,
 author={Hutsemekers, Damien and Manfroid, Jean and Opitom, Cyrielle and Jehin, Emmanuel and Krishnakumar, Aravind and Fernandes, Fernando Massa and Bannister, Michele and Bodewits, Dennis and Dorsey, Rosemary and La Forgia, Fiorangela and Murphy, Brian},
 title={{Origin and evolution of NiI and FeI in the coma of the interstellar comet 3I/ATLAS throughout its trajectory}},
 year={2026}, journal={arXiv e-prints}, eprint={2605.07652}, archivePrefix={arXiv}, url={https://arxiv.org/abs/2605.07652}}

@article{Bohlin2014,
 author={Bohlin, Ralph C. and Gordon, Karl D. and Tremblay, P.-E.},
 title={{Techniques and Review of Absolute Flux Calibration from the Ultraviolet to the Mid-Infrared}},
 year={2014}, journal={PASP}, volume={126}, pages={711--732}, doi={10.1086/677655}, eprint={1406.1707}, archivePrefix={arXiv}}

@misc{ATel17363,
 author={Bolin, Bryce T. and Abron, Laura-May and Belyakov, Matthew and Carvajal, Juan Pablo and Fremling, Christoffer and Hanu{\v s}, Josef and Ingebretsen, Carl and King, Piera and Luco, Baltasar and Noll, Keith S. and Puzia, Thomas H. and Rahatgaonkar, Rohan and Silva, Karleyne and Wong, Ian},
 title={{The Anti-Solar Tail of Comet 3I/ATLAS Seen in Deep Multi-band Imaging Taken at Gemini South}}, year={2025}, howpublished={The Astronomer's Telegram, 17363}, url={https://www.astronomerstelegram.org/?read=17363}}

@misc{ATel17503,
 author={Bolin, Bryce T. and Wong, Ian and Lemaux, Brian and Abron, Laura-May and Belyakov, Matthew and Candia, Pablo and Chiboucas, Kristin and Carrasco, Rodrigo and Firpo, Veronica and Fremling, Christoffer and Hanu{\v s}, Josef and Heo, Jeong-Eun and King, Piera and Noll, Keith S. and Silva, Karleyne},
 title={{Gemini pre-perihelion detection of CN/C3/C2 gas emission and dust fans in interstellar comet 3I/ATLAS}}, year={2025}, howpublished={The Astronomer's Telegram, 17503}, url={https://www.astronomerstelegram.org/?read=17503}}

@misc{ATel17561,
 author={Bolin, Bryce T. and Wong, Ian and Abron, Laura-May and Marshall, Jameeka and Farnocchia, Davide and Micheli, Marco and Lemaux, Brian and Peavy, Emily and Figura, Erin and Hung, Denise and Belyakov, Matthew and Fremling, Christoffer and Graykowski, Ariel and Hanu{\v s}, Josef and Ivanova, Oleksandra and Kolokolova, Ludmilla and Mocnik, Teo and Noll, Keith S. and Rodriguez, Matias and Tholen, David and Veres, Peter},
 title={{Gemini post-perihelion visible photometry and spectroscopy of interstellar comet 3I/ATLAS}}, year={2025}, howpublished={The Astronomer's Telegram, 17561}, url={https://www.astronomerstelegram.org/?read=17561}}

@misc{ATel17719,
 author={Bolin, Bryce T. and Wong, Ian and Abron, Laura-May and Lemaux, Brian and Alamos, Leila and Belyakov, Matthew and Chiboucas, Kristin and Fremling, Christoffer and Hanu{\v s}, Josef and Ivanova, Oleksandra and Kim, Eunchong and Kolokolova, Ludmilla and Miller, Jen and Noll, Keith S.},
 title={{December 2025 Gemini spectroscopic and photometric constraints on the activity and gas contents of 3I/ATLAS}}, year={2026}, howpublished={The Astronomer's Telegram, 17719}, url={https://www.astronomerstelegram.org/?read=17719}}

@misc{ATel17841,
 author={Bolin, Bryce T. and Wong, Ian and Abron, Laura-May and Lemaux, Brian and Alamos, Leila and Belyakov, Matthew and Chiboucas, Kristin and Figura, Erin and Fremling, Christoffer and Hanu{\v s}, Josef and Ivanova, Oleksandra and Kim, Eunchong and Kolokolova, Ludmilla and Miller, Jen},
 title={{2026 January Gemini imaging and spectroscopic monitoring of 3I/ATLAS}}, year={2026}, howpublished={The Astronomer's Telegram, 17841}, url={https://www.astronomerstelegram.org/?read=17841}}

@misc{ATel18098,
 author={Bolin, Bryce T. and Wong, Ian and Alcorn, Leo Y. and Belyakov, Matthew and Gomez, Percy and Hershley, Heather and Lundquist, Michael},
 title={{2026 May Keck/LRIS imaging of interstellar comet 3I/ATLAS}}, year={2026}, howpublished={The Astronomer's Telegram, 18098}, url={https://www.astronomerstelegram.org/?read=18098}}

@article{Ahearn1984,
 author={A'Hearn, M. F. and Schleicher, D. G. and Millis, R. L. and Feldman, P. D. and Thompson, D. T.},
 title={{Comet Bowell 1980b}}, year={1984}, journal={AJ}, volume={89}, pages={579--591}, doi={10.1086/113552}}

@article{FinsonProbstein1968,
 author={Finson, Michael L. and Probstein, Ronald F.},
 title={{A Theory of Dust Comets. I. Model and Equations}}, year={1968}, journal={ApJ}, volume={154}, pages={327--352}, doi={10.1086/149761}}

@article{Schleicher1998,
 author={Schleicher, D. G. and Millis, R. L. and Birch, P. V.},
 title={{Narrowband Photometry of Comet P/Halley: Variation with Heliocentric Distance, Season, and Solar Phase Angle}},
 year={1998}, journal={Icarus}, volume={132}, pages={397--417}, doi={10.1006/icar.1997.5902}}

@misc{Schleicher2010HM,
 author={Schleicher, D. G.}, title={{Composite Dust Phase Function for Comets}}, year={2010}, month=may,
 howpublished={Lowell Observatory}, url={https://comet.lowell.edu/comet/dustphase/}, note={Halley--Marcus composite phase function and tabulated values; accessed 2026-10-19}}

@article{Samarasinha2014,
 author={Samarasinha, Nalin H. and Larson, Stephen M.},
 title={Image enhancement techniques for quantitative investigations of morphological features in cometary comae: A comparative study},
 journal={Icarus}, year={2014}, volume={239}, pages={168--185},
 doi={10.1016/j.icarus.2014.05.028}}

@misc{CIE2013,
 author={Samarasinha, N. H. and Martin, M. P. and Larson, S. M.},
 title={Cometary Coma Image Enhancement Facility}, year={2013},
 url={https://cie.psi.edu/}, note={Documentation accessed 2026 October 19}}

@article{Burns1979,
 author={Burns, Joseph A. and Lamy, Philippe L. and Soter, Steven},
 title={Radiation forces on small particles in the solar system},
 journal={Icarus}, year={1979}, volume={40}, pages={1--48},
 doi={10.1016/0019-1035(79)90050-2}}

@article{Willmer2018,
 author={Willmer, Christopher N. A.},
 title={{The Absolute Magnitude of the Sun in Several Filters}},
 year={2018}, journal={ApJS}, volume={236}, pages={47},
 doi={10.3847/1538-4365/aabfdf}, eprint={1804.07788}, archivePrefix={arXiv},
 url={https://arxiv.org/abs/1804.07788}}

@article{Giorgini1996,
 author={Giorgini, J. D. and Yeomans, D. K. and Chamberlin, A. B. and Chodas, P. W. and Jacobson, R. A. and Keesey, M. S. and Lieske, J. H. and Ostro, S. J. and Standish, E. M. and Wimberly, R. N.},
 title={{JPL's On-Line Solar System Data Service}},
 year={1996}, journal={Bulletin of the American Astronomical Society}, volume={28}, number={3}, pages={1158},
 url={https://ssd.jpl.nasa.gov/horizons/manual.html}, note={Horizons system reference recommended in the official JPL manual}}

@article{Ivezic2002,
  author = {{Ivezi\'c}, {\v{Z}}. and Lupton, R. H. and {Juri\'c}, M. and Tabachnik, S. and Quinn, T. and Gunn, J. E. and Knapp, G. R. and Rockosi, C. M. and Brinkmann, J.},
  title = {Color Confirmation of Asteroid Families},
  journal = {The Astronomical Journal},
  year = {2002}, volume = {124}, pages = {2943--2948},
  doi = {10.1086/344077},
  eprint = {astro-ph/0208098}, archivePrefix = {arXiv},
  url = {https://faculty.washington.edu/ivezic/Publications/asteroids3.pdf}
}

@article{Ofek2012,
  author = {Ofek, Eran O.},
  title = {Sloan Digital Sky Survey Observations of Kuiper Belt Objects: Colors and Variability},
  journal = {The Astrophysical Journal},
  year = {2012}, volume = {749}, pages = {10},
  doi = {10.1088/0004-637X/749/1/10},
  eprint = {1202.1538}, archivePrefix = {arXiv},
  url = {https://arxiv.org/abs/1202.1538}
}

@article{Solontoi2012,
  author = {Solontoi, M. and {Ivezi\'c}, {\v{Z}}. and {Juri\'c}, M. and Becker, A. C. and Jones, L. and West, A. A. and Kent, S. and Lupton, R. H. and Claire, M. and Knapp, G. R. and Quinn, T. and Gunn, J. E. and Schneider, D. P.},
  title = {Ensemble Properties of Comets in the Sloan Digital Sky Survey},
  journal = {Icarus},
  year = {2012}, volume = {218}, pages = {571--584},
  doi = {10.1016/j.icarus.2011.10.008},
  eprint = {1202.3999}, archivePrefix = {arXiv},
  url = {https://faculty.washington.edu/ivezic/Publications/Solontoi2012.pdf}
}

@article{Tonry2012,
 author={Tonry, J. L. and Stubbs, C. W. and Lykke, K. R. and Doherty, P. and Shivvers, I. S. and Burgett, W. S. and Chambers, K. C. and Hodapp, K. W. and Kaiser, N. and Kudritzki, R.-P. and Magnier, E. A. and Morgan, J. S. and Price, P. A. and Wainscoat, R. J.},
 title={{The Pan-STARRS1 Photometric System}},
 year={2012}, journal={ApJ}, volume={750}, pages={99},
 doi={10.1088/0004-637X/750/2/99},
 eprint={1203.0297}, archivePrefix={arXiv}, url={https://arxiv.org/abs/1203.0297}}

@article{Meech2017,
 author={Meech, Karen J. and Weryk, Robert and Micheli, Marco and Kleyna, Jan T. and Hainaut, Olivier R. and Jedicke, Robert and Wainscoat, Richard J. and Chambers, Kenneth C. and Keane, Jacqueline V. and Petric, Andreea and Denneau, Larry and Magnier, Eugene and Berger, Travis and Huber, Mark E. and Flewelling, Heather and Waters, Chris and Schunova-Lilly, Eva and Chastel, Serge},
 title={{A brief visit from a red and extremely elongated interstellar asteroid}},
 year={2017}, journal={Nature}, volume={552}, pages={378--381},
 doi={10.1038/nature25020}, url={https://www.nature.com/articles/nature25020}}

@article{Schleicher2010CN,
  author = {Schleicher, David G.},
  title = {The Fluorescence Efficiencies of the {CN} Violet Bands in Comets},
  journal = {The Astronomical Journal},
  year = {2010}, volume = {140}, pages = {973--984},
  doi = {10.1088/0004-6256/140/4/973}
}

@article{Bromley2026Erratum,
  author = {Bromley, S. J. and Neff, B. and Loch, S. D. and Marler, J. P. and Orsz{\'a}gh, J. and Venkataramani, K. and Bodewits, D.},
  title = {Erratum: ``Atomic Iron and Nickel in the Coma of {C/1996 B2 (Hyakutake)}: Production Rates, Emission Mechanisms, and Possible Parents'' (2021, {PSJ}, 2, 228)},
  journal = {The Planetary Science Journal}, year = {2026}, volume = {7}, eid = {131},
  doi = {10.3847/PSJ/ae706d}
}

@misc{Bromley2026NiGrid,
  author = {Bromley, Steven},
  title = {Computed Fluorescence Efficiencies of Atomic {Ni} and Atomic {Fe}},
  year = {2026}, publisher = {Zenodo},
  doi = {10.5281/zenodo.20514283}, url = {https://zenodo.org/records/20514283}
}

@article{Ahearn1995,
  author = {A'Hearn, Michael F. and Millis, Robert C. and Schleicher, David G. and Osip, David J. and Birch, Peter V.},
  title = {The Ensemble Properties of Comets: Results from Narrowband Photometry of 85 Comets, 1976--1992},
  journal = {Icarus}, year = {1995}, volume = {118}, pages = {223--270},
  doi = {10.1006/icar.1995.1190}
}

@article{Ahuja2026,
 author={{Ahuja}, Goldy and {Ganesh}, Shashikiran and {Aravind}, K and {Jehin}, Emmanuel and {Ailawadhi}, Bhavya and {Agnihotri}, Vikrant K and {Arvind}, B and {Saraf}, Pallavi and {Singh}, Alka and {Sarkar}, Deekshya R and {Chandra}, Sunil and {Halder}, Prithish and {Sahu}, Devendra and {Sivarani}, T and {Bhardwaj}, Anil},
 year={2026},
 title={{Post-perihelion photometric and spectroscopic monitoring of the interstellar comet 3I/ATLAS}},
 journal={Monthly Notices of the Royal Astronomical Society},
 volume={552},
 pages={stag1800},
 doi={10.1093/mnras/stag1800}}

@article{SantanaRos2026,
 author={{Santana-Ros}, T. and {Sergeyev}, A. and {Ivanova}, O. and {Mykhailova}, S. and {Markkanen}, J. and {Luk'yanyk}, I. and {Kwiatkowski}, T. and {Oszkiewicz}, D. and {Penttil{\"{a}}}, A. and {Erasmus}, N. and {Ergashev}, K. and {Krugly}, Y.},
 year={2026},
 title={{The Inbound Gas and Dust Evolution of Interstellar Comet 3I/ATLAS from Optical Spectroscopy and Multiband Photometry*}},
 journal={The Astrophysical Journal},
 volume={1007},
 pages={182},
 doi={10.3847/1538-4357/ae8dbb}}

@article{Beniyama2025,
 author={Beniyama, Jin}, title={Simultaneous visible spectrophotometry of interstellar object 3I/ATLAS with Seimei/TriCCS}, year={2025}, journal={\pasj}, volume={77}, pages={L71}, doi={10.1093/pasj/psaf097}, eprint={2508.08829}, archivePrefix={arXiv}}

@article{Yang2020Ice,
 author={Yang, Bin and Kelley, Michael S. P. and Meech, Karen J. and Keane, Jacqueline V. and Protopapa, Silvia and Bus, Schelte J.},
 title={Searching for water ice in the coma of interstellar object 2I/Borisov}, year={2020}, journal={\aap}, volume={634}, pages={L6}, doi={10.1051/0004-6361/201937129}, eprint={1912.05318}, archivePrefix={arXiv}}

@article{Protopapa2018Ice,
 author={Protopapa, Silvia and Kelley, Michael S. P. and Yang, Bin and Bauer, James M. and Kolokolova, Ludmilla and Woodward, Charles E. and Keane, Jacqueline V. and Sunshine, Jessica M.},
 title={Icy Grains from the Nucleus of Comet C/2013 US10 (Catalina)}, year={2018}, journal={\apjl}, volume={862}, pages={L16}, doi={10.3847/2041-8213/aad33b}, eprint={1807.08215}, archivePrefix={arXiv}}

@article{Mastrapa2008,
 author={Mastrapa, R. M. E. and Bernstein, M. P. and Sandford, S. A. and Roush, T. L. and Cruikshank, D. P. and Dalle Ore, C. M.},
 title={Optical constants of amorphous and crystalline H2O-ice in the near infrared from 1.1 to 2.6 microns}, journal={Icarus}, year={2008}, volume={197}, pages={307--320}, doi={10.1016/j.icarus.2008.04.008}}

@misc{Mastrapa2020PDS,
 author={Mastrapa, R. M. E.}, title={Optical Constants of Water Ice: 150 K Crystalline Ice}, year={2020}, publisher={NASA Planetary Data System}, doi={10.26033/c30a-es84}}

@article{Mommert2019,
 author={Mommert, Michael and Kelley, Michael S. P. and de Val-Borro, Miguel and Li, Jian-Yang and Guzman, Giannina and Sip{\H o}cz, Brigitta and {\v D}urech, Josef and Granvik, Mikael and Grundy, Will and Moskovitz, Nick and Penttil{\"a}, Antti and Samarasinha, Nalin},
 title={sbpy: A Python module for small-body planetary astronomy},
 journal={Journal of Open Source Software},year={2019},volume={4},number={38},pages={1426},doi={10.21105/joss.01426}}

@article{SantanaRos2025,
 author={Santana-Ros, T. and Ivanova, O. and Mykhailova, S. and Erasmus, N. and Kami{\'n}ski, K. and Oszkiewicz, D. and Kwiatkowski, T. and Hus{\'a}rik, M. and Ngwane, T. S. and Penttil{\"a}, A.},
 title={Temporal Evolution of the Third Interstellar Comet 3I/ATLAS: Spin, Color, Spectra and Dust Activity},
 year={2025}, journal={\aap}, volume={702}, pages={L3},
 doi={10.1051/0004-6361/202556717}, eprint={2508.00808}, archivePrefix={arXiv}, primaryClass={astro-ph.EP}}

@article{Hoogendam2025KCWI,
 author={Hoogendam, W. B. and Shappee, B. J. and Wray, J. J. and Yang, B. and Meech, K. J. and Ashall, C. and Desai, D. D. and Hart, K. and Hinkle, J. T. and Hoffman, A. and Hu, E. M. and Jones, D. O. and Medler, K. and Pfeffer, C.},
 title={{Spatial Profiles of 3I/ATLAS CN and Ni Outgassing from Keck/KCWI Integral Field Spectroscopy}},
 year={2025}, journal={arXiv e-prints}, eid={arXiv:2510.11779}, eprint={2510.11779}, archivePrefix={arXiv}, doi={10.48550/arXiv.2510.11779}}

@article{Hoogendam2026SNIFS,
 author={Hoogendam, W. B. and Kuesters, D. and Shappee, B. J. and Aldering, G. and Wray, J. J. and Yang, B. and Meech, K. J. and Tucker, M. A. and Huber, M. E. and Auchettl, K. and Angus, C. R. and Desai, D. D. and Hinkle, J. T. and Kiyokawa, J. and Paek, G. S. H. and Romagnoli, S. and Shi, J. and Syncatto, A. and Ashall, C. and Dixon, M. and Hart, K. and Hoffman, A. M. and Jones, D. O. and Medler, K. and Pfeffer, C.},
 title={{University of Hawaii 88-inch Telescope Observations of the Interstellar Comet 3I/ATLAS: Spectrophotometric Blue-Sensitive Spectral Time Series Spanning Two Months from Discovery}},
 year={2026}, journal={The Open Journal of Astrophysics}, volume={9}, doi={10.33232/001c.165099}, eprint={2512.09020}, archivePrefix={arXiv}}

@article{Hoogendam2026Post,
 author={Hoogendam, Willem B. and Jones, David O. and Yang, Bin and Shappee, Benjamin J. and Wray, James J. and Meech, Karen J. and Ashall, Christopher and Desai, Dhvanil D. and Hinkle, Jason T. and Hoffman, Andrew M. and Medler, Kyle and Pfeffer, Cameron and Zhao, Ruining},
 title={{Postperihelion Integral Field Spectroscopy of the Interstellar Comet 3I/ATLAS}},
 year={2026}, journal={\apj}, volume={1003}, pages={245}, doi={10.3847/1538-4357/ae6243}, eprint={2601.16983}, archivePrefix={arXiv}}

@article{Hinkle2026JCMT,
 author={Hinkle, Jason T. and Yang, Bin and Meech, Karen J. and Hoffman, Andrew and Shappee, Benjamin J. and Hoogendam, W. B. and Wray, James J.},
 title={{JCMT Constraints on the Early-Time HCN and CO Emission and HCN Temporal Evolution of 3I/ATLAS}},
 year={2026}, journal={The Open Journal of Astrophysics}, volume={9}, doi={10.33232/001c.163250}, eprint={2512.02106}, archivePrefix={arXiv}}

@article{Cochran2012,
 author = {Cochran, Anita L. and Barker, Edwin S. and Gray, Candace L.},
 title = {Thirty Years of Cometary Spectroscopy from McDonald Observatory},
 journal = {Icarus},
 year = {2012},
 doi = {10.1016/j.icarus.2011.12.010},
 eprint = {1112.4770},
 archivePrefix = {arXiv}
}

@article{Jewitt2019Gault,
 author={Jewitt, D. and Kim, Y. and Luu, J. and Rajagopal, J. and Kotulla, R. and Ridgway, S. and Liu, W.},year={2019},title={Episodically Active Asteroid 6478 Gault},journal={ApJL},volume={876},pages={L19},doi={10.3847/2041-8213/ab1be8}}

@article{DeLeon2020Borisov,
 author={{de Le\'on}, J. and Licandro, J. and {de la Fuente Marcos}, C. and {de la Fuente Marcos}, R. and Lara, L. M. and Moreno, F. and Pinilla-Alonso, N. and Serra-Ricart, M. and {De Pr\'a}, M. and Tozzi, G. P. and Souza-Feliciano, A. C. and Popescu, M. and Scarpa, R. and {Font Serra}, J. and Geier, S. and Lorenzi, V. and Harutyunyan, A. and Cabrera-Lavers, A.},year={2020},title={Visible and near-infrared observations of interstellar comet 2I/Borisov with the 10.4-m GTC and the 3.6-m TNG telescopes},journal={MNRAS},volume={495},pages={2053--2062},doi={10.1093/mnras/staa1190}}

@article{GravHolman2004,
 author={Grav, T. and Holman, M. J.},year={2004},title={Near-Infrared Photometry of the Irregular Satellites of Jupiter and Saturn},journal={ApJL},volume={605},pages={L141--L144},doi={10.1086/420881}}

@article{Sykes2000,
 author={Sykes, M. V. and Cutri, R. M. and Fowler, J. W. and Tholen, D. J. and Skrutskie, M. F. and Price, S. and Tedesco, E. F.},year={2000},title={The 2MASS Asteroid and Comet Survey},journal={Icarus},volume={146},pages={161--175}}

@article{Jewitt2025HST,
 author={Jewitt, David and Hui, Man-To and Mutchler, Max and Kim, Yoonyoung and Agarwal, Jessica},
 title={Hubble Space Telescope Observations of the Interstellar Interloper 3I/ATLAS},
 year={2025}, journal={ApJL}, volume={990}, pages={L2}, doi={10.3847/2041-8213/adf8d8}}

@article{JewittLuu2025Development,
 author={Jewitt, David and Luu, Jane},
 title={Preperihelion Development of Interstellar Comet 3I/ATLAS},
 year={2025}, journal={ApJL}, volume={994}, pages={L3}, doi={10.3847/2041-8213/ae1832}}

@article{Seligman2025Discovery,
 author={Seligman, Darryl Z. and Micheli, Marco and Farnocchia, Davide and others},
 title={{Discovery and Preliminary Characterization of a Third Interstellar Object: 3I/ATLAS}},
 year={2025}, journal={ApJL}, volume={989}, pages={L36},
 doi={10.3847/2041-8213/adf49a}, eprint={2507.02757}, archivePrefix={arXiv}}

@article{Ren2026Tianwen,
 author={Ren, Xin and Yan, Wei and Zhao, Ruining and Wang, Shu and Gao, Xingye and Fu, Qiang and Zhang, Qing and Yang, Bin and Hui, Man-To and Xiao, Zhiyong and Liu, Xiaodong and Li, Cunhui and Tian, Renhao and Liu, Wenguang and Wang, Dong and Liu, Shaoran and Ren, Cong and Dong, Jie and Zhu, Xinbo and Xie, Pan and Li, Jian-Yang and Geng, Yan and Liu, Jianjun},
 title={Interstellar Object 3I/ATLAS Observed from Mars by China's Tianwen-1 Spacecraft},
 year={2026}, journal={ApJL}, volume={1003}, pages={L10},
 doi={10.3847/2041-8213/ae61b3}, eprint={2603.10350}, archivePrefix={arXiv}}

@article{Liu2026DustIce,
 author={Liu, Bin and Zhang, Xiaoping and Liu, Xiaodong and Hui, Man-To},
 title={Dust Activity Evolution and Constraints on Icy Grains in the Interstellar Comet 3I/ATLAS},
 year={2026}, journal={ApJL}, volume={1008}, pages={L26},
 doi={10.3847/2041-8213/ae9a89}, eprint={2610.21206}, archivePrefix={arXiv}}

@article{Moreno2026Dust,
 author={Moreno, Fernando and Serra-Ricart, Miquel and Licandro, Javier and Guti{\'e}rrez, Pedro J. and Lara, Luisa M. and Mariblanca-Escalona, Irene and Alarcon, Miguel R.},
 title={Dust colour, phase behaviour, and Monte Carlo modelling of interstellar Comet 3I/ATLAS from 4 au Pre- to 4 au post-perihelion},
 year={2026}, journal={MNRAS}, volume={550}, pages={stag1164}, doi={10.1093/mnras/stag1164}, eprint={2606.18751}, archivePrefix={arXiv}}

@article{Ye2025Prediscovery,
 author={Ye, Quanzhi and Kelley, Michael S. P. and Hsieh, Henry H. and Bellm, Eric C. and Chen, Tracy X. and Dekany, Richard and Drake, Andrew and Groom, Steven L. and Helou, George and Kulkarni, Shrinivas R. and Prince, Thomas A. and Riddle, Reed},
 title={Prediscovery Activity of New Interstellar Object 3I/ATLAS: Rapid Brightening from 6 to 4 au},
 year={2025}, journal={ApJL}, volume={993}, pages={L31}, doi={10.3847/2041-8213/ae147b}, eprint={2509.08792}, archivePrefix={arXiv}}

@article{Chandler2026Rubin,
 author={Chandler, Colin Orion and Bernardinelli, Pedro H. and Juri{\'c}, Mario and Singh, Devanshi and Hsieh, Henry H. and Sullivan, Ian and Jones, R. Lynne and others},
 title={{NSF-DOE Vera C. Rubin Observatory Observations of Interstellar Comet 3I/ATLAS (C/2025 N1)}},
 journal={ApJL}, year={2026}, volume={1001}, pages={L35},
 doi={10.3847/2041-8213/ae4b3a}, eprint={2507.13409}, archivePrefix={arXiv}}

@article{Drahus2018Tumbling,
 author={Drahus, Micha{\l} and Guzik, Piotr and Waniak, Wac{\l}aw and Handzlik, Barbara and Kurowski, Sebastian and Xu, Siyi},
 title={{Tumbling motion of 1I/`Oumuamua and its implications for the body's distant past}},
 year={2018}, journal={Nature Astronomy}, volume={2}, pages={407--412},
 doi={10.1038/s41550-018-0440-1}, eprint={1712.00437}, archivePrefix={arXiv}}

@article{JewittLuu2019Borisov,
 author={Jewitt, David and Luu, Jane},
 title={{Initial Characterization of Interstellar Comet 2I/2019 Q4 (Borisov)}},
 year={2019}, journal={ApJL}, volume={886}, pages={L29},
 doi={10.3847/2041-8213/ab530b}, eprint={1910.02547}, archivePrefix={arXiv}}

@article{Guzik2020Borisov,
 author={Guzik, Piotr and Drahus, Micha{\l} and Rusek, Krzysztof and Waniak, Wac{\l}aw and Cannizzaro, Giacomo and Pastor-Marazuela, In{\'e}s},
 title={{Initial characterization of interstellar comet 2I/Borisov}},
 year={2020}, journal={Nature Astronomy}, volume={4}, pages={53--57},
 doi={10.1038/s41550-019-0931-8}, eprint={1909.05851}, archivePrefix={arXiv}}

@article{Fitzsimmons2018Oumuamua,
 author={Fitzsimmons, Alan and Snodgrass, Colin and Rozitis, Ben and Yang, Bin and Hyland, Meabh and Seccull, Tom and Bannister, Michele T. and Fraser, Wesley C. and Jedicke, Robert and Lacerda, Pedro},
 title={{Spectroscopy and thermal modelling of the first interstellar object 1I/2017 U1 `Oumuamua}},
 year={2018}, journal={Nature Astronomy}, volume={2}, pages={133--137}, doi={10.1038/s41550-017-0361-4}, eprint={1712.06552}, archivePrefix={arXiv}}

@article{Micheli2018Oumuamua,
 author={Micheli, Marco and Farnocchia, Davide and Meech, Karen J. and others},
 title={{Non-gravitational acceleration in the trajectory of 1I/2017 U1 (`Oumuamua)}},
 year={2018}, journal={Nature}, volume={559}, pages={223--226}, doi={10.1038/s41586-018-0254-4}}

@article{Jewitt2020BorisovNucleus,
 author={Jewitt, David and Hui, Man-To and Kim, Yoonyoung and Mutchler, Max and Weaver, Harold and Agarwal, Jessica},
 title={{The Nucleus of Interstellar Comet 2I/Borisov}},
 year={2020}, journal={ApJL}, volume={888}, pages={L23}, doi={10.3847/2041-8213/ab621b}, eprint={1912.05422}, archivePrefix={arXiv}}

@article{Bodewits2020BorisovCO,
 author={Bodewits, Dennis and Noonan, John W. and Feldman, Paul D. and Bannister, Michele T. and Farnocchia, Davide and Harris, Walter M. and Li, Jian-Yang and Mandt, Kathleen E. and Parker, Joel Wm. and Xing, Zexi},
 title={{The carbon monoxide-rich interstellar comet 2I/Borisov}},
 year={2020}, journal={Nature Astronomy}, volume={4}, pages={867--871}, doi={10.1038/s41550-020-1095-2}, eprint={2004.08972}, archivePrefix={arXiv}}

@article{Cordiner2020BorisovCO,
 author={Cordiner, Martin A. and Milam, Stefanie N. and Biver, Nicolas and Bockel{\'e}e-Morvan, Dominique and Roth, Nathan X. and others},
 title={{Unusually high CO abundance of the first active interstellar comet}},
 year={2020}, journal={Nature Astronomy}, volume={4}, pages={861--866}, doi={10.1038/s41550-020-1087-2}, eprint={2004.09586}, archivePrefix={arXiv}}

@article{Yang2021BorisovPebbles,
 author={Yang, Bin and Li, Aigen and Cordiner, Martin A. and Chang, Chin-Shin and Hainaut, Olivier R. and Williams, Jonathan P. and Meech, Karen J. and Keane, Jacqueline V. and Villard, Eric},
 title={{Compact pebbles and the evolution of volatiles in the interstellar comet 2I/Borisov}},
 year={2021}, journal={Nature Astronomy}, volume={5}, pages={586--593}, doi={10.1038/s41550-021-01336-w}, eprint={2104.13545}, archivePrefix={arXiv}}

@article{GuzikDrahus2021Ni,
 author={Guzik, Piotr and Drahus, Micha{\l}},
 title={{Gaseous atomic nickel in the coma of interstellar comet 2I/Borisov}},
 year={2021}, journal={Nature}, volume={593}, pages={375--378}, doi={10.1038/s41586-021-03485-4}, eprint={2105.09305}, archivePrefix={arXiv}}

@article{Hainaut2012,
 author={Hainaut, O. R. and Boehnhardt, H. and Protopapa, S.},
 title={Colours of minor bodies in the outer solar system. II. A statistical analysis revisited},
 journal={Astronomy and Astrophysics}, year={2012}, volume={546}, pages={A115},
 doi={10.1051/0004-6361/201219566}
}

@article{Raymond2018Ejection,
 author={Raymond, Sean N. and Armitage, Philip J. and Veras, Dimitri and Quintana, Elisa V. and Barclay, Thomas},
 title={Implications of the interstellar object 1I/'Oumuamua for planetary dynamics and planetesimal formation},
 journal={Monthly Notices of the Royal Astronomical Society}, year={2018}, volume={476}, pages={3031--3038}, doi={10.1093/mnras/sty468}
}

@article{MoroMartin2018Origin,
 author={{Moro-Mart\'in}, Amaya},
 title={Origin of 1I/'Oumuamua. I. An Ejected Protoplanetary Disk Object?},
 journal={The Astrophysical Journal}, year={2018}, volume={866}, pages={131}, doi={10.3847/1538-4357/aadf34}, eprint={1810.02148}, archivePrefix={arXiv}
}

@article{Rawat2026Methane,
 author={{Rawat}, Akshat and {Acharyya}, Kinsuk},
 year={2026},
 title={{Delayed Methane Detection in 3I/ATLAS: Galactic Cosmic Ray--Driven Subsurface Stratification and Interstellar Exposure Age Constraints}},
 journal={The Astrophysical Journal},
 volume={1010},
 pages={106},
 doi={10.3847/1538-4357/aea83c}}

@article{Rice2026Ejection,
 author={{Rice}, Malena and {DeRocco}, William and {Raymond}, Sean N.},
 year={2026},
 title={{The Dynamics of Planetary Ejection}},
 journal={Publications of the Astronomical Society of the Pacific},
 volume={138},
 pages={082001},
 doi={10.1088/1538-3873/ae9343}}

@article{Mura2026Profiles,
 author={{Mura}, A. C. and {Kawakita}, H. and {La Forgia}, F. and {Cremonese}, G. and {Lazzarin}, M. and {Cambianica}, P. and {Farina}, A. and {Kobayashi}, H. and {Manzini}, F. and {Munaretto}, G. and {Ochner}, P. and {Shinnaka}, Y. and {Tsujimoto}, K.},
 year={2026},
 title={{Spatial profiles and scale lengths of nickel, iron, and CN on the interstellar comet 3I/ATLAS}},
 journal={Astronomy \& Astrophysics},
 volume={712},
 pages={A9},
 doi={10.1051/0004-6361/202660649}}

@article{Lee2026Mass,
 author={{Lee}, Eve J. and {DeRocco}, William and {Hadden}, Sam and {Gaudi}, B. Scott},
 year={2026},
 title={{The Persistent Missing Mass Problem in Planet Formation}},
 journal={The Astrophysical Journal Letters},
 volume={1005},
 pages={L18},
 doi={10.3847/2041-8213/ae796f}}

@article{Voitko2026V2,
 author={Voitko, Anhelina and Markkanen, Johannes and Ivanova, Oleksandra and Skorov, Yuri and Kleshchonok, Valerii and Shubina, Olena and Luk'yanyk, Igor and Humes, Oriel},
 year={2026},
 title={{Activity and dust properties of comet C/2020 V2 (ZTF)}},
 journal={Monthly Notices of the Royal Astronomical Society},
 volume={549},
 pages={stag967},
 doi={10.1093/mnras/stag967}}

@article{Yunerman2026Disks,
 author={{Yunerman}, Elizabeth S. and {Price}, Ellen M. and {{\"{O}}berg}, Karin I.},
 year={2026},
 title={{Icy Volatile Enhancements in Evolving Protoplanetary Disks}},
 journal={The Astrophysical Journal},
 volume={1004},
 pages={63},
 doi={10.3847/1538-4357/ae60fe}}

@article{Hartman2026HATPI,
 author={{Hartman}, Joel D. and {Bakos}, G{\'{a}}sp{\'{a}}r {\'{A}}. and {Jord{\'{a}}n}, Andr{\'{e}}s and {Thiele}, Sarah and {Csubry}, Zolt{\'{a}}n and {Talens}, Geert Jan and {B{\'{o}}di}, Attila and {Pigai}, S{\'{a}}ndor and {Domsa}, Istv{\'{a}}n and {Keyes}, Anthony and {Suc}, Vincent and {Gaitan}, Adriana and {Thibault}, Antoine},
 year={2026},
 title={{HATPI Preperihelion Time-series Photometry of the Interstellar Comet 3I/ATLAS}},
 journal={The Astronomical Journal},
 volume={171},
 pages={270},
 doi={10.3847/1538-3881/ae4a99}}

@article{Kawakita2026Spectra,
 author={{Kawakita}, Hideyo and {Tsujimoto}, Ko and {Shinnaka}, Yoshiharu and {Kobayashi}, Hitomi and {Ootsubo}, Takafumi and {Watanabe}, Jun-ichi},
 year={2026},
 title={{Postperihelion Optical Low-resolution Spectroscopy of Interstellar Comet 3I/ATLAS}},
 journal={The Astrophysical Journal Letters},
 volume={1000},
 pages={L60},
 doi={10.3847/2041-8213/ae4c5b}}

@article{Wesolowski2026Photometry,
 author={{Weso{\l}owski}, Marcin and {Ryske}, Jorma and {Carson}, Peter and {Gritsevich}, Maria and {{(PASP Editorial Office)}}},
 year={2026},
 title={{Photometric Analysis of the Interstellar Comet 3I/ATLAS During its Pre-perihelion Period}},
 journal={Publications of the Astronomical Society of the Pacific},
 volume={138},
 pages={024402},
 doi={10.1088/1538-3873/ae3eb5}}

@article{Lazzarin2026CN,
 author={{Lazzarin}, M. and {Mura}, A. C. and {La Forgia}, F. and {Cremonese}, G. and {Cambianica}, P. and {Munaretto}, G. and {Farina}, A. and {Mazzotta Epifani}, E. and {Ieva}, S. and {Dotto}, E.},
 year={2026},
 title={{Preperihelion CN Production Rate of Interstellar Comet 3I/ATLAS}},
 journal={The Astrophysical Journal Letters},
 volume={998},
 pages={L30},
 doi={10.3847/2041-8213/ae3dd7}}

@article{Maggiolo2025GCR,
 author={{Maggiolo}, Romain and {Dhooghe}, Frederik and {Gronoff}, Guillaume P. and {de Keyser}, Johan and {Cessateur}, Ga{\"{e}}l},
 year={2026},
 title={{Interstellar Comet 3I/ATLAS: Evidence for Galactic Cosmic-Ray Processing}},
 journal={The Astrophysical Journal Letters},
 volume={996},
 pages={L34},
 doi={10.3847/2041-8213/ae2fff}}

@article{SerraRicart2025Jet,
 author={{Serra-Ricart}, M. and {Licandro}, J. and {Alarcon}, M. R.},
 year={2025},
 title={{Pre-perihelion detection of a wobbling high-latitude jet in the interstellar comet 3I/ATLAS}},
 journal={Astronomy \& Astrophysics},
 volume={705},
 pages={L3},
 doi={10.1051/0004-6361/202558072}}

@article{MartinezPalomera2025TESS,
 author={{Martinez-Palomera}, Jorge and {Tuson}, Amy and {Hedges}, Christina and {Dotson}, Jessie and {Barclay}, Thomas and {Powell}, Brian},
 year={2025},
 title={{Prediscovery TESS Observations of Interstellar Object 3I/ATLAS}},
 journal={The Astrophysical Journal Letters},
 volume={994},
 pages={L51},
 doi={10.3847/2041-8213/ae1f91}}

@article{Opitom2025MUSE,
 author={{Opitom}, Cyrielle and {Snodgrass}, Colin and {Jehin}, Emmanuel and {Bannister}, Michele T and {Bufanda}, Erica and {Deam}, Sophie E and {Dorsey}, Rosemary C and {Ferrais}, Marin and {Hmiddouch}, Said and {Knight}, Matthew M and {Kokotanekova}, Rosita and {Leicester}, Brayden and {Marsset}, Micha{\"{e}}l and {Murphy}, Brian and {Okoth}, Vincent and {Ridden-Harper}, Ryan and {Vander Donckt}, Mathieu and {Ferellec}, L{\'{e}}a and {Hutsem{\'{e}}kers}, Damien and {Lippi}, Manuela and {Manfroid}, Jean and {Benkhaldoun}, Zouhair},
 year={2025},
 title={{Snapshot of a new interstellar comet: 3I/ATLAS has a red and featureless spectrum}},
 journal={Monthly Notices of the Royal Astronomical Society: Letters},
 volume={544},
 pages={L31-L36},
 doi={10.1093/mnrasl/slaf095}}

@article{Sahu2025Icy,
 author={{Sahu}, Snehalata and {G{\"{a}}nsicke}, Boris T and {Williams}, Jamie T and {Koester}, Detlev G and {Farihi}, Jay and {Desch}, Steven J and {Gentile Fusillo}, Nicola Pietro and {Veras}, Dimitri and {Raymond}, Sean N and {Belmonte}, Maria Teresa},
 year={2025},
 title={{Discovery of an icy and nitrogen-rich extrasolar planetesimal}},
 journal={Monthly Notices of the Royal Astronomical Society},
 volume={543},
 pages={223-232},
 doi={10.1093/mnras/staf1424}}

@article{Gray2025Polarization,
 author={{Gray}, Zuri and {Bagnulo}, Stefano and {Borisov}, Galin and {Kwon}, Yuna G. and {Cellino}, Alberto and {Kolokolova}, Ludmilla and {Dorsey}, Rosemary C. and {Fedorets}, Grigori and {Granvik}, Mikael and {MacLennan}, Eric and {Mu{\~{n}}oz}, Olga and {Bendjoya}, Philippe and {Devog{\`{e}}le}, Maxime and {Ieva}, Simone and {Penttil{\"{a}}}, Antti and {Muinonen}, Karri},
 year={2025},
 title={{Extreme Negative Polarization of New Interstellar Comet 3I/ATLAS}},
 journal={The Astrophysical Journal Letters},
 volume={992},
 pages={L29},
 doi={10.3847/2041-8213/ae0c08}}

@article{Kareta2025IRTF,
 author={{Kareta}, Theodore and {Champagne}, Chansey and {McClure}, Lucas and {Emery}, Joshua and {Sharkey}, Benjamin N. L. and {Bauer}, James and {Connelley}, Michael S. and {Rayner}, John and {Thomas}, Cristina A. and {Reddy}, Vishnu and {Firgard}, Megan},
 year={2025},
 title={{Near-discovery Observations of Interstellar Comet 3I/ATLAS with the NASA Infrared Telescope Facility}},
 journal={The Astrophysical Journal Letters},
 volume={990},
 pages={L65},
 doi={10.3847/2041-8213/adfbdf}}

@article{Puzia2025SOAR,
 author={{Puzia}, Thomas H. and {Rahatgaonkar}, Rohan and {Carvajal}, Juan Pablo and {Nayak}, Prasanta K. and {Luco}, Baltasar},
 year={2025},
 title={{Spectral Characteristics of Interstellar Object 3I/ATLAS from SOAR Observations*}},
 journal={The Astrophysical Journal Letters},
 volume={990},
 pages={L27},
 doi={10.3847/2041-8213/adfa0b}}

@article{AlvarezCandal2025XSHOOTER,
 author={{Alvarez-Candal}, A. and {Rizos}, J. L. and {Lara}, L. M. and {Santos-Sanz}, P. and {Gutierrez}, P. J. and {Ortiz}, J. L. and {Morales}, N.},
 year={2025},
 title={{X-SHOOTER spectrum of comet 3I/ATLAS: Insights into a distant interstellar visitor}},
 journal={Astronomy \& Astrophysics},
 volume={700},
 pages={L10},
 doi={10.1051/0004-6361/202556338}}

@article{FuenteMarcos2025GTC,
 author={{de la Fuente Marcos}, R. and {Alarcon}, M. R. and {Licandro}, J. and {Serra-Ricart}, M. and {de Le{\'{o}}n}, J. and {de la Fuente Marcos}, C. and {Lombardi}, G. and {Tejero}, A. and {Cabrera-Lavers}, A. and {Guerra Arencibia}, S. and {Ruiz Cejudo}, I.},
 year={2025},
 title={{Assessing interstellar comet 3I/ATLAS with the 10.4 m Gran Telescopio Canarias and the Two-meter Twin Telescope}},
 journal={Astronomy \& Astrophysics},
 volume={700},
 pages={L9},
 doi={10.1051/0004-6361/202556439}}

@article{Morate2021,
 author={{Morate}, David and {Marcio Carvano}, Jorge and {Alvarez-Candal}, Alvaro and {De Pr{\'{a}}}, M{\'{a}}rio and {Licandro}, Javier and {Galarza}, Andr{\'{e}}s and {Mahlke}, Max and {Solano-M{\'{a}}rquez}, Enrique and {Cenarro}, Javier and {Crist{\'{o}}bal-Hornillos}, David and {Hern{\'{a}}ndez-Monteagudo}, Carlos and {L{\'{o}}pez-Sanjuan}, Carlos and {Mar{\'{i}}n-Franch}, Antonio and {Moles}, Mariano and {Varela}, Jes{\'{u}}s and {V{\'{a}}zquez Rami{\'{o}}}, H{\'{e}}ctor and {Alcaniz}, Jailson and {Dupke}, Renato and {Ederoclite}, Alessandro and {Sodr{\'{e}}}, Laerte and {Angulo}, Raul E. and {Jim{\'{e}}nez-Esteban}, Francisco M. and {Siffert}, Beatriz B. and {{J-PLUS Collaboration}}},
 year={2021},
 title={{J-PLUS: A first glimpse at the spectrophotometry of asteroids}},
 journal={Astronomy \& Astrophysics},
 volume={655},
 pages={A47},
 doi={10.1051/0004-6361/202038477}}

@article{Popescu2016,
 author={{Popescu}, M. and {Licandro}, J. and {Morate}, D. and {de Le{\'{o}}n}, J. and {Nedelcu}, D. A. and {Rebolo}, R. and {McMahon}, R. G. and {Gonzalez-Solares}, E. and {Irwin}, M.},
 year={2016},
 title={{Near-infrared colors of minor planets recovered from VISTA-VHS survey (MOVIS)}},
 journal={Astronomy \& Astrophysics},
 volume={591},
 pages={A115},
 doi={10.1051/0004-6361/201628163}}

@article{Popescu2018,
 author={{Popescu}, M. and {Licandro}, J. and {Carvano}, J. M. and {Stoicescu}, R. and {de Le{\'{o}}n}, J. and {Morate}, D. and {Boacă}, I. L. and {Cristescu}, C. P.},
 year={2018},
 title={{Taxonomic classification of asteroids based on MOVIS near-infrared colors}},
 journal={Astronomy \& Astrophysics},
 volume={617},
 pages={A12},
 doi={10.1051/0004-6361/201833023}}

@article{Cardelli1989,
 author={{Cardelli}, Jason A. and {Clayton}, Geoffrey C. and {Mathis}, John S.},
 year={1989},
 title={{The relationship between infrared, optical, and ultraviolet extinction}},
 journal={The Astrophysical Journal},
 volume={345},
 pages={245},
 doi={10.1086/167900}}

@inproceedings{CastelliKurucz2003,
 author={Castelli, F. and Kurucz, R. L.},year={2003},
 title={{New Grids of ATLAS9 Model Atmospheres}},
 booktitle={Modelling of Stellar Atmospheres},series={IAU Symposium},volume={210},pages={A20},
 editor={Piskunov, N. and Weiss, W. W. and Gray, D. F.},
 eprint={astro-ph/0405087},archivePrefix={arXiv}}

@article{Bolin2025KY26,
 author={Bolin, Bryce T. and Fremling, Christoffer and Belyakov, Matthew and others},
 title={{Keck and Gemini Characterization of Hayabusa2\# Rendezvous Target 1998 KY$_{26}$}},
 year={2025}, journal={AJ}, volume={169}, pages={303},
 doi={10.3847/1538-3881/adccbe}, eprint={2501.17156}, archivePrefix={arXiv}}

@article{Bolin2023Trojans,
 author={Bolin, B. T. and Fremling, C. and Morbidelli, A. and others},
 title={{Keck, Gemini, and Palomar 200-inch visible photometry of red and very-red Neptunian Trojans}},
 year={2023}, journal={MNRAS}, volume={521}, pages={L29--L33},
 doi={10.1093/mnrasl/slad018}, eprint={2302.04280}, archivePrefix={arXiv}}

@article{Bolin2025YR4,
 author={Bolin, Bryce T. and Hanu\v{s}, Josef and Denneau, Larry and others},
 title={{The Discovery and Characterization of Earth-crossing Asteroid 2024 YR$_4$}},
 year={2025}, journal={ApJL}, volume={984}, pages={L25},
 doi={10.3847/2041-8213/adc910}, eprint={2503.05694}, archivePrefix={arXiv}}

@article{Bolin2025PT5,
 author={Bolin, Bryce T. and Denneau, Larry and Abron, Laura-May and others},
 title={{The Discovery and Characterization of Minimoon 2024 PT$_5$}},
 year={2025}, journal={ApJL}, volume={978}, pages={L37},
 doi={10.3847/2041-8213/ada1d0}, eprint={2411.08029}, archivePrefix={arXiv}}

@article{Beniyama2025PT5,
 author={Beniyama, Jin and Bolin, Bryce T. and Sergeyev, Alexey V. and others},
 title={{Multi-epoch spectro-photometric characterization of the minimoon 2024 PT$_5$ in the visible and near-infrared}},
 year={2025}, journal={A\&A}, volume={700}, pages={A183},
 doi={10.1051/0004-6361/202555633}, eprint={2507.10527}, archivePrefix={arXiv}}

@article{Tonry3IATLASPhotometry2025,
 author={Tonry, John L. and Denneau, Larry and Alarc{\'o}n, Miguel R. and Clocchiatti, Alejandro and Erasmus, Nicolas and Fitzsimmons, Alan and Licandro, Javier and Meech, Karen J. and Siverd, Robert and Weiland, Henry},
 title={{ATLAS Photometry of Interstellar Object 3I/ATLAS}},
 journal={The Astrophysical Journal Letters},
 year={2025},
 volume={995},
 pages={L15},
 doi={10.3847/2041-8213/ae1f12}
}

@article{PaekCNComa2026,
 author={Paek, Gregory S. H. and Im, Myungshin and Jeong, Mankeun and Choi, Hyeonho and Bach, Yoonsoo P. and Ishiguro, Masateru and Lim, Bumhoo and Chang, Seo-Won and Kim, Ji Hoon and Geem, Jooyeon and Hoogendam, Willem B.},
 title={{Pre-perihelion Emergence of the CN Gas Coma in 3I/ATLAS Temporally and Spatially Resolved by the 7-dimensional Telescope}},
 year={2026}, journal={\apj}, volume={1000}, pages={53},
 doi={10.3847/1538-4357/ae47f0}, eprint={2602.12930}, archivePrefix={arXiv}
}

@article{JewittComaProfiles1987,
 author={Jewitt, David C. and Meech, Karen J.},
 title={{Surface Brightness Profiles of 10 Comets}},
 journal={ApJ}, year={1987}, volume={317}, pages={992--1001},
 doi={10.1086/165347}}

@incollection{FitzsimmonsInterstellarObjects2024,
 author = {Fitzsimmons, Alan and Meech, Karen J. and Matr{\`a}, Luca and Pfalzner, Susanne},
 title = {Interstellar Objects and Exocomets},
 booktitle = {Comets III},
 editor = {Meech, Karen J. and Combi, Michael R. and Bockel{\'e}e-Morvan, Dominique and Raymond, Sean N. and Zolensky, Michael},
 publisher = {University of Arizona Press},
 year = {2024},
 pages = {731--766},
 eprint = {2303.17980},
 archivePrefix = {arXiv},
 primaryClass = {astro-ph.EP},
 url = {https://pure.qub.ac.uk/en/publications/interstellar-objects-and-exocomets/}
}

@article{PerleyLPipe2019,
  author = {Perley, Daniel A.},
  title = {Fully Automated Reduction of Longslit Spectroscopy with the Low Resolution Imaging Spectrometer at the Keck Observatory},
  journal = {Publications of the Astronomical Society of the Pacific},
  year = {2019},
  volume = {131},
  number = {1002},
  pages = {084503},
  doi = {10.1088/1538-3873/ab215d},
  eprint = {1903.07629},
  archivePrefix = {arXiv},
  primaryClass = {astro-ph.IM}
}

@article{JewittCometColors1986,
 author={Jewitt, David and Meech, Karen J.},
 title={{Cometary Grain Scattering versus Wavelength, or, ``What Color Is Comet Dust?''}},
 year={1986}, journal={ApJ}, volume={310}, pages={937--952},
 doi={10.1086/164745}}

@article{JewittCometColors1988,
 author={Jewitt, David and Meech, Karen J.},
 title={{The Absence of a Color-Distance Trend in Comets}},
 year={1988}, journal={AJ}, volume={96}, pages={1723--1730},
 doi={10.1086/114924}}

@article{HannerCometInfrared1984,
 author={Hanner, M. S. and Tokunaga, A. T. and Veeder, G. J. and A'Hearn, M. F.},
 title={{Infrared Photometry of the Dust in Comets}},
 year={1984}, journal={AJ}, volume={89}, pages={162--169}, doi={10.1086/113495},
 url={https://adsabs.harvard.edu/pdf/1984AJ.....89..162H}}

@article{PicazzioMcNaughtNIR2010,
 author={Picazzio, Enos and Figueredo, Elysandra and {de Almeida}, Amaury Augusto and {Mendes de Oliveira}, Claudia and Churyumov, Klim Ivanovich},
 title={{Near Infrared Photometry of Comet C/2005 E2 (McNaught)}},
 year={2010}, journal={Proceedings of the International Astronomical Union},
 volume={5}, number={S263}, pages={285--288},
 doi={10.1017/S1743921310001948}}

@article{HainautMBOSSColors2002,
 author={Hainaut, O. R. and Delsanti, A. C.},
 title={{Colors of Minor Bodies in the Outer Solar System: A Statistical Analysis}},
 year={2002}, journal={A\&A}, volume={389}, pages={641--664},
 doi={10.1051/0004-6361:20020431}}
